\documentclass[]{interact}
\usepackage{epstopdf}% To incorporate .eps illustrations using PDFLaTeX, etc.
\usepackage[caption=false]{subfig}% Support for small, `sub' figures and tables

\usepackage[numbers,sort&compress]{natbib}% Citation support using natbib.sty
\bibpunct[, ]{[}{]}{,}{n}{,}{,}% Citation support using natbib.sty
\makeatletter% @ becomes a letter
\def\NAT@def@citea{\def\@citea{\NAT@separator}}% Suppress spaces between citations using natbib.sty
\makeatother% @ becomes a symbol again

\usepackage{color}
\usepackage{amsmath}
\usepackage{graphicx}
\usepackage{booktabs}
\usepackage{subfig}
\usepackage{miller}
\usepackage{placeins}
\usepackage{bibunits}
\usepackage{soul}
\usepackage[utf8]{inputenc}

\usepackage{geometry}
\usepackage{multicol}
\usepackage{changepage}
\usepackage{pdflscape}
\usepackage{xspace}
\usepackage{array}
\usepackage{float}
\newcommand{\angstrom}{\text{\normalfont\AA}}

\newcommand{\nhv}{n\textsubscript{H/Vac}}

\begin{document}
\begin{bibunit}[ieeetr]

\title{Comparative study of H accumulation in differently oriented grains of Cu}

\author{
\name{
A. Lopez-Cazalilla\textsuperscript{a}\thanks{CONTACT A. Lopez-Cazalilla. Email: alvaro.lopezcazalilla@helsinki.fi}, Catarina Serafim\textsuperscript{a,b}
F. Djurabekova\textsuperscript{a}, Ana Teresa Perez-Fontenla\textsuperscript{b}, Sergio Calatroni\textsuperscript{b}, Walter Wuensch\textsuperscript{b}}
\affil{\textsuperscript{a}Department of Physics, P.O. Box 43, FI-00014 University of Helsinki, Finland;
}\affil{\textsuperscript{b}CERN, European Organization for Nuclear Research, 1211 Geneva, Switzerland;
}}

\maketitle

\begin{abstract}
When metal surfaces are exposed to hydrogen ion irradiation, the light ions are expected to penetrate deep into the material and dissolve in the matrix. However, these atoms are seen to cause significant modification of surfaces, indicating that they accumulate in vicinity of the surface. The process known as blistering may reduces the vacuum dielectric strength above the metal surface, which shows a dense population of surface blisters. 

In this paper, we investigate how a bubble can grow under the pressure exerted by hydrogen atoms on the walls of the bubble and how this affect to the surface of Cu, whether an external electric field is applied or not.
\end{abstract}

\begin{keywords}
Copper; Surface; Blisters; Hydrogen; Molecular dynamics;
\end{keywords}
%%%%%%%%%%%%%%%%%%%% After discussion%%%%%%%%%%%%%%%%%%%%%
%%% Experiments show that the blister accumulation is taking place in the (100) grains, but not on the rest. Apparently the fluences are similar, but we need to verify that the fluxes are similar than in the previous experiments. Anyway, the fact that are not created at the same fluence in ones and not in other is rather interesting. We can explore the possibility of channeling, for that we can run our in-house calculations (by Kai). But there is some bibliography on the topic:
%%% Sergio suggested me this reference some time ago: "Moreno, D., Eliezer, D. On the blister formation in copper alloys due to the helium ion implantation. Metall Mater Trans A 28, 755–762 (1997). https://doi.org/10.1007/s11661-997-0062-1". Here, we see how, for He not H, the penetration of He and the formation of blisters differ on the crystallographic orientation. Also at higher energies than 10 keV-He, no blister is observed in any of the orientations. But focusing on the the 10 keV-He case, the results show that for (110) surface there is no blister formation at all, and the results of 100 and 111 differ between them in the size of the blisters, 

\section{Introduction}
\label{sec:introd}

The mechanisms behind the blistering process has been deeply studied in many materials due to its importance in the proper functioning of the devices which surfaces are exposed to hydrogen and noble gases \cite{Hblisteringreview}. These light species accumulate in the lattice defects as well as in surfaces and grain boundaries \cite{Bai10, Smi15, Yan15}. It has been shown that for hydrogen, the vacancies created in the bulk can accommodate more than one implanted ion \cite{Ste11,ganchenkova2014,Korzhavyi2014} when the surfaces are irradiated with it. The hydrogen that is accumulated leads to an increase of the size of the empty regions and more ions can occupy this space \cite{Hen05b}, forming bubbles approximately 1.5 nm \cite{thomson2015} to larger sizes. These defects are eventually observed in the exposed to irradiation surface, and are known as "blisters" \cite{Wan01,All20,Joh82,nakahara1989,taskaev2018}. 

Blistering is observed in tungsten, which is the most promising plasma-facing material for the future nuclear fusion power plants due to its high melting point, good thermal conductivity, low thermal expansion, significant strength at elevated temperatures and also its high sputtering threshold energy~\cite{barabash1999,nygren2011,alvarez2011,kaufmann2007}. However, it is still susceptible to suffer from it~\cite{Wan01,causey2001,Hen05b,thomson2015}. Blistering has been also observed in a number of materials such as V \cite{pesch1980}, Nb \cite{schober1993}, carbon steel \cite{ju1985,iino1978,ravi1990}, Fe \cite{lee1987,lee1985,lee1989}, Al\cite{kamada1988,kamada1989, milacek1968a,milacek1968b,daniels1971,xie2015,sznajder2018} and Cu \cite{Joh82,nakahara1989,nakahara1991,taskaev2018,fukui1995,robinson2017}.

Moreover, the disruption of operation in accelerating devices such as the future Compact Linear Collider (CLIC) \cite{CLIC} due to breakdowns \cite{barengolts2018,kyritsakis2018}, damaging the surface, is a bottleneck in the functioning of these systems. Norlund et al. suggested \cite{Nord2012} and others further developed~\cite{Engelberg2018} that there is a correlation of the dependence of the breakdown rate (number of breakdowns per pulse) on the accelerating field gradient in radio-frequency (RFQ) accelerating Cu structures. A similar correlation was established in the direct-current (DC) experiments with different materials \cite{descoeudres2009,Des10}, where the materials appeared to rank according to their crystallographic lattice structure. Despite many materials showed much higher minimal breakdown field, Cu was select as the functional material for the CLIC \cite{clicreport2018} due to its abundance, good conductivity and mechanical properties \cite{descoeudres2009}.

Previously, it was shown that the accumulation of implanted ions follows to pressurize gas as a consequence of rising density, then it leads to plastic deformation via bubble growth emitting dislocations. The emission of dislocations is due to exerted pressure by the gas to bubble walls \cite{Lop21}. It is also known that continuous plastic deformation in metals changes their properties inducing hardening, hindering dislocation activity. On the other hand, blistering is seen to grow continuously with the increase of the ion beam fluence \cite{nakahara1989,nakahara1991,taskaev2018}. However, we know that this is eventually noticed in the surface, and the dependence on the crystallographic orientation arises. However, the role of the grains and their orientation were not discussed. The grain orientation and its position from the surface, are determinant to understand the response of it under shear stress \cite{SUN2021116474}, according to Sun et al. They found that for \hkl[100]-oriented grains, at a given strain, the dislocation system behaves similarly in the surface or bulk, however for the \hkl[110] and \hkl[111] within a certain range, do behave differently whether is a bulk or surface grain. In Ref. \cite{hansen2011}, they deeply studied the relationship between the orientation, size and type of strain under the grain is to determine its response.  
%%%% In Ref. \cite{LI2022143082} they claim that the influence of H in the dislocation network is not so relevant in coarse-grained Cu.

%%%% https://iopscience.iop.org/article/10.1088/1361-6587/ac36e6, for the Stress-driven surface swell and exfoliation of copper as the plasma-facing materials in NBI ICP source

Computational studies on the effect of the stress on copper has been developed in different contexts. In Cu, Molecular dynamics (MD) has been used to study mechanical properties\cite{heino1998} for the different crystallographic orientations; also the effect of several sizes and shapes of voids under tensile stress \cite{wang2018,Lin2012} and triaxial stress \cite{YANG2016917,seppala2004,seppala2005} considering the void coalescence. However, the combined effect of bubbles (H in this case) and different surface orientations has not been studied using MD. 

In a previous work \cite{Lop21}, it was proved the emission of prismatic loops under hydrostratic stress from a pressurized bubble, showing that a number of shear loops surrounding the bubble can eventually form a prismatic loop, which is emitted not necessarily smaller than the void cross-section. Following these findings, we propose a new setup of simulations for reveal the mechanisms behind the surface modification under different surface orientations over a H bubble, considering the interaction of dislocations with the close Cu surface. 

%Moreover, an experimental analysis of a preirradiated Cu sample {Catarina} %Moreover, we analyze the stress distribution due to the gas pressure accumulated in bubbles and the presence of a nearby surface.

%%\ALC{The diffusion of H in Cu has been measured using molecular dynamics (MD) previously \cite{acharjee2019,sami2021}, resulting in higher diffusion at high temperatures and showing that H tends to escape into vacancies.}

\section{Methods}
\label{sec:methods}

\subsection{Computational simulations}%%% Computational methods, and not

PARCAS MD code \cite{nor95} was used to perform the simulations. Mishin et al. developed the embedded atom model (EAM) interatomic potential \cite{Mis01b} (\emph{Mishin} potential), which was used to described the Cu-Cu interactions. Besides, following the previous work \cite{Lop21}, we include the description EAM formalism \cite{foiles1987,baskes1987,angelo1995} for the H-H interactions and purely repulsive ZBL potential \cite{ZBL} for Cu-H. Doing this, we can assure that all the pressure is exerted to the walls of the void.

We performed simulations using periodic boundary conditions in the x-y direction and open surface in the z direction. The simulations were carried out at 600 K using the Berendsen thermostat \cite{ber84} (NVT) to scale the velocities of the atoms. At the bottom of the cell, a fixed layer was applied in order to prevent cell movement. The simulation time was 100 ps.

\begin{figure}[H]
\begin{center} 
\subfloat[]{\includegraphics[width=.33\linewidth]{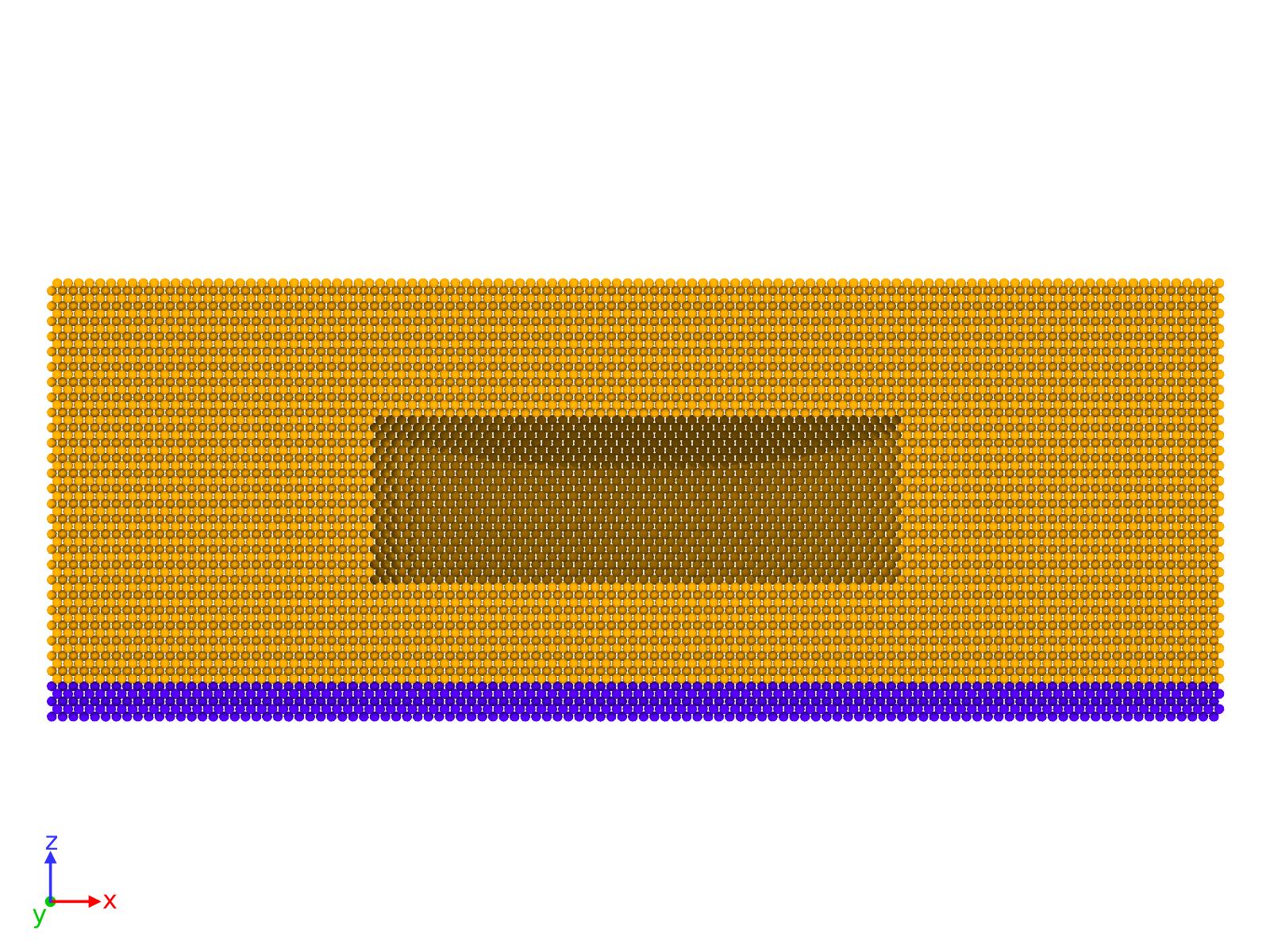}}
\subfloat[]{\includegraphics[width=.33\linewidth]{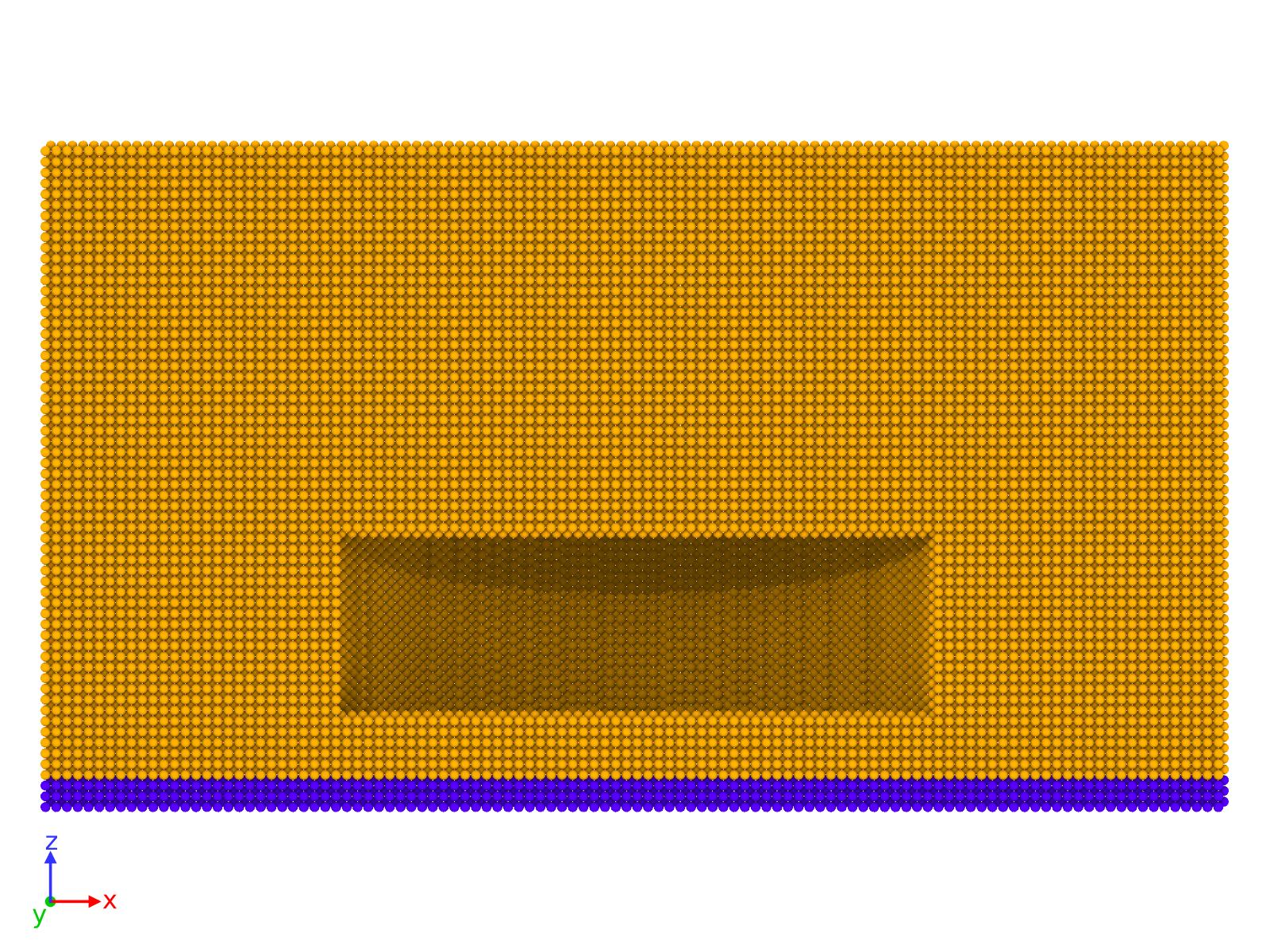}} 
\subfloat[]{\includegraphics[width=.33\linewidth]{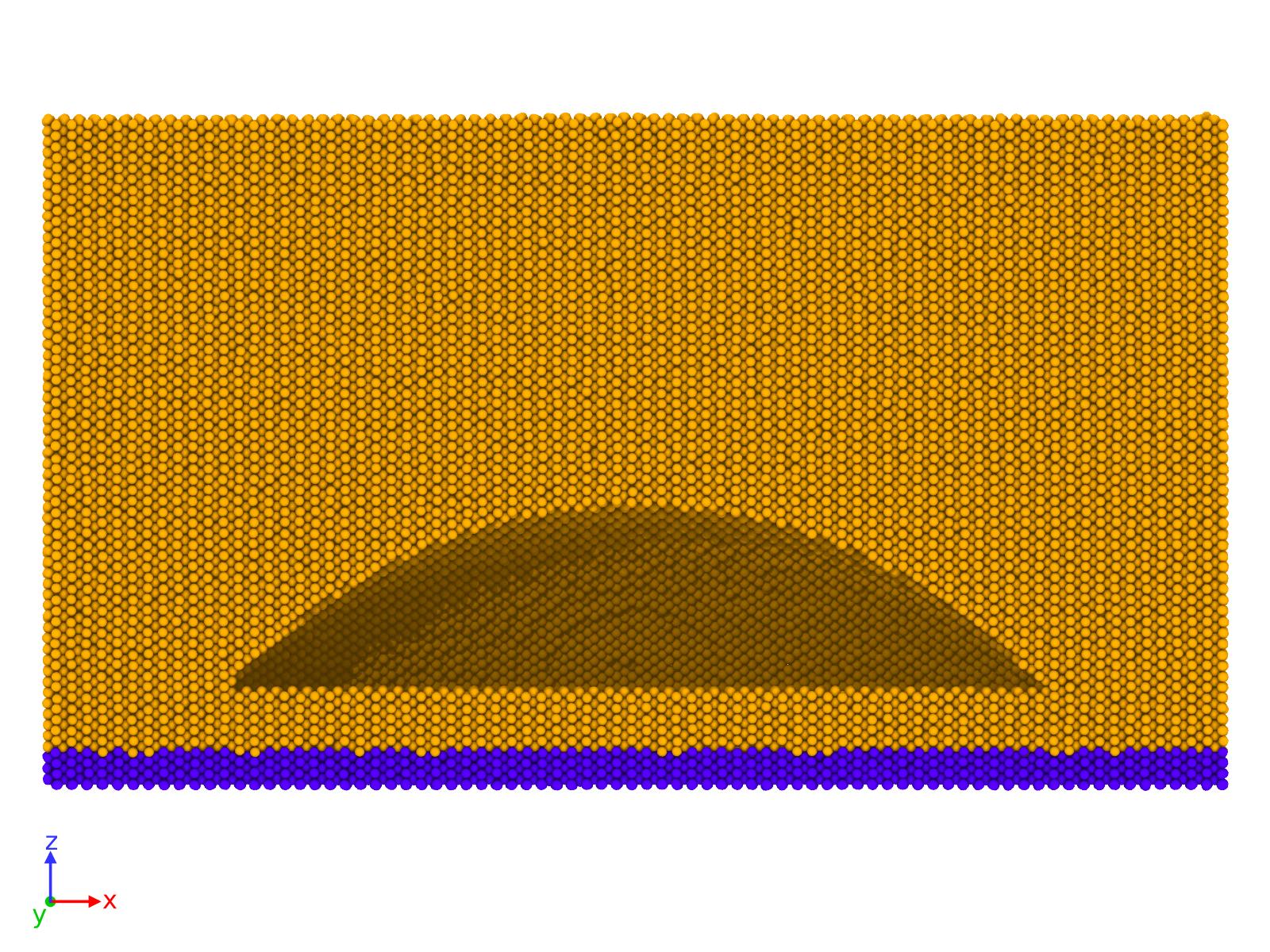}} 
\end{center}
\caption{\label{fig:initial_cells} Slice in the y-direction of the different  simulation cells. In blue the fixed layer at the bottom. (a) Disk-shaped void in the smaller cell, (b) disk-shaped void and (c) hemisphere-shaped void. }
\end{figure}

Our study is based in two types of voids: a disk-shaped void and a hemisphere-like void (see Figure \ref{fig:initial_cells}). These shapes were selected in order to mimic the different shapes that are observed experimentally. These voids were inserted in differently oriented cells. In the case of the disk-shaped voids, we additionally simulated using voids closer to the surface at lower pressures. These cells were created using different orientations in the z-direction, i.e. the open surfaces, corresponding to \hkl(1 1 0), \hkl(1 0 0), and \hkl(1 1 1).  With this, we analysed the effect of the hydrostatic pressure in the different surfaces. The volume of all the voids introduced is the same, only changing their shapes. In Table \ref{tab:cells_info} the dimensions and number of atoms in cells are presented. 

\begin{table}[H]
\resizebox{\textwidth}{!}{\begin{tabular}{|l|l|l|l|}
\hline
  & Smaller cell disk-void \hkl(110)     & Smaller cell disk-void \hkl(100)    & Smaller cell disk-void \hkl(111) \\ \hline 
Cell dimensions (Å) & $(281.9,289.4,101)$ & $(279.3,279.3,105)$  & $(278.2,279.9,104.4)$ \\ \hline
Number of Cu atoms & 648956 & 646924  & 664755 \\ \hline 
& Disk-void \hkl(110)     & Disk-void \hkl(100)    & Disk-void \hkl(111)  \\ \hline 
Cell dimensions (Å) & $(281.9,289.4,157.2)$  & $(284.5,287,153.6)$  & $(284.6,294.7,154.4)$ \\ \hline
Number of Cu atoms & 1036156 & 1014067  & 1048588\\ \hline & Hemisphere-void \hkl(110)    & Hemisphere-void \hkl(100)    & Hemisphere-void \hkl(111)   \\ \hline
Cell dimensions (Å) & $(283.7,291.1,160.9)$ & $(288.3,288.3,157)$  & $(288.7,295.2,158.3)$ \\ \hline
Number of Cu atoms & 1033830 & 1017695  & 1051281 \\ \hline 
\end{tabular}}
\caption{\label{tab:cells_info} Information about the cells used in the MD study. The cell dimensions are x, y and z respectively.}
\end{table}

%%We have tested different pressures (densities, \nhv (meaning number of H atoms per missing Cu atoms)) inside all the voids, located at 35.8, 30.7 and 31.3 Å from the \hkl(1 1 0), \hkl(1 0 0) and \hkl(1 1 1) surfaces respectively in the small cell cases. Using these cells, we simulated different H pressures, but it is only at 20 kbar (\nhv $\approx$1.2) when we indeed observe some changes in the structure.
We have tested different pressures (densities, \nhv (meaning number of H atoms per missing Cu atoms)) inside all the voids, located under the surfaces. In the smaller cells cases, we simulated different H pressures, but it is only at 20 kbar (\nhv $\approx$1.2) when we indeed observe some changes in the structure. Densities of \nhv=1.2 and 2 (20 and 30 kbar respectively \cite{Lop21}) were introduced tried in the larger cells with the disk-shaped void, so we can compare the effect of the bubble when is located at different distances. In the hemisphere-shaped voids, we introduced only \nhv= 2, because the lower density did not induce any remarkable change in the disk-shaped ones.

The volume of the voids, regardless the shape, is similar in all the cases.

In previous works \cite{Lop21}, we estimated that the solubility of H in Cu at 600 K and 20 kbar \cite{Fromm1984} is very reduced. Hence, assume that most of the H will end up accumulating in a void, it is a fair approximation for our purpose: the response of the material under this hydrostatic stress. 

For visualization of the results, we use the Open Visualization Tool (OVITO) \cite{ovito}. We analyzed the stacking faults in the cells by using the centrosymmetry (CS) parameter \cite{centrosymmetryparam} and the dislocation extraction algorithm (DXA) available in OVITO\cite{dxaanalysis}.

MDRANGE code \cite{Nor94b} is used to estimate the penetration depth of 45 keV H ions in Cu. 10\,000 cases are simulated in order to collect statistics. Contrarily to some conventional binary collisions approximation methods \cite{SRIMbook}, the different crystallographic orientations can be considered in the calculation \cite{JUSSILA2018113}.

\section{Results}
\label{sec:results}

\subsection{Experiments}
\label{sec:res_experiments}

\begin{figure}[H]
\begin{center} 
\includegraphics[width=0.6\columnwidth]{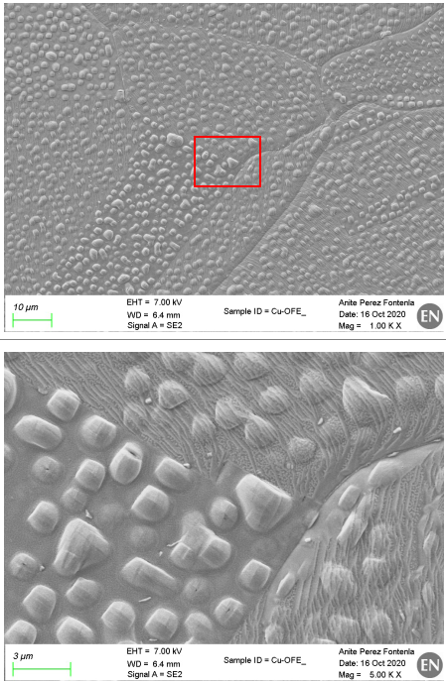} 
\end{center} 
\caption{\label{fig:exper}SEM images of the Cu surface irradiated by H$^-$ ions up to a  fluence of $\sim 2.5\times 10^{19}$ H$^-$/cm$^2$. Imaging was performed with Field Emission Gun Scanning Electron Microscope (FEG-SEM) Sigma (from ZEISS) using the secondary electron detector at 10 keV and various magnifications. The images represent two different locations, with two different levels of magnification.}
\end{figure}

Figure \ref{fig:exper} shows the Cu surface exposed to the H$^-$ ion irradiation for 40 hours reaching the fluence of $\sim 2.5\times10^{19}$ cm$^{-2}$. The image clearly shows formation of blisters with either rounded shape or having lids of random, but geometrically well defined shapes, in some cases showing signs of an opening at their apex, probably after having burst open. In some places, the blisters are self-organized in rows along specific crystallographic directions. These features indicate that the dislocation-mediated mechanism, whose linear nature, may explain the geometrically recognizable forms and alignment of the bubbles in a self-organized manner.

This observation motivated us to study the effect of hydrogen accumulation in voids, forming bubbles, following the effect that these bubbles may cause in the surface.

\begin{figure}[H]
\begin{center} 
\includegraphics[width=\linewidth]{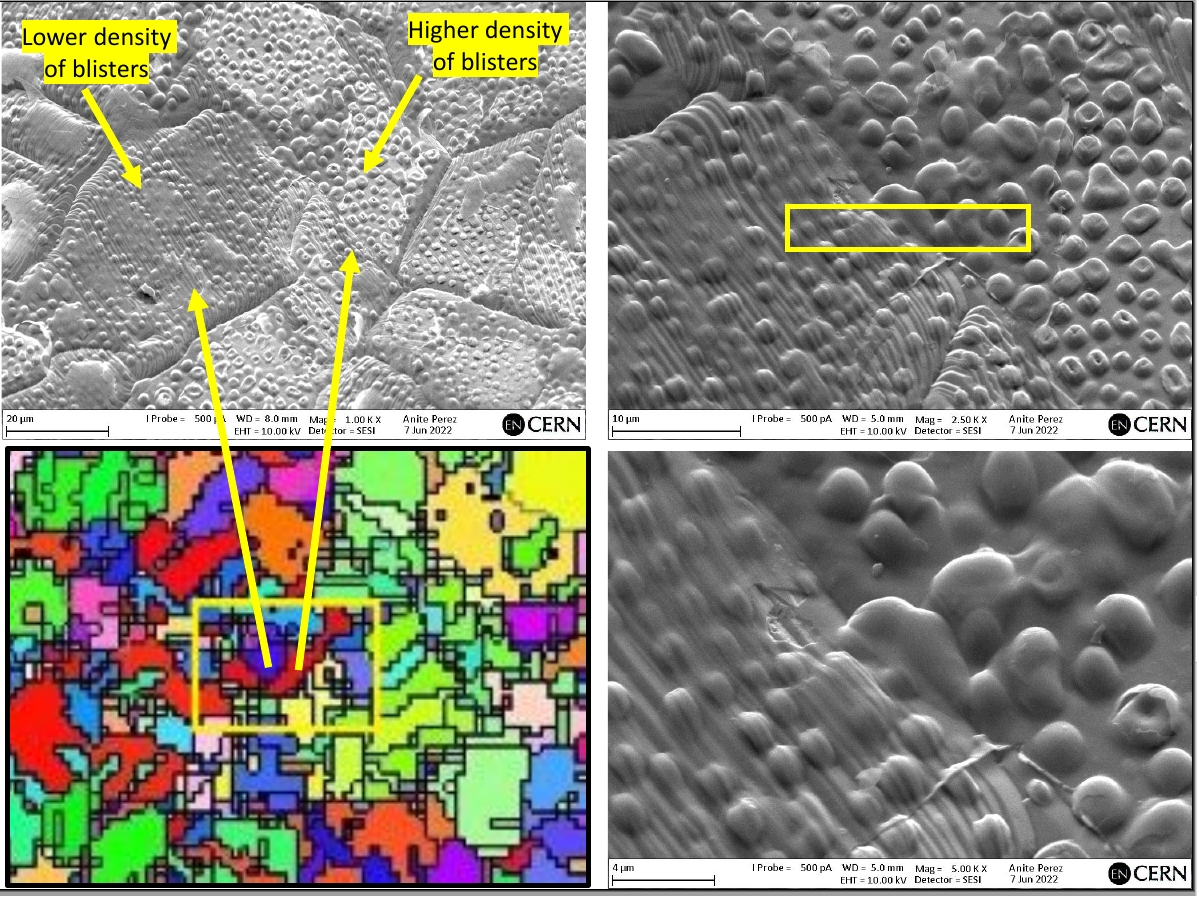}
\end{center}
\caption{\label{fig:exp_1} 
Capture of surface in between two grains. \hkl(100) in red and \hkl(111) in blue} 
\end{figure}
%\subsection{Penetration depth}
%\label{sec:penetration_depth}

In Figure \ref{fig:exp_1} we observe that the density of protrusions in the \hkl(100) grain is higher than in the \hkl(111), however the formation of them is observed in both.

\subsection{Disk-shape bubble - Small cell}
\label{sec:disk-shape-void-small}

We simulated lower density (\nhv) in a disk-shaped bubble for a smaller cell differently oriented (see Figure \ref{fig:initial_cells} (a))

\begin{figure}[H]
\begin{center} 
\subfloat[]{\includegraphics[width=.33\linewidth]{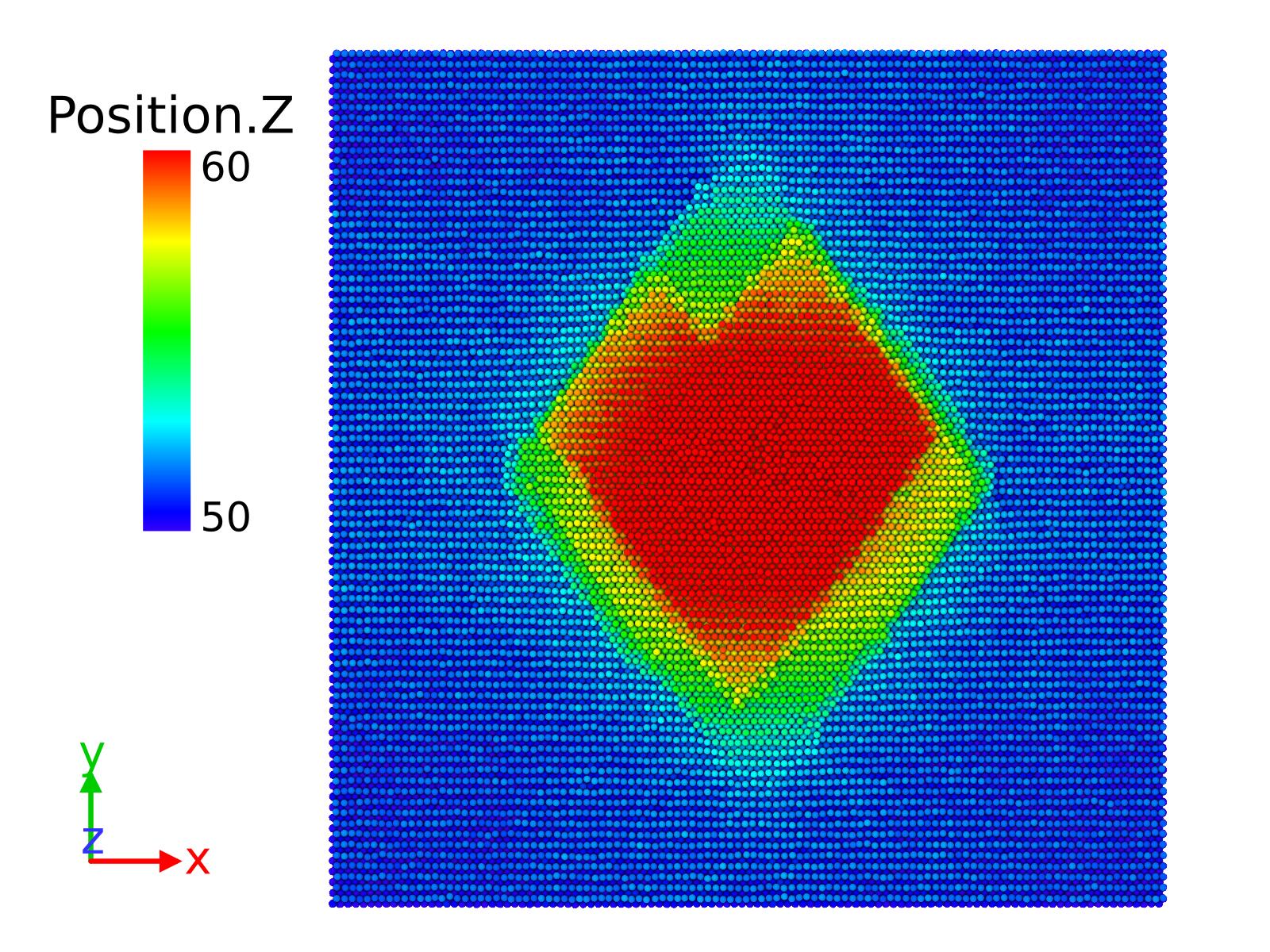}} 
\subfloat[]{\includegraphics[width=.33\linewidth]{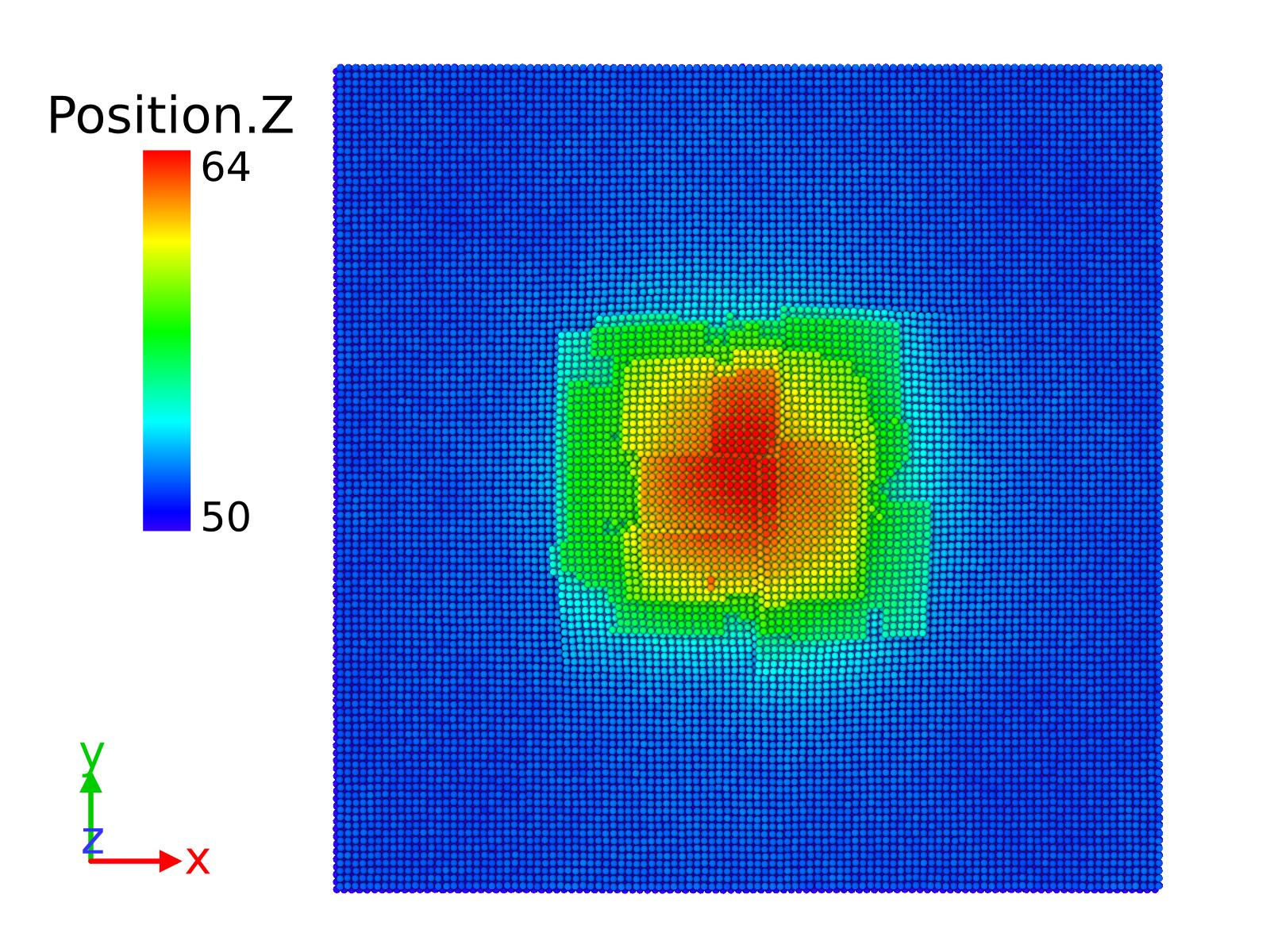}} 
\subfloat[]{\includegraphics[width=.33\linewidth]{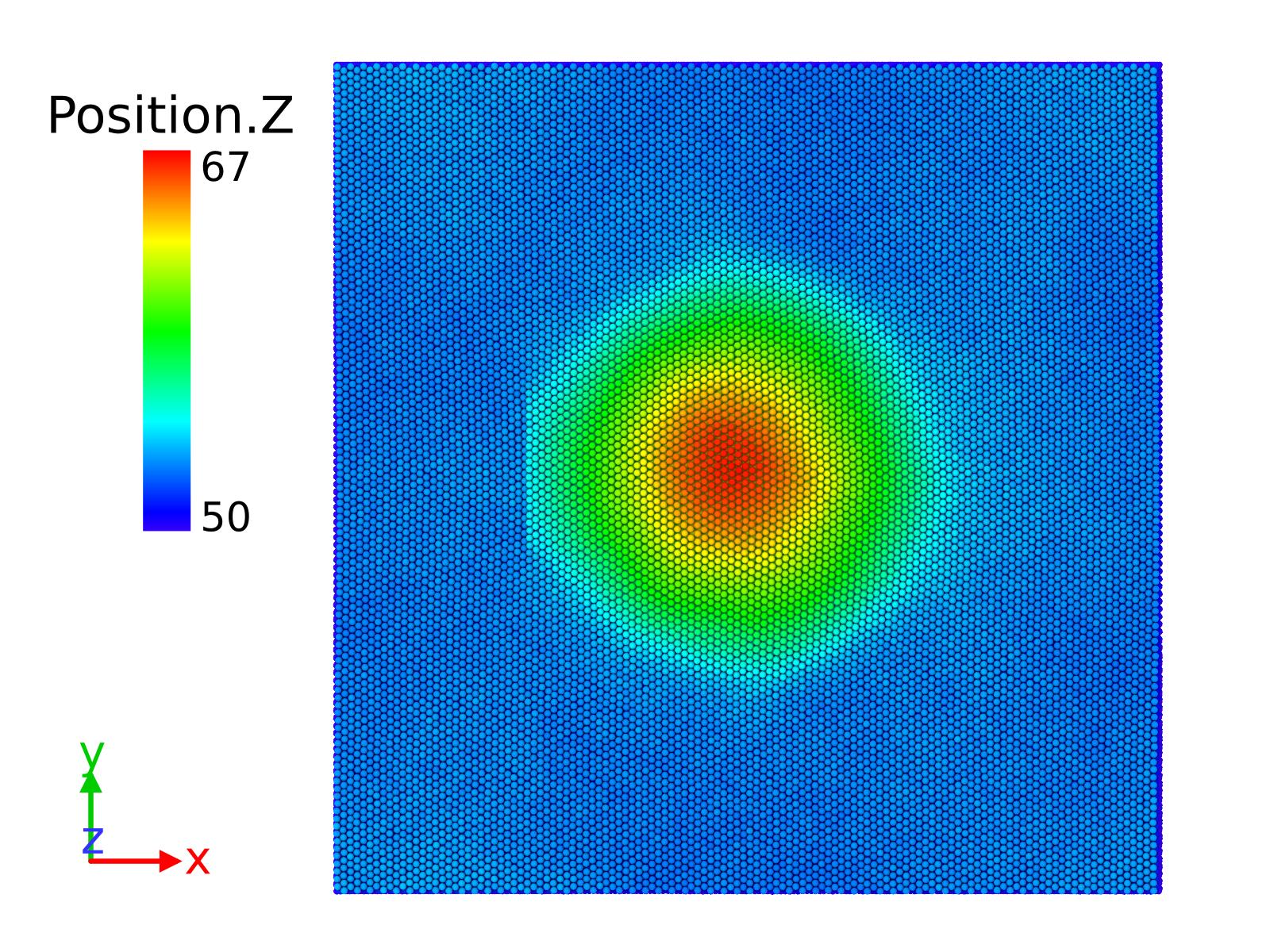}} 
\\
\subfloat[]{\includegraphics[width=.33\linewidth]{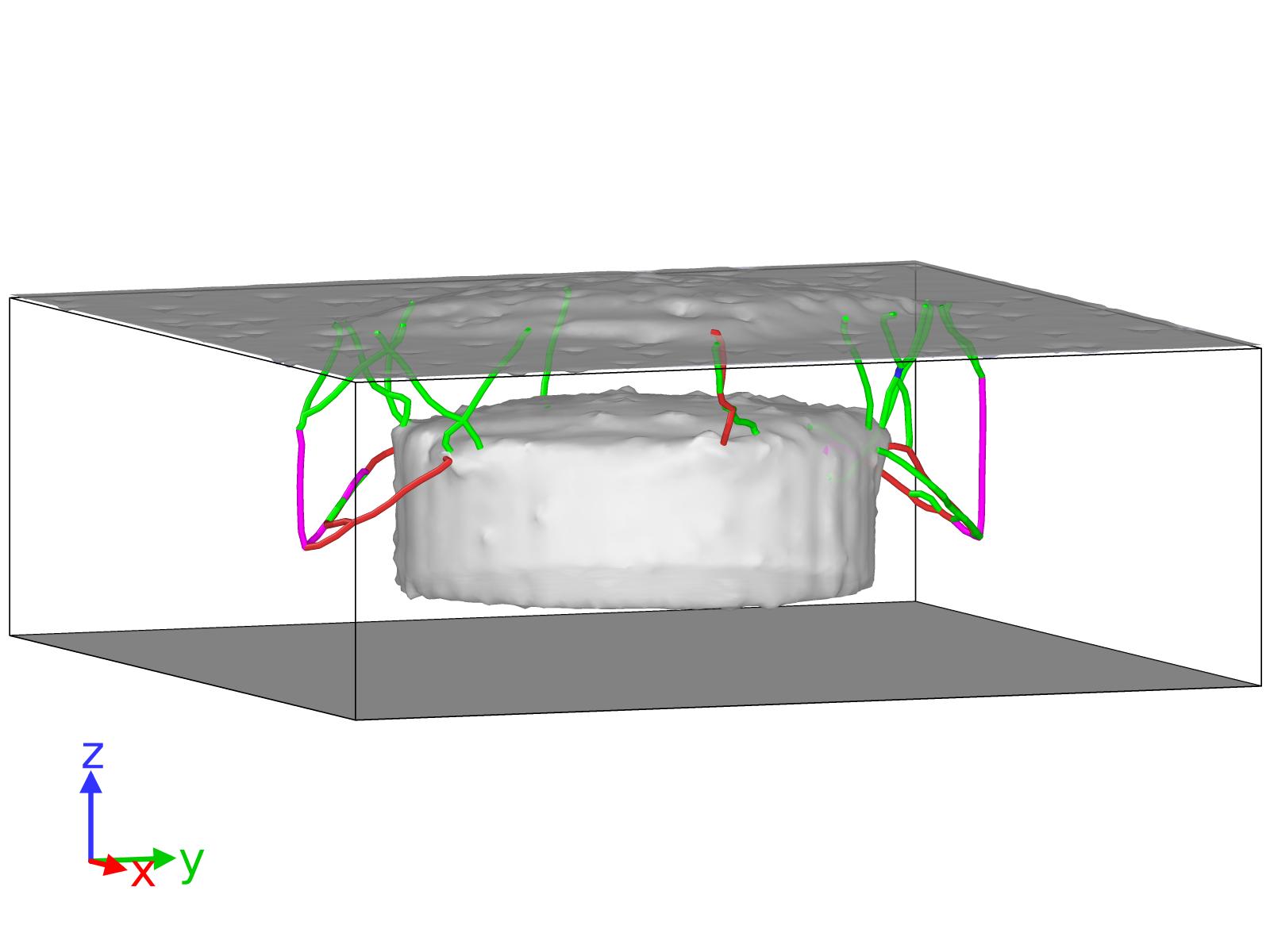}} 
\subfloat[]{\includegraphics[width=.33\linewidth]{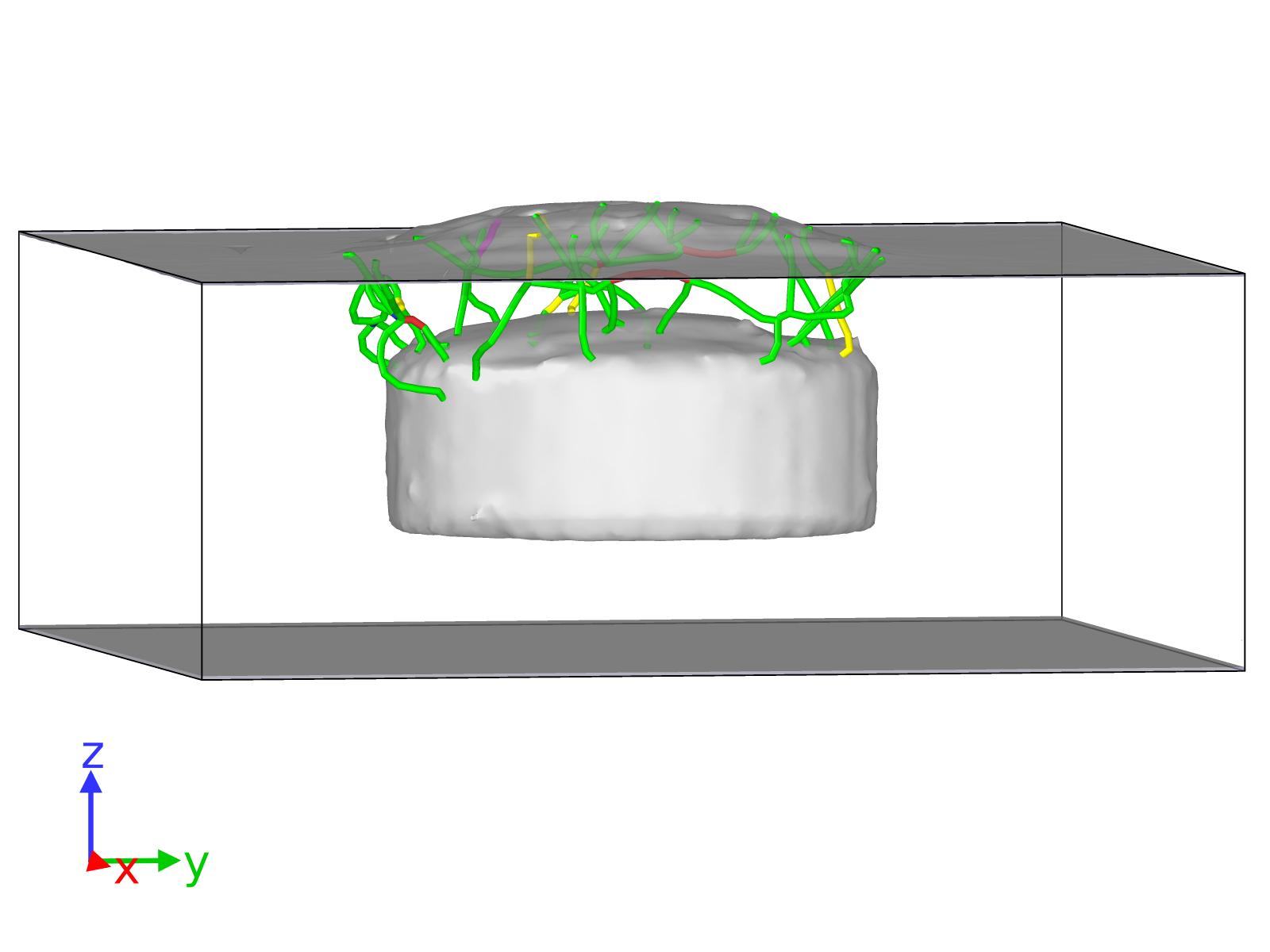}}
\subfloat[]{\includegraphics[width=.33\linewidth]{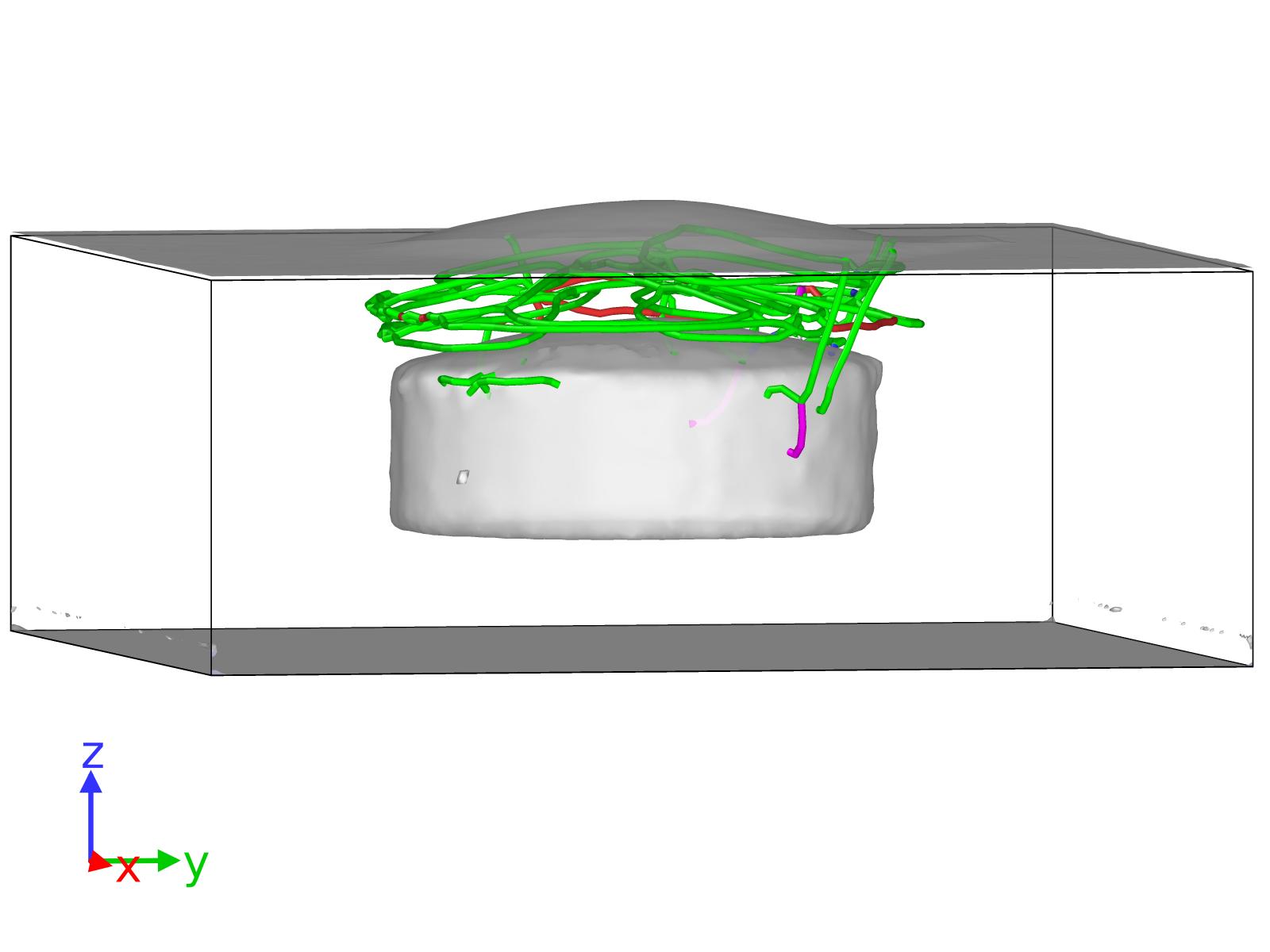}} 
\end{center}
\caption{\label{fig:disk_protrusions} The final configurations (after 100 ps) of the small cell surfaces (a,b,c) and dislocation systems (d,e,f) using \hkl(110), \hkl(100) and \hkl(111) surfaces orientation, respectively. The density inserted in the disk-shape void is \nhv= 1.2. The dislocation lines identified by DXA \cite{dxaanalysis} are shown in different colors corresponding to the type of the dislocation. Green color shows Shockley partials, blue shows perfect partials, red shows "others" dislocations, pink shows stair-rod, yellow shows Hirth and cyan shows Frank partial dislocations.}
\end{figure}

In Figure \ref{fig:disk_protrusions} we see that regardless the orientation of the surface, the protrusion is always created, but the shapes of these protrusions are different in all the cases. In the case of \hkl(110) surface (Figure \ref{fig:disk_protrusions} (a)) we see that the result is a diamond-shape protrusion. In the \hkl (100) surface (Figure \ref{fig:disk_protrusions} (b)), we observe an almost square shape as a product of the dislocation movement. For the \hkl(111) surface (Figure \ref{fig:disk_protrusions} (c)) we clearly see that the shape of the protrusion is circular. As we can check in Figures \ref{fig:disk_protrusions} (d-f), planes of stacking faults are formed differently between the disk void and the surface, where most of the stress in concentrated. We clearly see that most of the dislocations created are Shockley partials. The pressure originated by the bubble makes easier to slip the planes in those directions, creating the different orientations of the stacking faults planes respect to the surface. This leads us to see their relationship with the shape of the protrusion (Figures \ref{fig:disk_protrusions} (a-c)). 
In the \hkl(110) case (Figure \ref{fig:disk_protrusions} (a)), we realise that we have two superposed parallelograms where the lower angle is about 71$^\circ$, similarly as the one found by Pohjonen et al. \cite{Poh13}. It corresponds to the angle between two different \hkl{111} oriented planes, which are perpendicular to the \hkl(110) in the surface. Shockley dislocations are formed surrounding the staking faults and gliding towards the surface. We can see in Figure \ref{fig:disk_protrusions} (d) the nucleation of stair-rods dislocations in two Shockley partials, which interact with the surface configuring the acute corners of the rhomboid protrusion.

In the \hkl(100) case (Figure \ref{fig:disk_protrusions} (e)), we observe that the squared protrusion is induced by Shockley partial dislocations, that surround \hkl{111}-oriented planes that induce the changes in the surface. We know that out of the six \hkl{100} planes, four of them have a 90$^{\circ}$ angle respect the \hkl(111) orientation, that explains the shape of the squared-shape protrusion created by the upcoming dislocations from the edge of the bubble.

\begin{figure}[H]
\begin{center} 
\includegraphics[width=0.3\columnwidth]{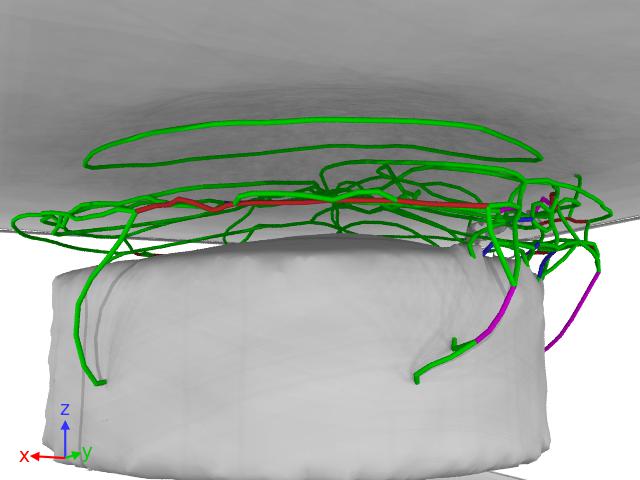} 
\end{center}
\caption{\label{fig:prismatic_loop_111} Evolution of the dislocations in the \hkl(111) surface after 60 ps for n$_{\frac{H}{Cu}}$= 1.2. The dislocation lines identified by DXA \cite{dxaanalysis} are shown in different colors corresponding to the type of the dislocation. Green color shows Shockley partials, blue shows perfect partials, red shows "others" dislocations, pink shows stair-rod, yellow shows Hirth and cyan shows Frank partial dislocations.}
\end{figure}

In the \hkl(111) case (Figure \ref{fig:disk_protrusions} (f)), we see that there are Shockley partial dislocations parallel to the surface orientation, that explains the circular shape of the protrusion. We also observed a prismatic loop that right under the surface (see Figure \ref{fig:prismatic_loop_111}).

%The emission of that prismatic loop is partially responsible for the shape of the protrusion. However, as we can see in Figure \ref{fig:higher_surface_disk} (b), the shape of the protrusion is more pronounced than in the Figure \ref{fig:prismatic_loop_111}. The reason of this difference is that in the \nhv= 1.2 case, the pressure induced is not enough to develop larger Shockley partial dislocations ($\frac{1}{6}$\hkl[-2-11] and $\frac{1}{6}$\hkl[21-1]) that surround the stacking faults plane that make the surface flip in that direction. This would create one of the sides of the triangular protrusion quite noticeable in the Figure \ref{fig:higher_surface_disk} (b). However, the no production of that marked protrusion at this H concentration is consistent with the results shown in Ref. \cite{king1981}, where the threshold-surface is energy is about twice higher for \hkl(111) than for the other two orientations.

%The H concentration introduced in these larger bubbles is less than in the smaller bubbles presented in Ref. \cite{Lop21}. So, being the surface considerably closer to the surface, the effect of the the surface plays a role in the development of the dislocations and the eventual protrusion creation. 

\subsection{Disk-shape bubble}
\label{sec:disk-shape-void}

In this cell, the distance from the top of the bubble to the surface of the cell is twice larger than in the previous cell, and in this case, we introduced two densities, \nhv=1.2 and 2. We observe that, when increasing the distance of the void from the surface, at the lower H concentration, the effect is negligible. This changes when we introduce 2 H atoms per Cu missing atoms (\nhv=2) for 100 ps.

\begin{figure}[H]
\begin{center}
\subfloat[\hkl(110)]{\includegraphics[width=.3\linewidth]{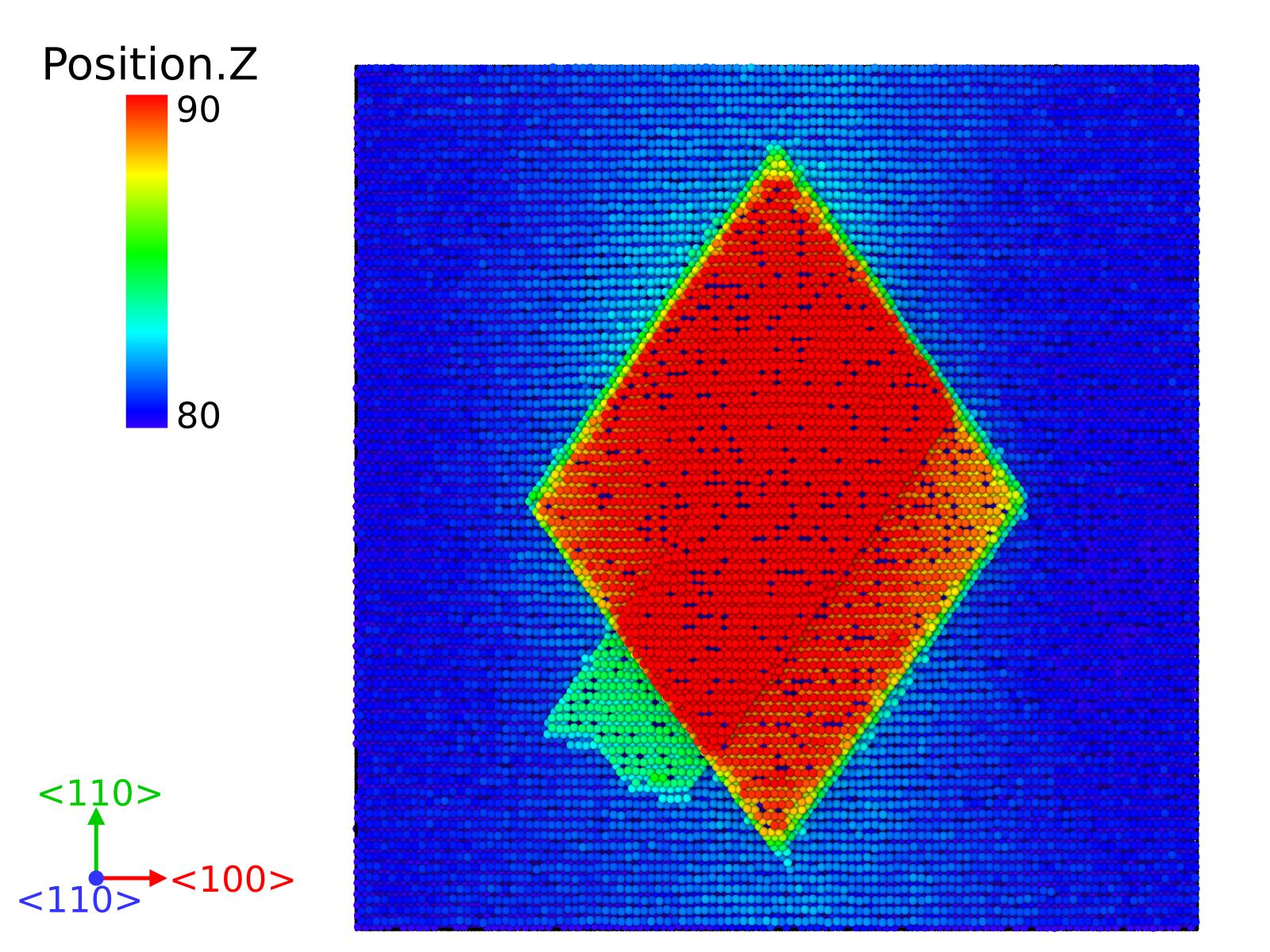}}
\subfloat[\hkl(100)]{\includegraphics[width=.3\linewidth]{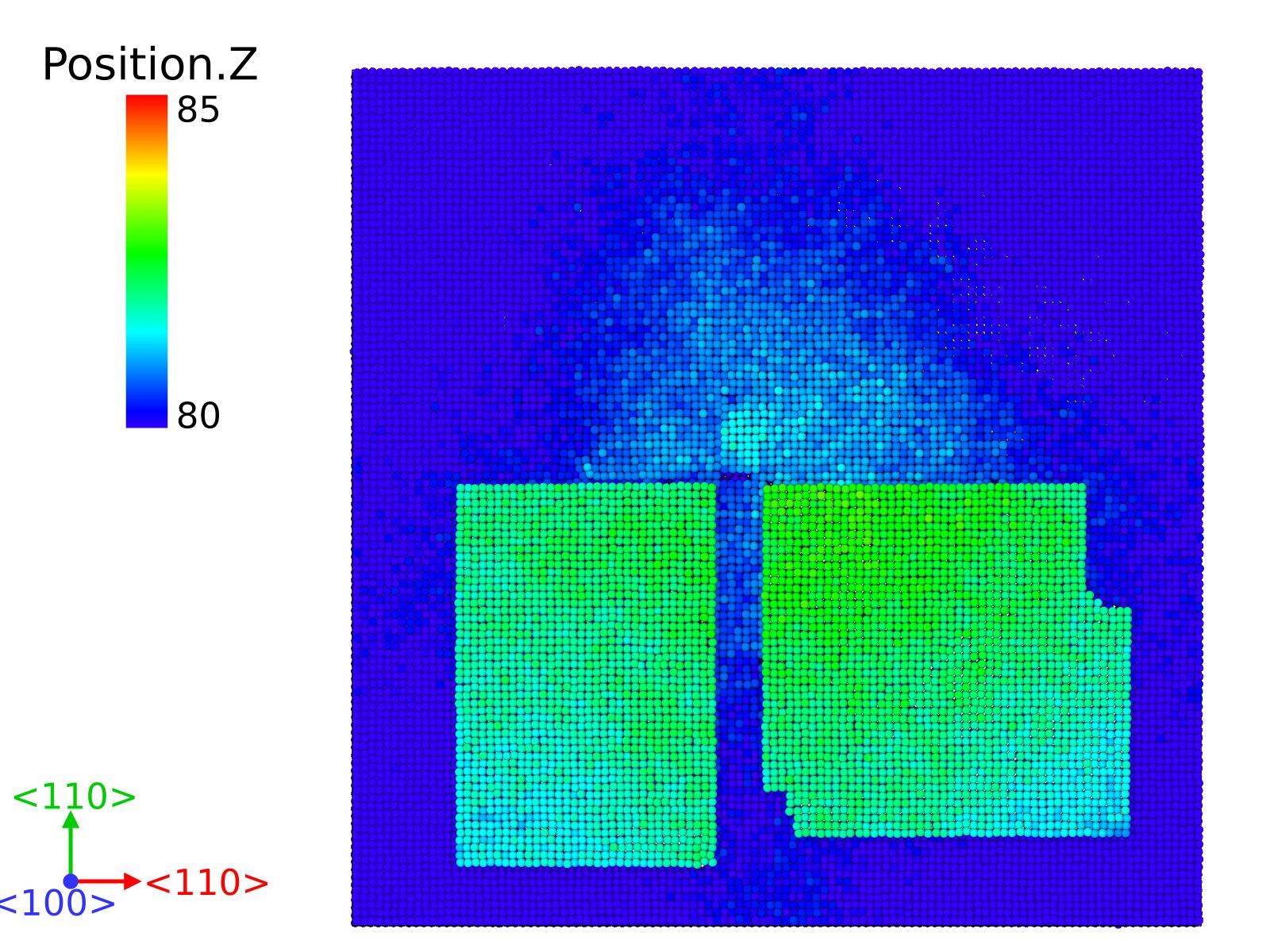}}
\subfloat[\hkl(111)]{\includegraphics[width=.3\linewidth]{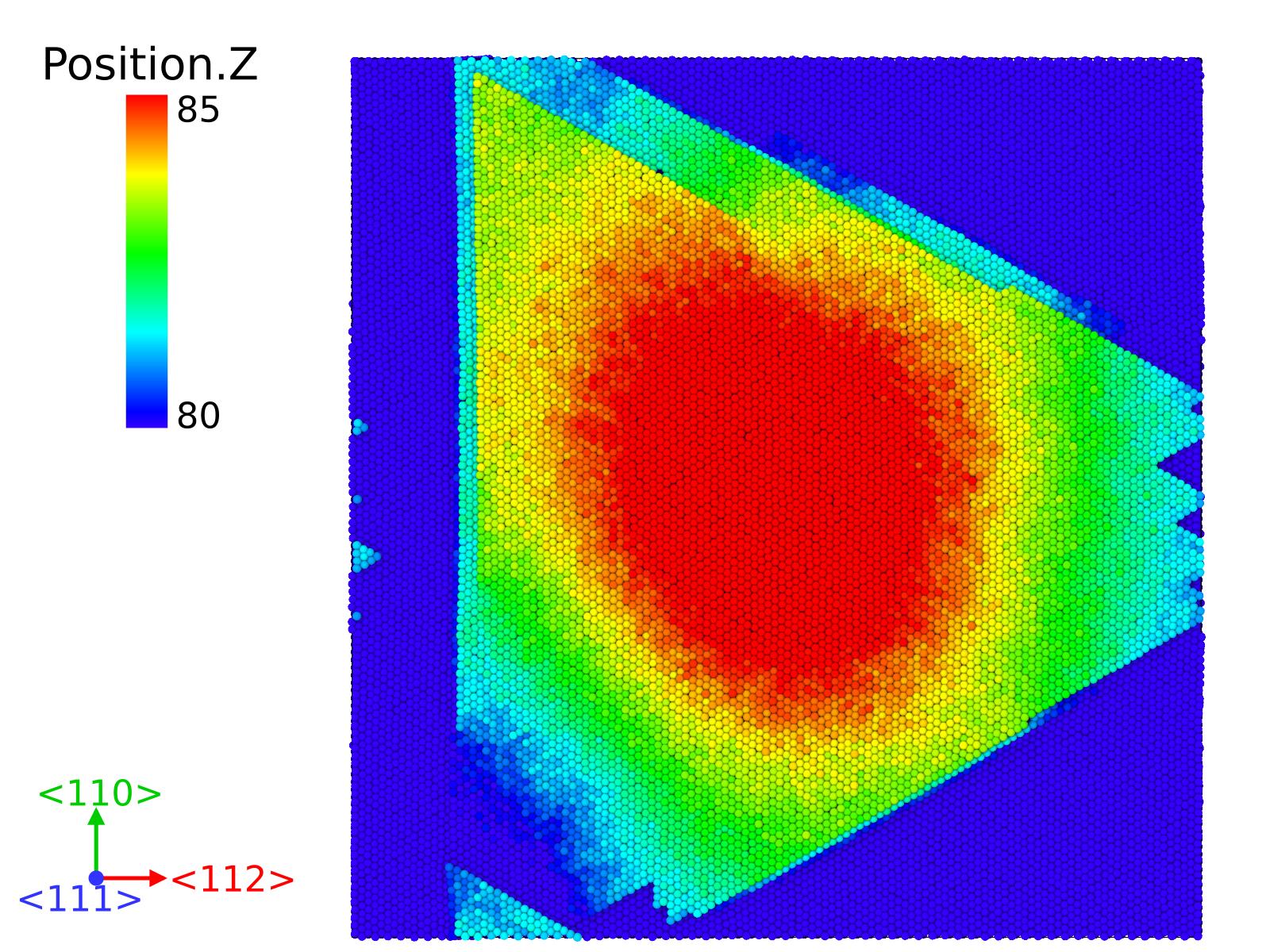}}
\end{center}
\caption{\label{fig:higher_surface_disk} Evolution of the (a) \hkl(110), (b) \hkl(100 )and (c) \hkl(111) surfaces after 100 ps for \nhv = 2.}
\end{figure}

\begin{figure}[H]
\begin{center} 
%\subfloat[\hkl(110)]{\includegraphics[width=.5\linewidth]{figures/disk_void_small_cell/disloc_higher_110_n1p2_100ps.jpg}}
\subfloat[\hkl(110)]{\includegraphics[width=.5\linewidth]{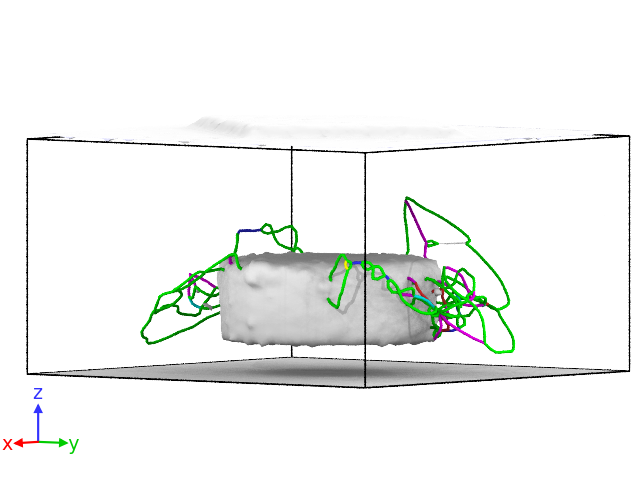}}
%\subfloat[\hkl(100)]{\includegraphics[width=.5\linewidth]{figures/disk_void_small_cell/disloc_higher_100_n1p2_100ps.jpg}}
\subfloat[\hkl(100)]{\includegraphics[width=.5\linewidth]{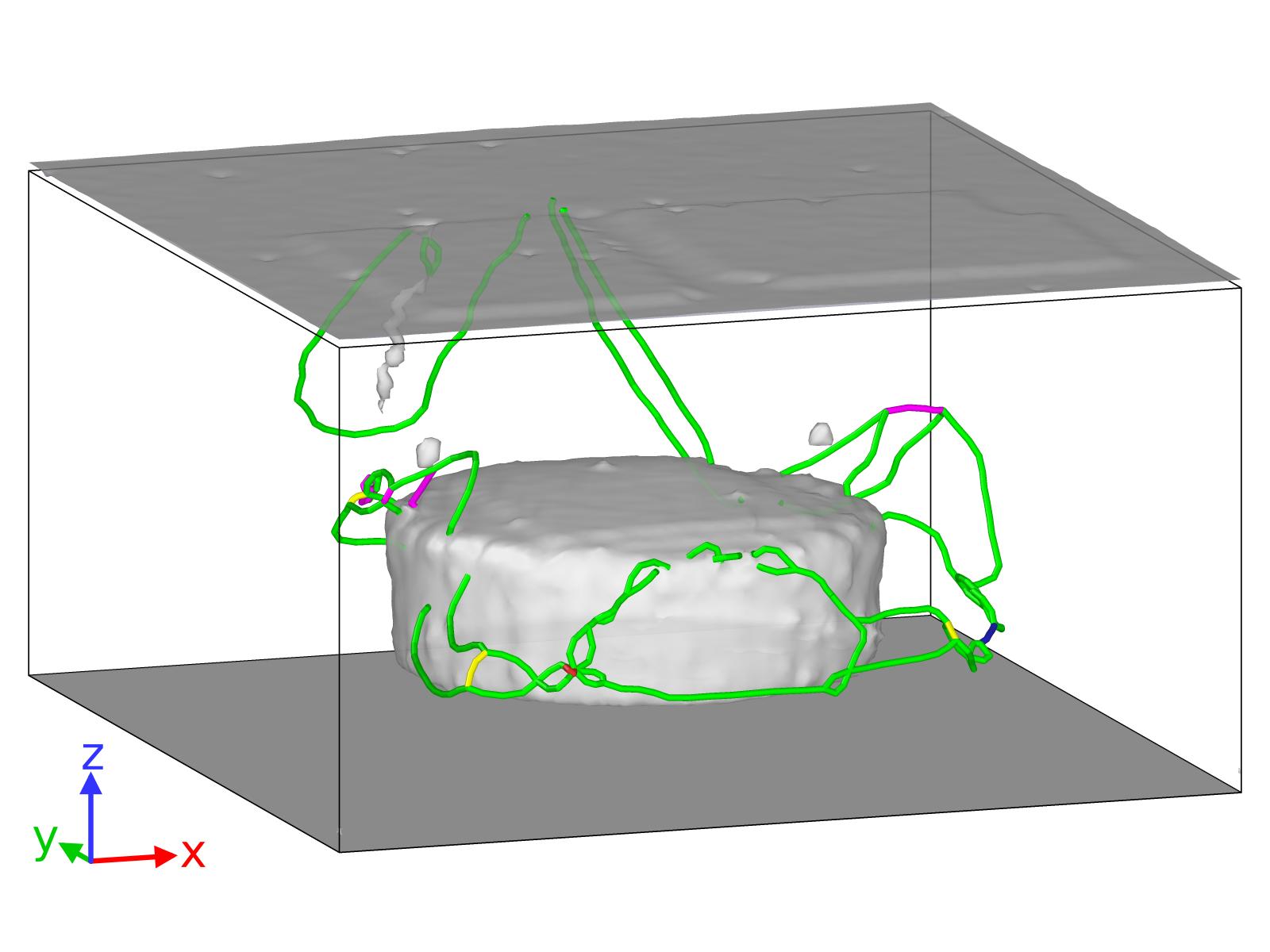}}\\
%\subfloat[\hkl(111)]{\includegraphics[width=.5\linewidth]{figures/disk_void_small_cell/disloc_higher_111_n1p2_100ps.jpg}}
\subfloat[\hkl(111)]{\includegraphics[width=.5\linewidth]{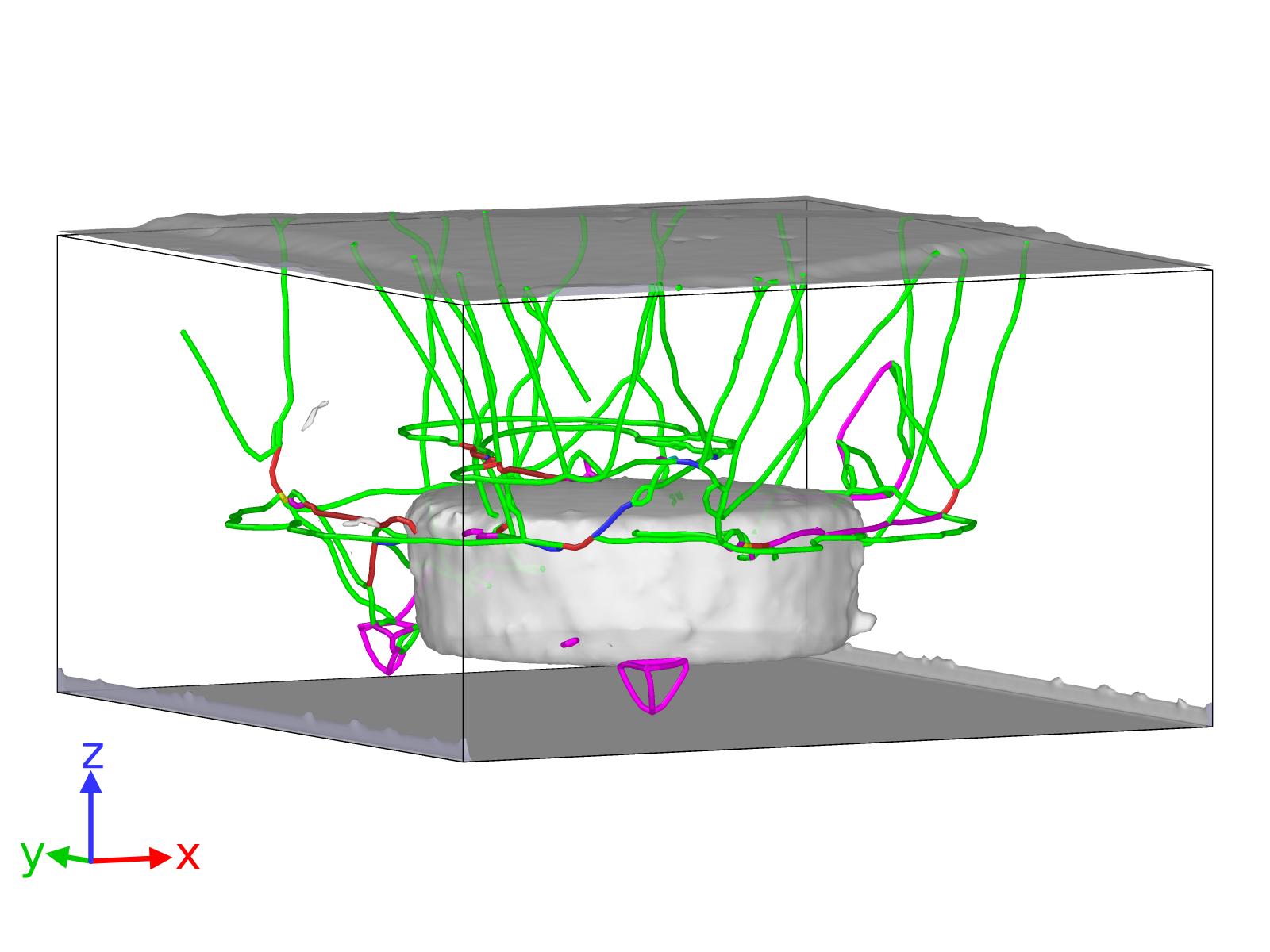}}
\end{center}
\caption{\label{fig:higher_surface_disk_disloc} Configuration of the cells for \nhv = 2 in \hkl(110) case (a), \hkl(100) case (b) and \hkl(111) (c) after 100 ps. The dislocation lines identified by DXA \cite{dxaanalysis} are shown in different colors corresponding to the type of the dislocation. Green color shows Shockley partials, blue shows perfect partials, red shows "others" dislocations, pink shows stair-rod, yellow shows Hirth and cyan shows Frank partial dislocations.}
\end{figure}

\begin{figure}[H]
\begin{center} 
\subfloat[]{\includegraphics[width=.5\linewidth]{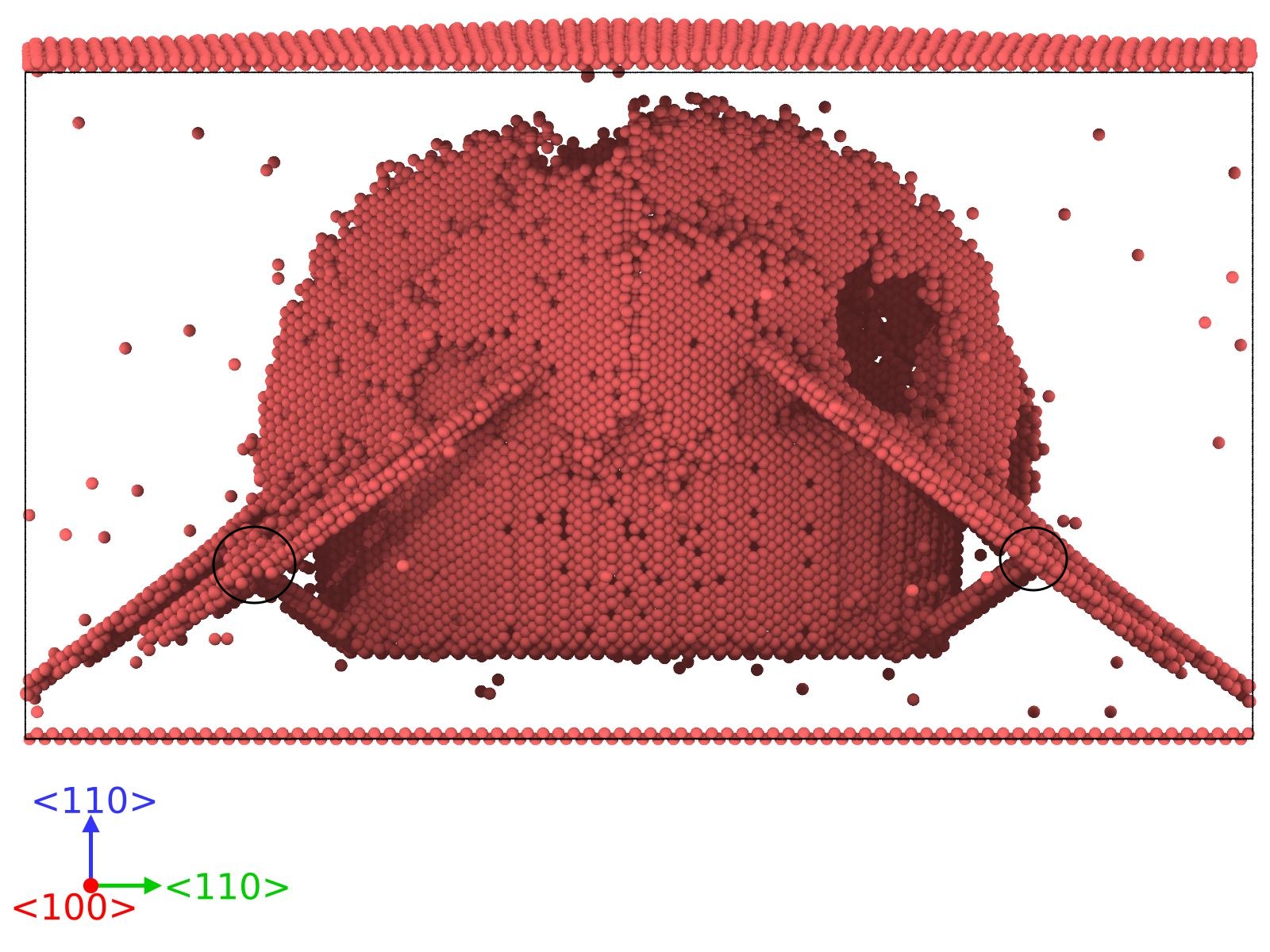}}
\subfloat[]{\includegraphics[width=.5\linewidth]{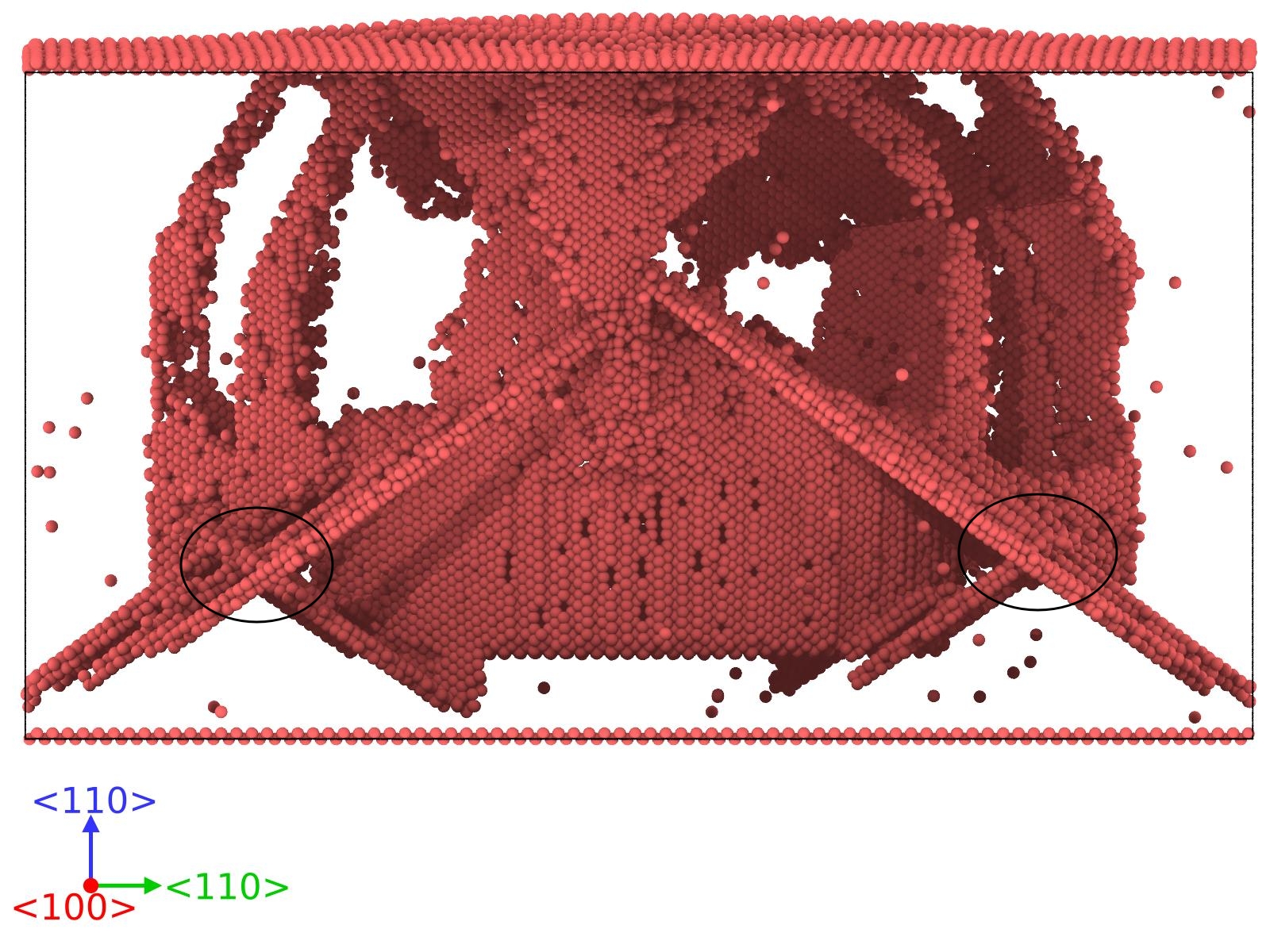}}\\
\subfloat[]{\includegraphics[width=.5\linewidth]{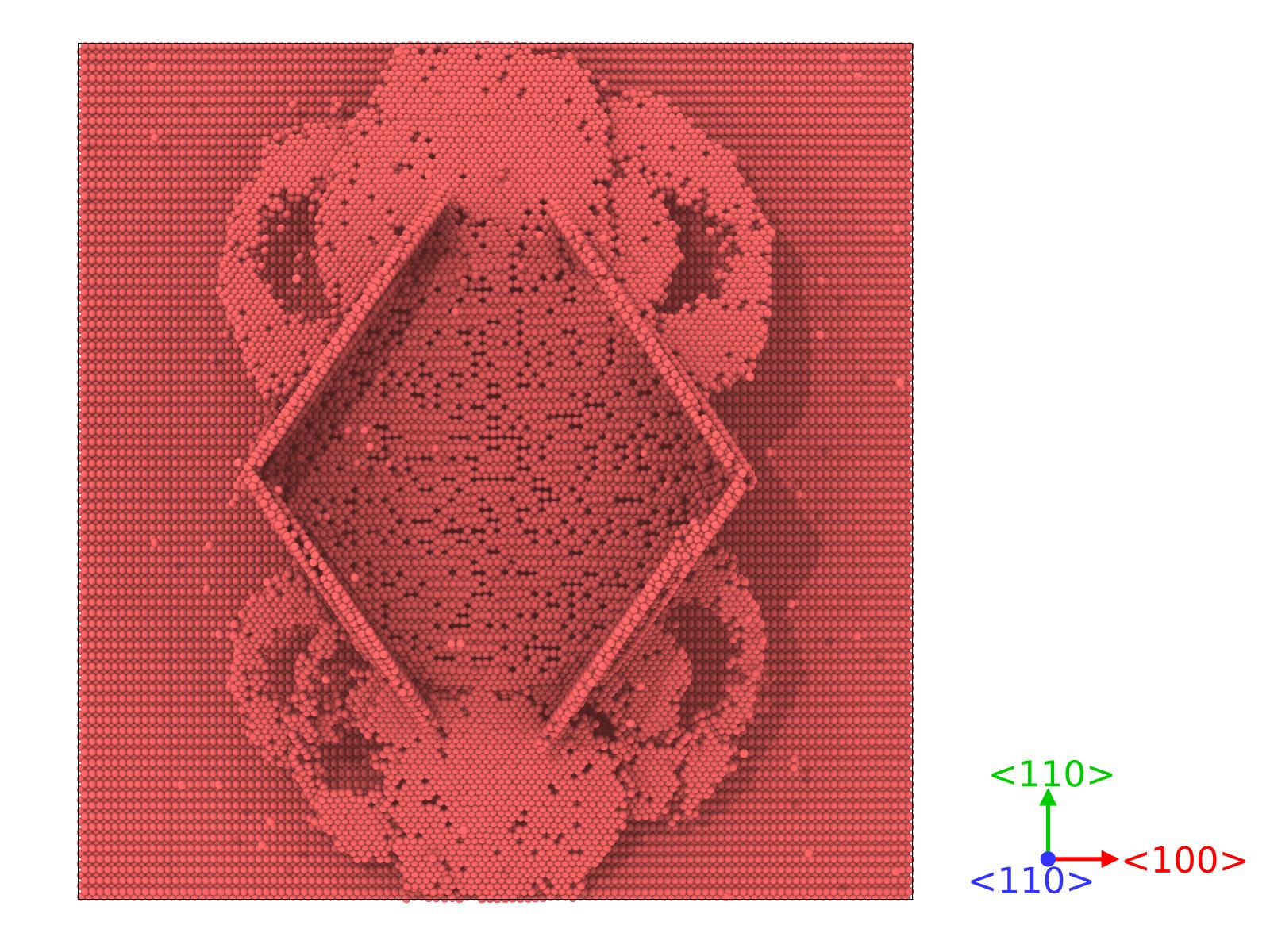}}
\subfloat[]{\includegraphics[width=.5\linewidth]{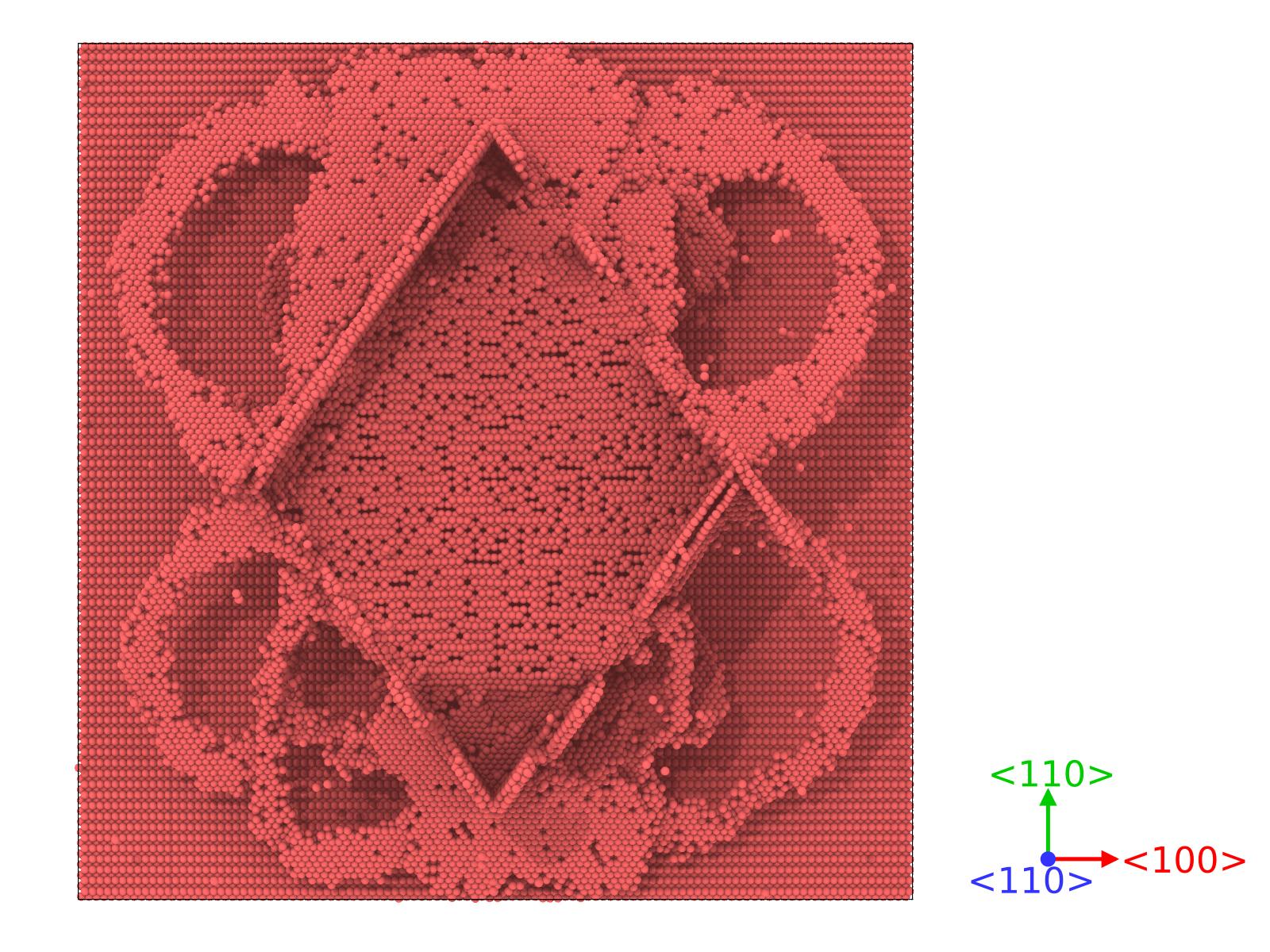}}
\end{center}
\caption{\label{fig:SF_110} 
Configuration of lateral stacking faults expanding from top and bottom border of the disk-shape bubble (a,b) and stacking faults growing from the top surface of the bubble (c,d) (the atoms from the surface have been removed for clarity) for \hkl(110) surface at (a,c) 7.5 ps and (b,d) 10 ps. The black circles represent the meeting point of different stacking faults.} 
\end{figure}

In Figure \ref{fig:higher_surface_disk} (a) we observe that the surface yields similarly as in Figure \ref{fig:disk_protrusions} (a), corresponding to the \hkl(110) surface. Following the evolution of the simulation (see Figure \ref{fig:higher_surface_disk_disloc} (a) and high110-n2-disloc.mp4), we observed that in the first steps of the simulation, a number of shear dislocations are created and they induce the protrusion in the surface (see the stacking faults created in Figure \ref{fig:SF_110}), similarly as they were created in the case exposed in Figure \ref{fig:disk_protrusions} (d). We can see that the rhomboid protrusion is delimited by the \hkl<112> directions. The angles of the rhomboid are 71$^{\circ}$ and 109$^{\circ}$. However, we observed that using this larger cell with a concentration \nhv = 1.2 , nothing remarkable happened in the surface, despite the generation of dislocations beneath it (see high110-n1p2-disloc.mp4). We only noticed that at the end of the simulation, some shear dislocations that were created during the initial stages, are dissipated by then. We observe that in the shorter cell, the surface is sufficiently close to the void for generating the protrusion (see Figure \ref{fig:disk_protrusions} (a)). %%%% larger distance to the surface, it implies that less chance to the surface to be modified due to this.

For the \hkl(100) surface shown in Figure \ref{fig:higher_surface_disk} (b), the shape of the protrusion created is similar to the one created in Figure \ref{fig:disk_protrusions} (b) (see high100-n2-disloc.mp4). The mechanism behind the surface modification is the same as in the smaller cell with lower H density. We clearly observe how the shape of this protrusion is the product of the superposition of several squares. These squares are delimited by the Shockley partials that reach the surface, forming the borders of the protrusions oriented in the different \hkl<110> directions. The dislocations glide in the \hkl{111} planes due to the bubble exerted pressure, and when interact with the surface, they move upwards some atom layers as can be seen in Figure \ref{fig:higher_surface_disk_disloc} (b), creating the final protrusion. In the \nhv = 1.2 case, no dislocation survives (see high100-n1p2-disloc.mp4).

The \hkl(111) surface can be seen in Figure \ref{fig:higher_surface_disk} (c), where we observe a triangle shape protrusion, which differs from the circular shape observe in Figure \ref{fig:disk_protrusions} (c). However, we also notice that the height of the atoms in the center marked by the triangle, is higher. This shows that the protrusion created in Figure \ref{fig:disk_protrusions} (c), due to the smaller pressure inside the void, may be not fully developed. Increasing the pressure and the distance to the surface, the process is more complex (see Figure \ref{fig:higher_surface_disk_disloc} (c) and high111-n2-disloc.mp4). The dislocations (Shockley partials) cut the surface creating that triangle shape. We also observe that in the center, the protrusion is higher as product of the loop that grew below it. In the \nhv = 1.2 case for the \hkl(111) surface, some dislocations are created but not enough to modify the surface (see high111-n1p2-disloc.mp4).

Following the evolution of \hkl(110) case, we observe that initially (after 2.5 ps) some dislocations are formed in the top border of the bubble. These partial dislocations ($\frac{1}{6}\hkl<112>$) surround stacking faults that grow differently: some of them continue to the bottom of the cell (from the top surface of the bubble) and meet other plane which prevent the growing of the latter plane upwards (see Figure \ref{fig:SF_110} (a,b), black circles). In Figure \ref{fig:SF_110} (c,d), we see how the stacking fault planes grow perpendicular to the bubble flat surface. Their expansion from the border form the eventual diamond shape that is observed in Figure \ref{fig:higher_surface_disk} (a).

\begin{figure}[H]
\begin{center} 
\subfloat[]{\includegraphics[width=.5\linewidth]{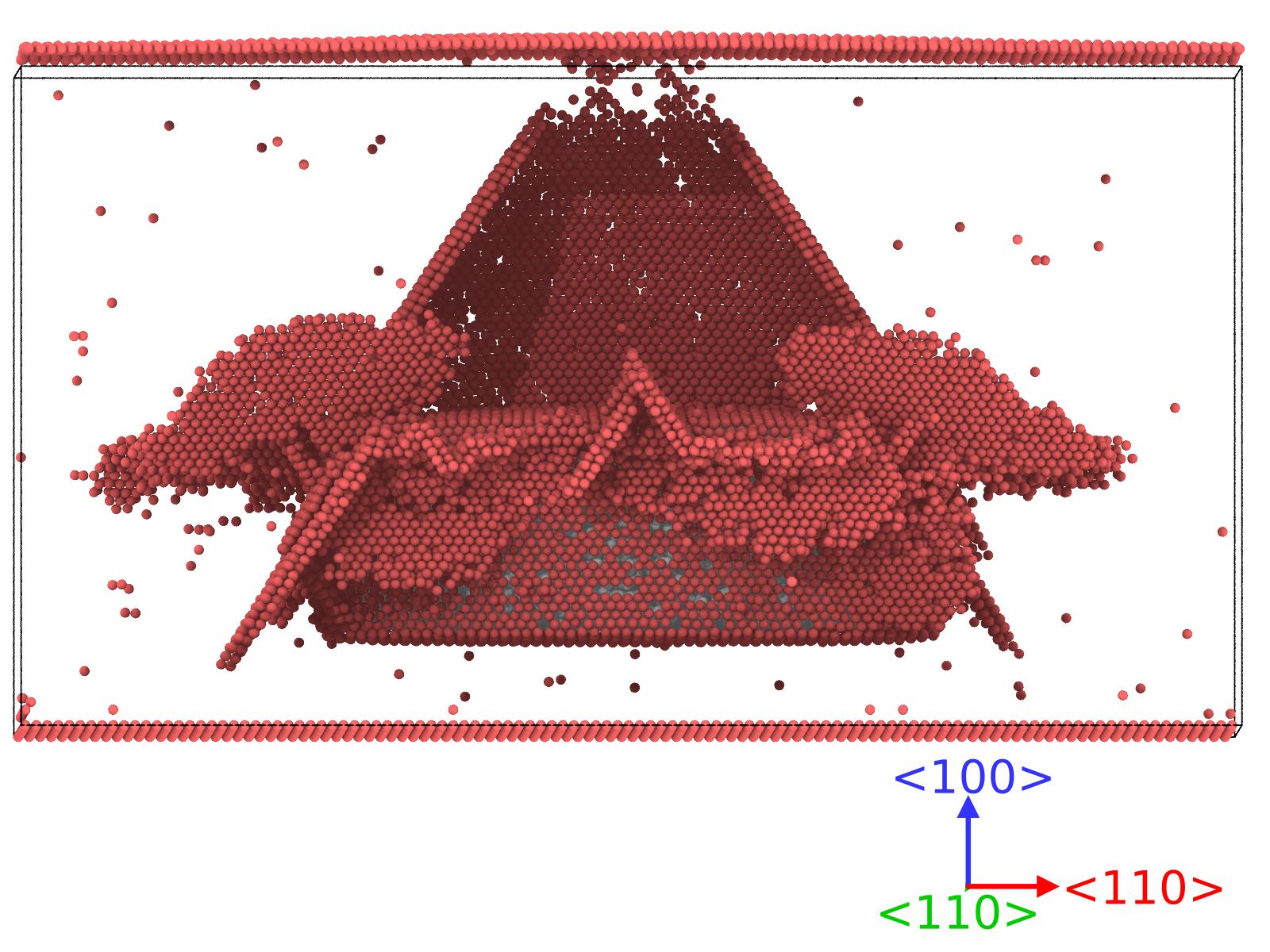}}
\subfloat[]{\includegraphics[width=.5\linewidth]{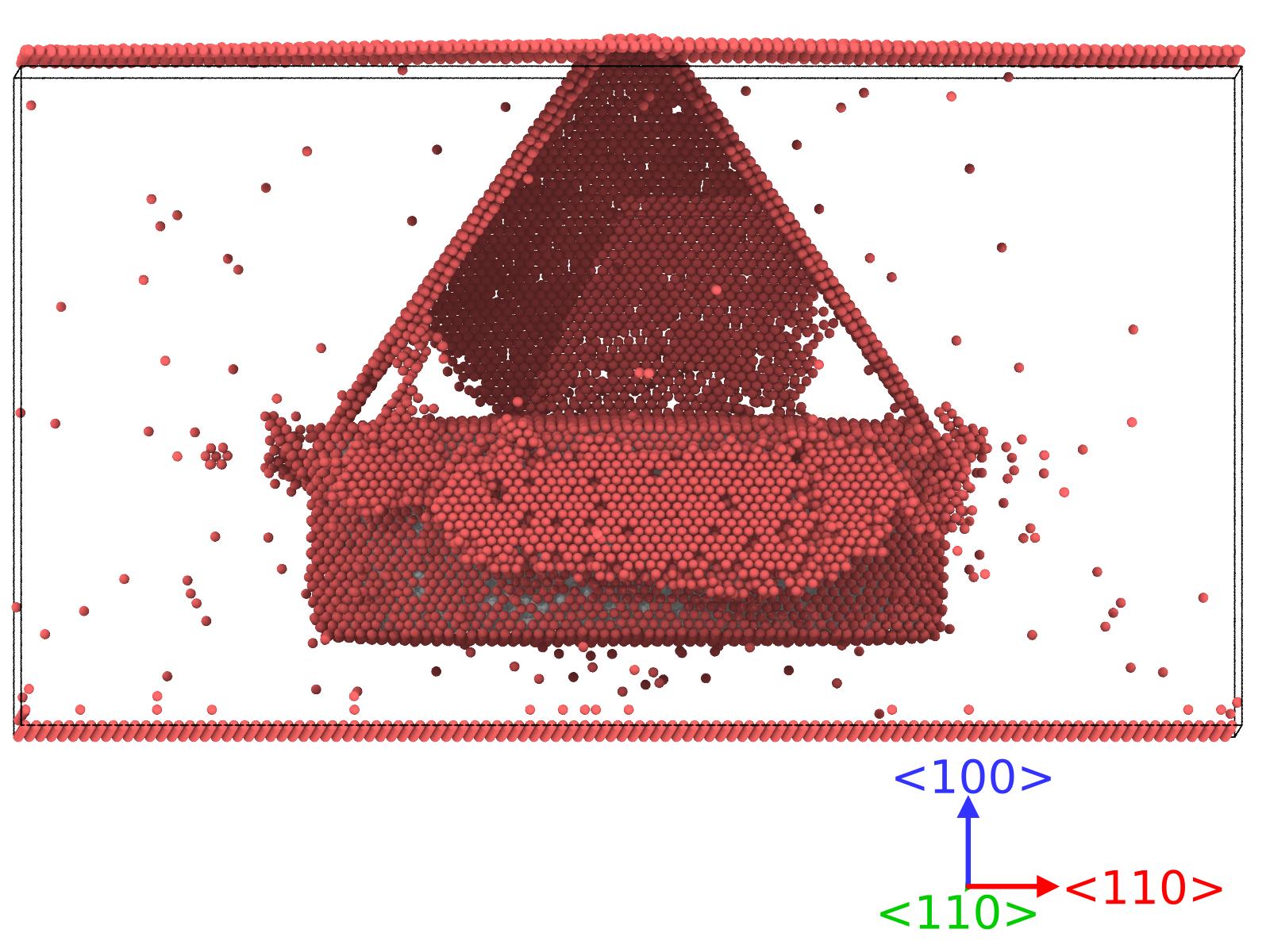}}
\end{center}
\caption{\label{fig:SF_100} 
Configuration of the partial stacking fault octahedron formed in the \hkl(100) surface case at (a) 10 ps and (b) 22.5 ps. Only those atoms with CSP greater than 5 are shown.} 
\end{figure}

The \hkl(100) case shows how the top border of the bubble starts to emit partial dislocations, leaving stacking faults planes that grows to the surface of the cell. In this case we clearly see the formation of the upper part of the stacking fault octahedron (Figure \ref{fig:SF_100}), which yields due to the high pressure; and the expansion of the dislocations forming it induces the change formed in the surface at the end of the simulation (see Figure \ref{fig:higher_surface_disk} (b)).

\begin{figure}[H]
\begin{center} 
\subfloat[]{\includegraphics[width=.5\linewidth]{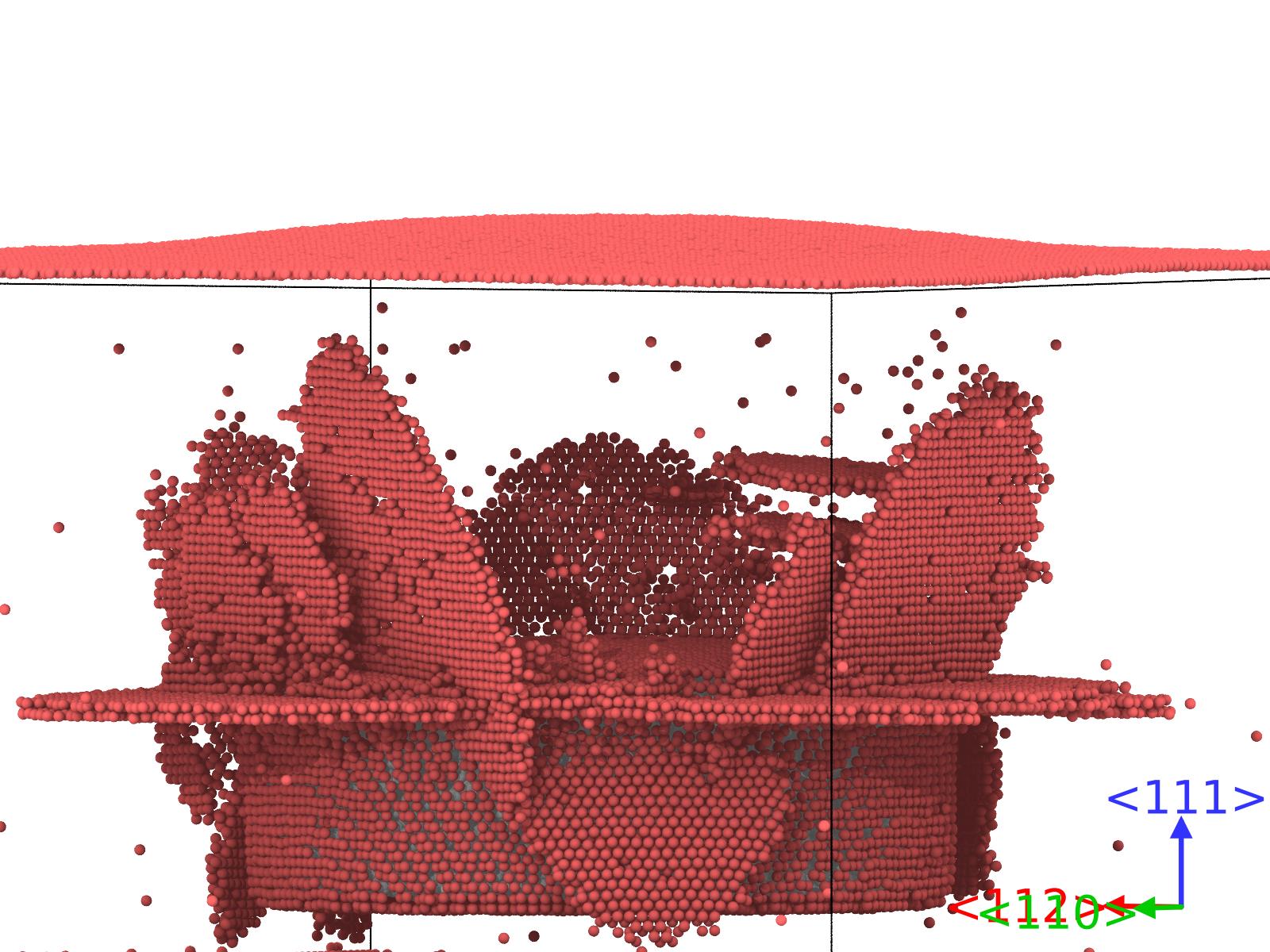}}
\subfloat[]{\includegraphics[width=.5\linewidth]{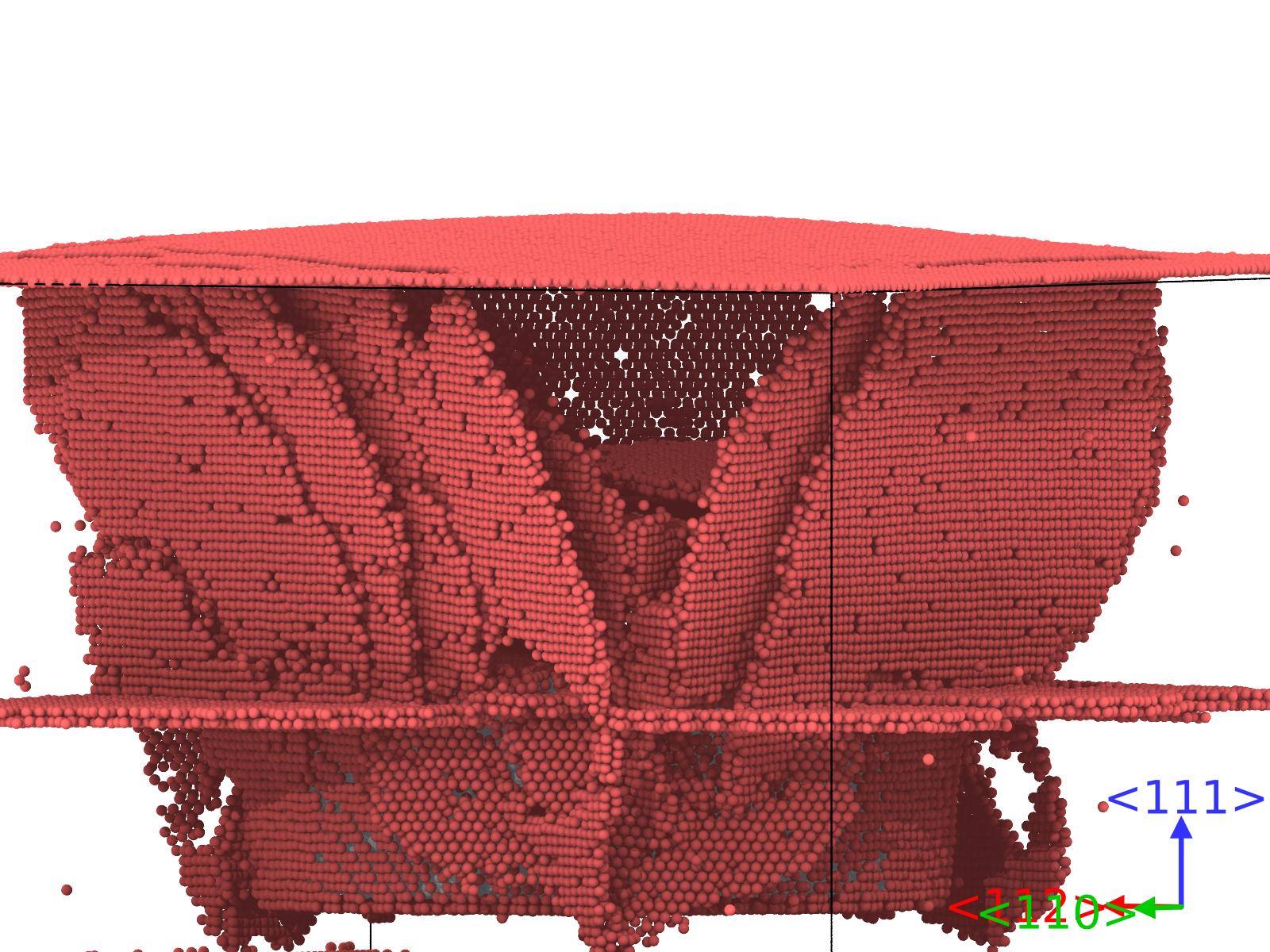}}
\end{center}
\caption{\label{fig:SF_111} 
Configuration of the partial stacking fault octahedron formed in the \hkl(111) surface case at (a) 7.5 ps and (b) 10 ps. Only those atoms with CSP greater than 5 are shown.} 
\end{figure}

In the \hkl(111) case, the effect of the pressurized bubble creates stacking faults at the border of the bubble parallel to surface of the bubble, but also towards the surface as can be observed in Figure \ref{fig:SF_111}. We notice that these partial dislocation leading the planes to the surface are the responsible for the the triangle shape observed in the Figure \ref{fig:higher_surface_disk} (c).

\subsection{Hemisphere-shape bubble}
\label{sec:hemisphere-shape-void}

We study in a similar cell as in the Section \ref{sec:disk-shape-void} the effect of a hemispherical bubble at \nhv=2.

\begin{figure}[H]
\begin{center}
\subfloat[\hkl(110)]{\includegraphics[width=.3\linewidth]{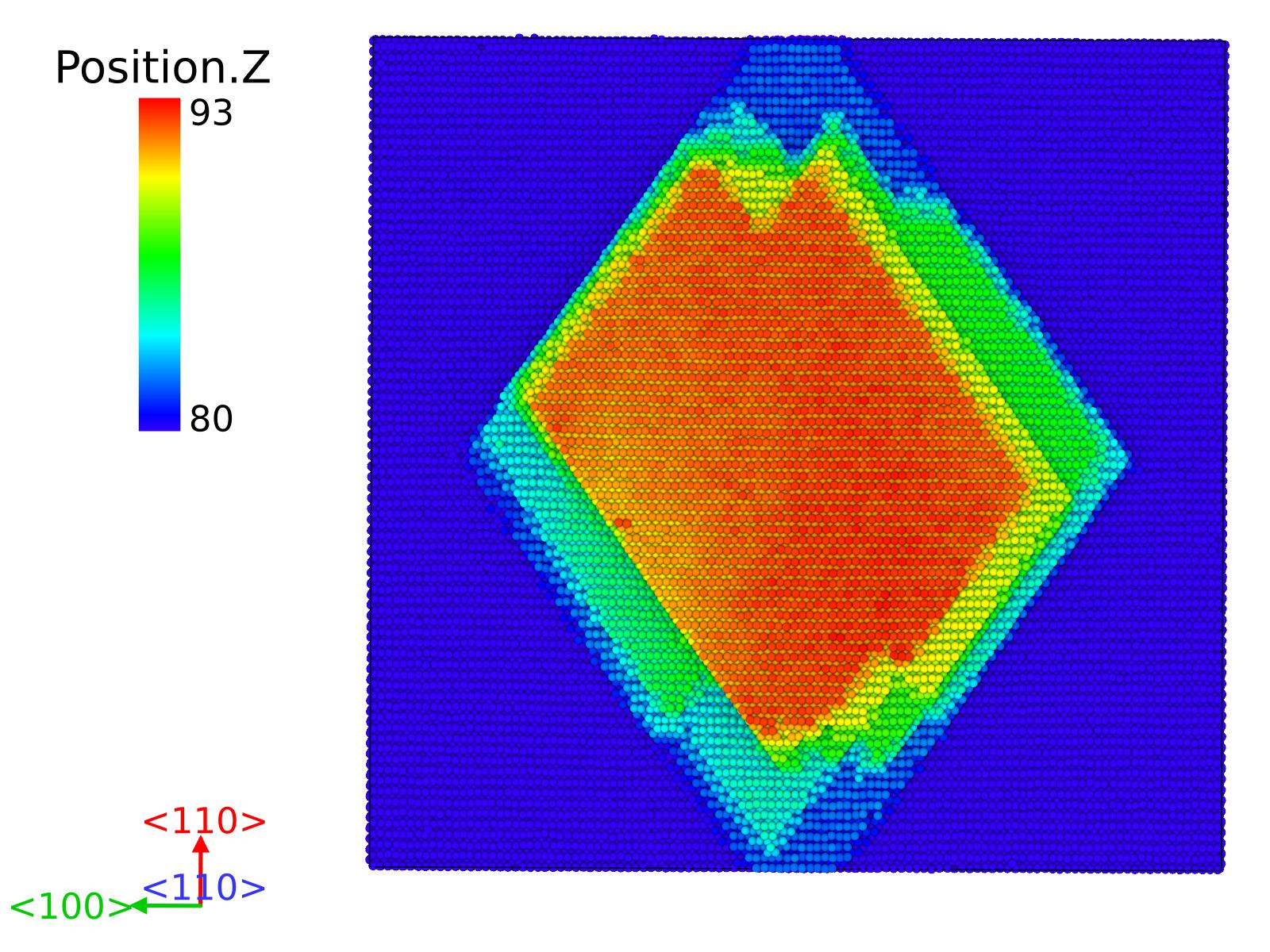}}
\subfloat[\hkl(100)]{\includegraphics[width=.3\linewidth]{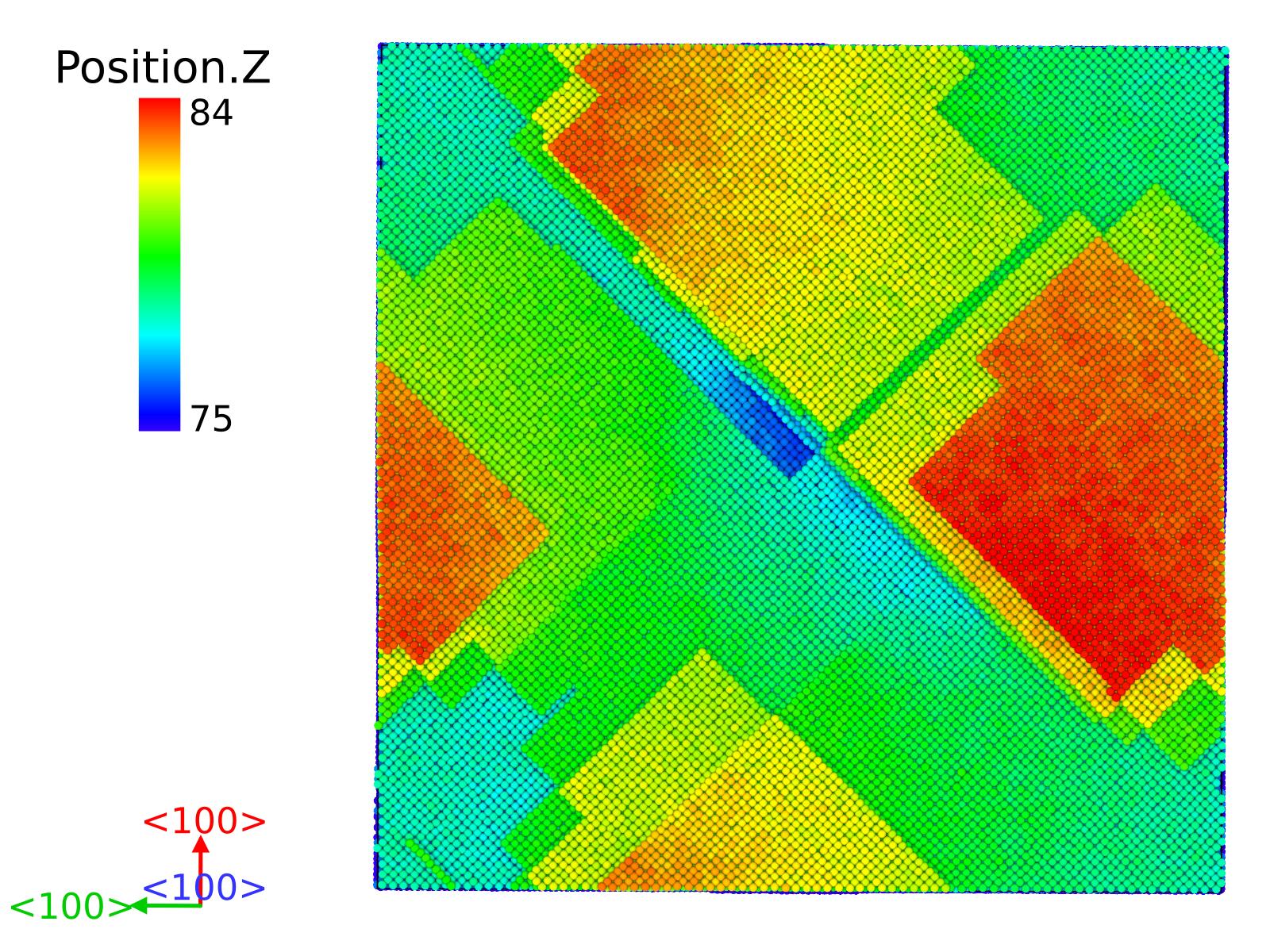}}
\subfloat[\hkl(111)]{\includegraphics[width=.3\linewidth]{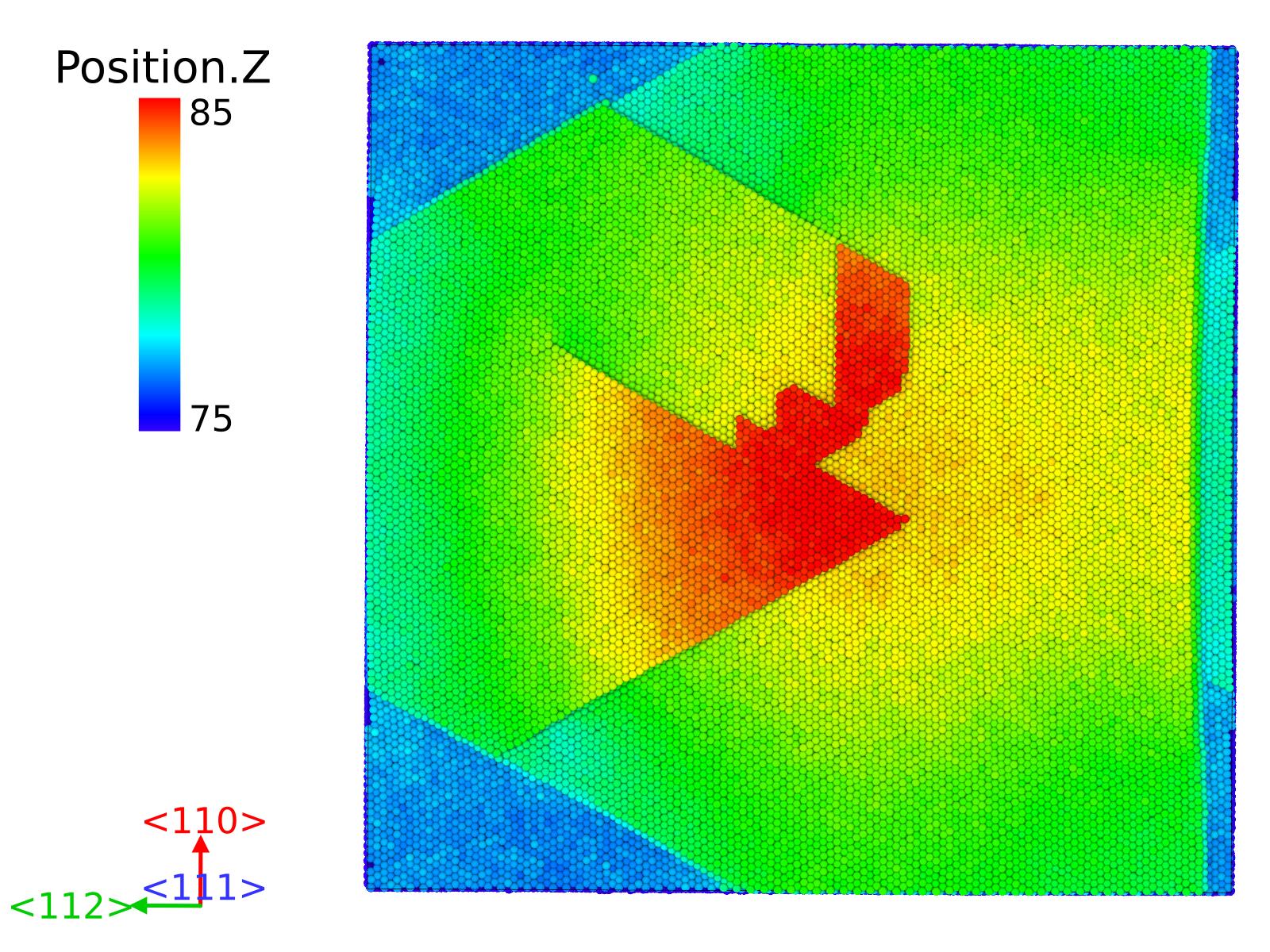}}
\end{center}
\caption{\label{fig:higher_surface_hemisphere} Evolution of the (a) \hkl(110), (b) \hkl(100) and (c) \hkl(111) surfaces after 100 ps for \nhv = 2 in the hemispheric bubble.}
\end{figure}

In Figure \ref{fig:higher_surface_hemisphere} we consider the final form of the differently oriented surfaces under the hemispheric bubble. We clearly observe that shape of the protrusions are similar as in Figure \ref{fig:higher_surface_disk}, however there are some remarkable differences. We see that in the \hkl(110) surface (Figure \ref{fig:higher_surface_hemisphere} (a)) the hemispheric bubble creates a clear superposition of rhomboid protrusions as the one observed in the disk bubble (see Figure \ref{fig:higher_surface_disk} (a)). In the \hkl(100) surface (Figure \ref{fig:higher_surface_hemisphere} (b)), we see several protrusions in the with square shape, as a product of the dislocations emitted from the bubble in the \hkl<110> directions. In Figure \ref{fig:higher_surface_hemisphere} (c) we notice that as in the disk-bubble case for the \hkl(111) case, we see the formation of triangular protrusions in the surface, where reach the larger height in the center(see Figure \ref{fig:disk_protrusions} (c)).

\begin{figure}[H]
\begin{center} 
%\subfloat[\hkl(110)]{\includegraphics[width=.5\linewidth]{figures/disk_void_small_cell/disloc_higher_110_n1p2_100ps.jpg}}
\subfloat[\hkl(110)]{\includegraphics[width=.5\linewidth]{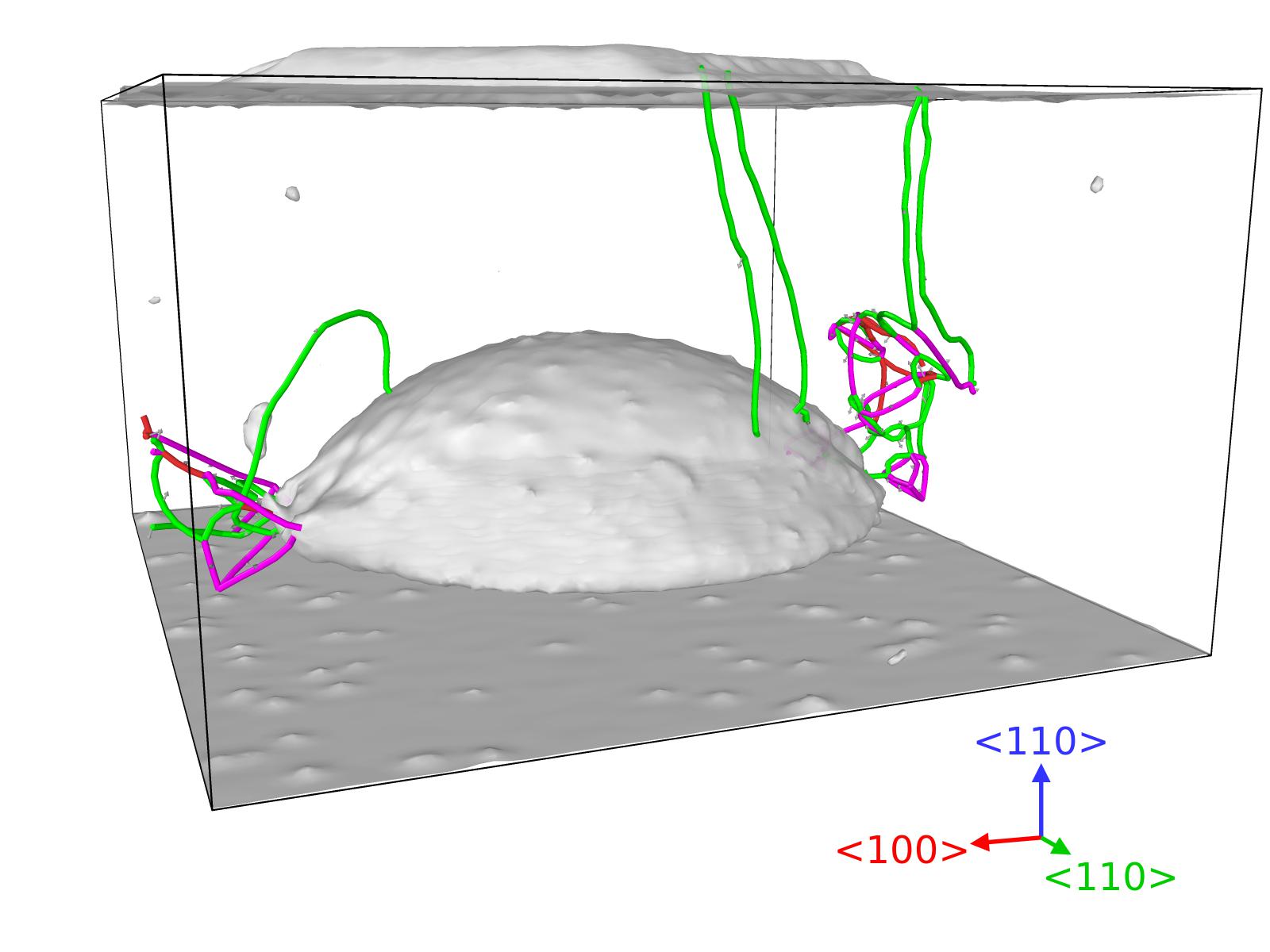}}
%\subfloat[\hkl(100)]{\includegraphics[width=.5\linewidth]{figures/disk_void_small_cell/disloc_higher_100_n1p2_100ps.jpg}}
\subfloat[\hkl(100)]{\includegraphics[width=.5\linewidth]{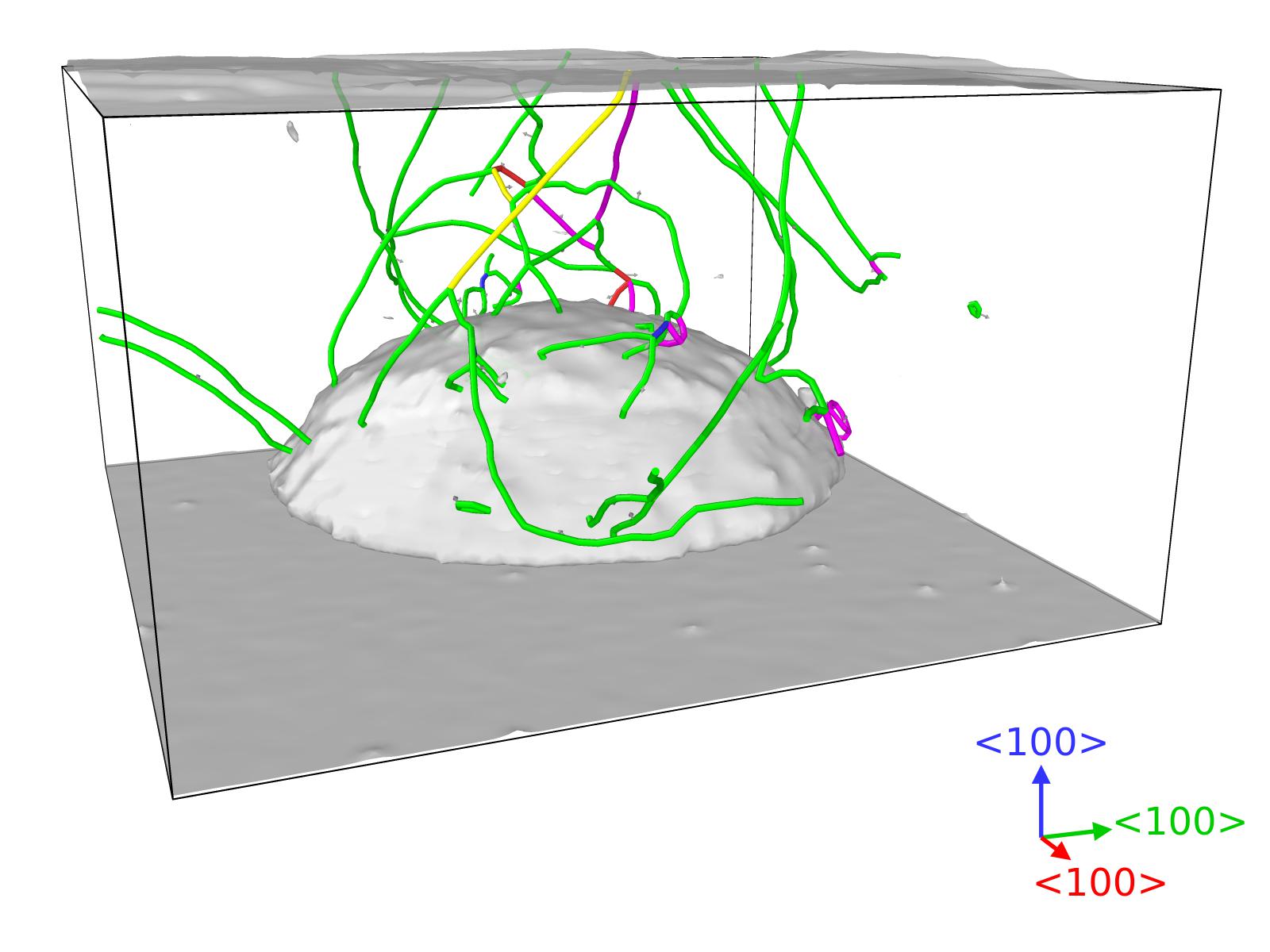}}\\
%\subfloat[\hkl(111)]{\includegraphics[width=.5\linewidth]{figures/disk_void_small_cell/disloc_higher_111_n1p2_100ps.jpg}}
\subfloat[\hkl(111)]{\includegraphics[width=.5\linewidth]{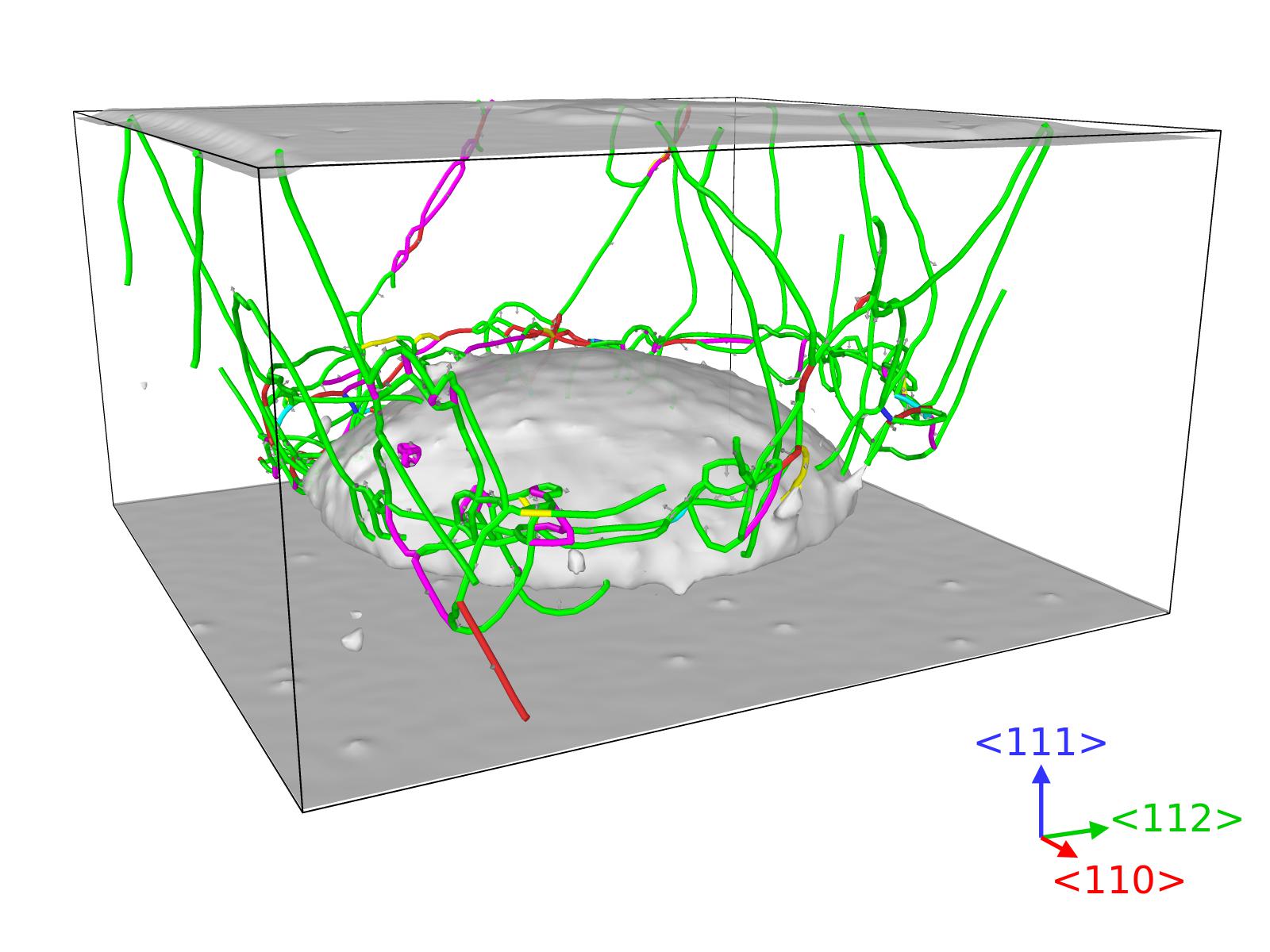}}
\end{center}
\caption{\label{fig:higher_surface_hemisphere_disloc} Configuration of the hemispheric bubble cells for \nhv = 2 in \hkl(110) case (a), \hkl(100) case (b) and \hkl(111) (c) after 100 ps. The dislocation lines identified by DXA \cite{dxaanalysis} are shown in different colors corresponding to the type of the dislocation. Green color shows Shockley partials, blue shows perfect partials, red shows "others" dislocations, pink shows stair-rod, yellow shows Hirth and cyan shows Frank partial dislocations.}
\end{figure}

In this case we also follow the dislocation network created in these cells. We notice that the part of the bubble of maximum stress is the bottom edge of the bubble, where generally all the dislocations starts at the beginning of the simulations. 

In the \hkl(110) case (see movie 110-hemisphere-n2.mp4 and Figures \ref{fig:higher_surface_hemisphere} (a) and \ref{fig:higher_surface_hemisphere_disloc} (a)), we see how the leading partial dislocations start to develop from the edge of the bubble and grow towards the surface. These dislocations expand in the four different \hkl{111} planes that eventually form the rhomboid shape of the protrusions, once they are absorbed by the surface. In this case, we observe that the process occur with more strength than in the disk bubble case, since several parallel atomic planes in the \hkl{111} planes creates the final rhomboid protrusion.

Considering the \hkl(100) case (see movie 100-hemisphere-n2.mp4 and Figures \ref{fig:higher_surface_hemisphere} (b) and \ref{fig:higher_surface_hemisphere_disloc} (b)), we observe that most of the dislocations start at the bottom edge, but some of them develop from parts higher in the bubble. These dislocations try to for the upper part of the stacking fault octahedron, but due to the high pressure, it does not entirely form, contrarily to the disk bubble case (see Figure \ref{fig:SF_100}). These dislocations form the eventual squared protrusion in the \hkl(100) surface.

The case corresponding to the \hkl(111) surface (see movie 111-hemisphere-n2.mp4 and Figures \ref{fig:higher_surface_hemisphere} (c) and \ref{fig:higher_surface_hemisphere_disloc} (c)), we observe that the leading partials starts to grow from the bubble, in some cases parallel to the surface from the upper part of the void and, at the same time, others start from the edge. These latter dislocations will form the triangular shape shown in Figures \ref{fig:higher_surface_hemisphere} (c), which in this case, is left a part of the atoms in the center of the protrusion slightly more elevated than the rest, reminding to the shape observed in the smaller cell containing a disk-shaped bubble (see Figure \ref{fig:disk_protrusions} (c)).

\section{Discussion}
\label{sec:discussion}

%\ALC{Introduce some discussion on this. Check paper Orientation dependence of dislocation structure in surface grain of pure copper deformed in tension. For knowing which grains are more likely to yield:$https://doi.org/10.1016/j.actamat.2020.11.016$}

% Probably the part of the penetration depth can be brought here quite nicely comparing the formation of voids at 100 and 111, and explain why they will form later in the 110 grains. In case Catarina finds some diamond shape protrusions would be outstanding, but with this, I think that we have enough. Also the information on volume, dislocation density and stress, might be a bit too much for the paper.

%At the sight of these results, we use MDRANGE code \cite{Nor94b} to calculate the penetration depth of 45 keV H on Cu on the different crystallographic orientations. We simulated 10000 cases in each direction.%, which are selected to be in the range of [0$^{\circ}$-89$^{\circ}$] and [0$^{\circ}$-90$^{\circ}$] with increment of 1$^{\circ}$, respectively.

\begin{figure}[H]
\begin{center} 
\subfloat[]{\includegraphics[width=0.5\columnwidth]{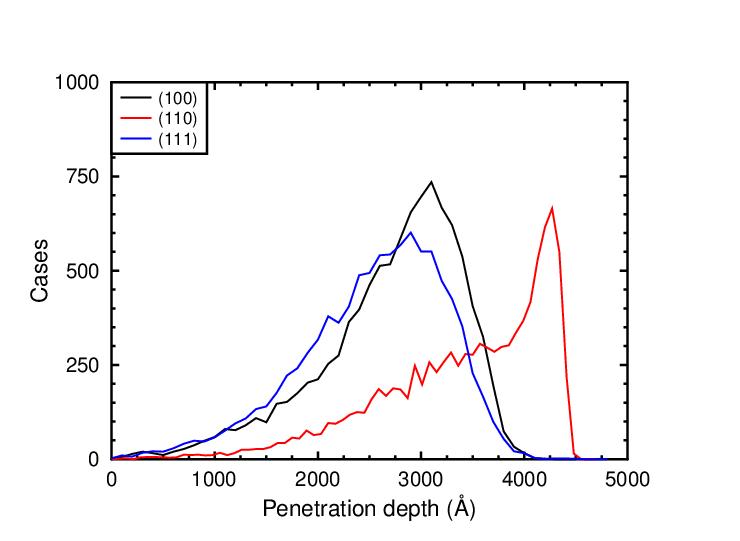}}
\subfloat[]{\includegraphics[width=0.5\columnwidth]{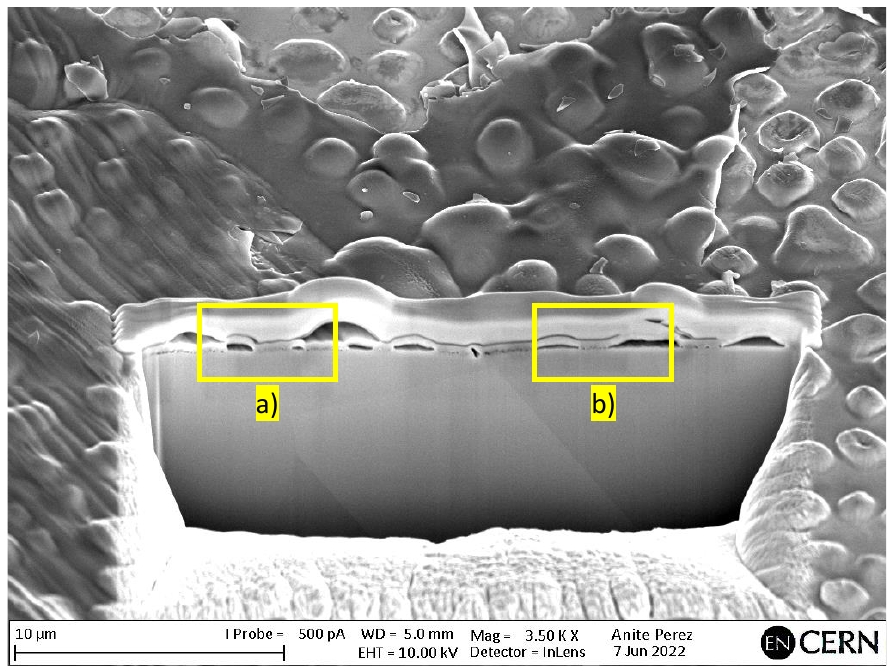}}\\
\subfloat[]{\includegraphics[width=0.5\columnwidth]{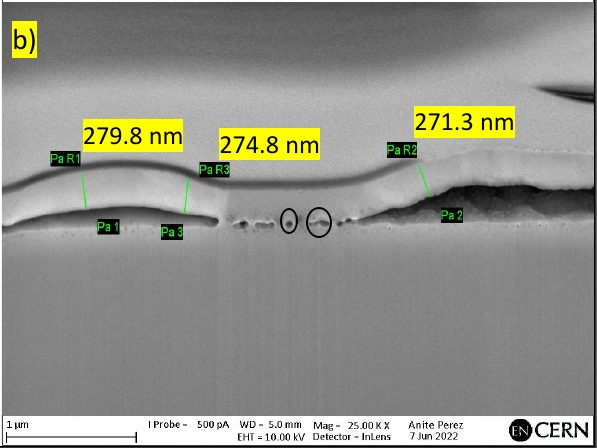}}
\subfloat[]{\includegraphics[width=0.5\columnwidth]{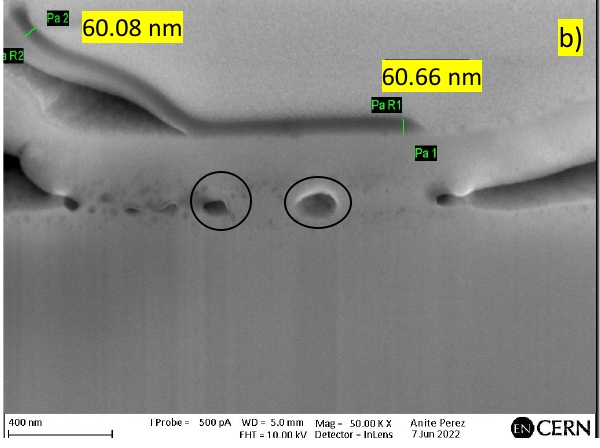}}
\end{center} 
\caption{\label{fig:penetration_blister} (a) Penetration depth of 45 keV H on the different crystallographic directions of Cu. (b) Blisters formed in the surface of two grains. \hkl(100) in b) and \hkl(111) in a). (c-d) Zoom-in of b) region. Small bubbles shown in black circles.}
\end{figure}

In Figure \ref{fig:penetration_blister} (a) we observe how the penetration depth is larger in the \hkl<110> direction (4\,056 \angstrom). For \hkl<100> (2\,710 \angstrom)and \hkl<111> (2\,775 \angstrom), we see that they are lower than for \hkl<110>, but rather similar to each other. We verify that in the Cu sample, the blisters formed close to the surface are at roughly the same distance in the \hkl(100) and \hkl(111) grains ($\sim$3\,000 Å). We see that in the \hkl<111> direction, the ions loose energy faster than in the \hkl<100>, which may explain the difference in the blister/protrusion density between the two grains, since the voids generated in the former can be smaller. This means that higher fluences are needed to obtain larger bubbles in order to reach higher densities of protrusions as in \hkl<100>-oriented grains. The similar penetration depth of H at this energy is consistent with Moreno et al. work \cite{moreno1997}, where they found a similar stopping power for He in differently oriented Cu.

Moreover, we see that the shapes of the protrusions observed in Figure \ref{fig:penetration_blister} (b) are in agreement with the obtained using MD for \hkl(100) and \hkl(111). In the case of the \hkl(100) grain, is easy to identify the square shape among the protrusions formed in the surface (see Figures \ref{fig:disk_protrusions} (b) and  \ref{fig:higher_surface_hemisphere} (b)). In some cases we observe that the protrusions has merged, producing slightly different shapes. In the case of the \hkl(111) grain, we observe rather circular shapes (see Figures \ref{fig:exp_1} and \ref{fig:penetration_blister} (b)), which are in agreement with the ones predicted by MD (Figures \ref{fig:disk_protrusions} (c) and  \ref{fig:higher_surface_hemisphere} (c)). 

Furthermore, we observe how the effect in the surface is similar regardless of the shape of the bubble. We see that in the both forms studied, the yielding points of the bubbles are those where the change of curvatures is larger i.e., the borders joining the flat surfaces with the curved ones. The growth of the bubble is done via emission of dislocations \cite{Lop21}, and these dislocations interact with the surface inducing the protrusions that are experimentally observed. The simulated bubbles are considerably smaller than the most noticeable ones observed in Figure \ref{fig:penetration_blister} (b). However, the size of the smaller bubbles marked in Figures \ref{fig:penetration_blister} (c-d), correspond to the ones simulated in this work.

{We observe that the emission of dislocations for the disk-shaped bubble occurs at \nhv=2, however, in previous study \cite{Lop21} it occurred at \nhv=3, while only minor defects were created around the spherical-shaped bubble. Many variables can change this evolution: the shape of the void might be determinant. The disk contains flat surfaces in two sides, and we have seen that the dislocations are created at the borders, where the change of curvature is larger. Moreover, the effect of the surface close to the bubble might affect this result compared with the results shown in Ref. \cite{Lop21}.}

In the disk-shaped bubble simulations, the border acts as the yielding point of the bubble, regardless the surface orientation, and the dislocations are emitted mostly from there.

%\ALC{Disk vs Hemisphere: We see how the surface of the bubble exposed to the cell surface is playing a determinant role, since it makes the protrusions to be wider, nevertheless depending on the orientation of the surface, less pronounced.}

\section{Conclusions}
\label{sec:conclusions}

In this work we have shown experimentally the effect of 45 keV H$^-$ on Cu, performing the EBSD analysis and identifying the effect of this irradiation on differently oriented grains.

Furthermore, using MD, we simulated the effect of H in different bubble shapes, pointing out the mechanisms which lead the bubbles to grow via dislocations emission and modifying the surface, creating protrusions. The protrusions shown computationally are similar in shape to the observed experimentally.

We have seen that the shape of the bubble, regardless of small differences, they create similar effect in the surface.

We have developed a computational model to show the effect of pressurized bubbles under differently oriented surfaces, which agrees considerably well with the experimental evidence.

\section*{Acknowledgements}

 Computer time granted by the IT Center for Science -- CSC -- Finland and the Finnish Grid and Cloud Infrastructure (persistent identifier urn:nbn:fi:research-infras-2016072533) is gratefully acknowledged. 
%\bibliography{mybib.bib}

%\putbib[mybib]

\end{bibunit}

\newpage

\section*{Supplementary material for "\textit{Hydrogen accumulation in copper}"}
\setcounter{figure}{0}
\renewcommand{\thefigure}{S\arabic{figure}}
\setcounter{section}{0}
\renewcommand{\thesection}{S\arabic{section}}
\setcounter{table}{0}
\renewcommand{\thetable}{S\arabic{table}}
\setcounter{page}{1}
\subsection{Volume \& Dislocation analysis}
\label{sec:vol_disloc_stress_analysis}
%%% Methods
The evolution of the volume change in the bubble, defined as $\Delta V = \frac{V_{final}-V_{initial}}{V_{initial}}$, is followed using the \textit{Construct Surface Mesh} from OVITO. For that, we do not consider the H atoms in the bubble. We also compute the strain using \textit{Atomic Strain} feature from OVITO, using a cutoff of 2.8 \angstrom. For the histograms, the \textit{Spatial binning} feature was used applying 50 bins in each of the 2 dimensions represented and 1 in the perpendicular to the plane. The mean value in the bin is represented.

\begin{figure}[H]
\begin{center} 
%\subfloat[Disk-bubble]{\includegraphics[width=.5\linewidth]{figures/volume_disloc_density/vol_change_disk.eps}}
%\subfloat[Disk-bubble]{\includegraphics[width=.5\linewidth]{figures/volume_disloc_density/disloc_evol_disk.eps}}\\
%\subfloat[Hemisphere-bubble]{\includegraphics[width=.5\linewidth]{figures/volume_disloc_density/vol_change_hemi.eps}}
%\subfloat[Hemisphere-bubble]{\includegraphics[width=.5\linewidth]{figures/volume_disloc_density/disloc_evol_hemi.eps}}
\subfloat[]{\includegraphics[width=.5\linewidth]{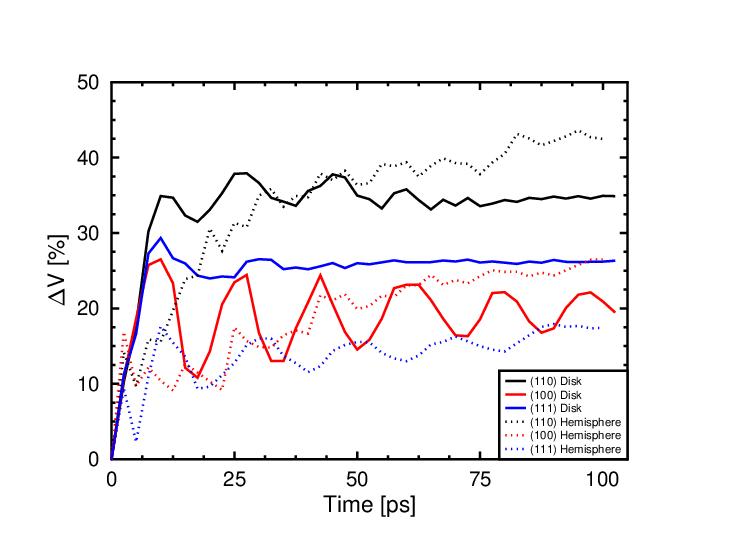}}
\subfloat[]{\includegraphics[width=.5\linewidth]{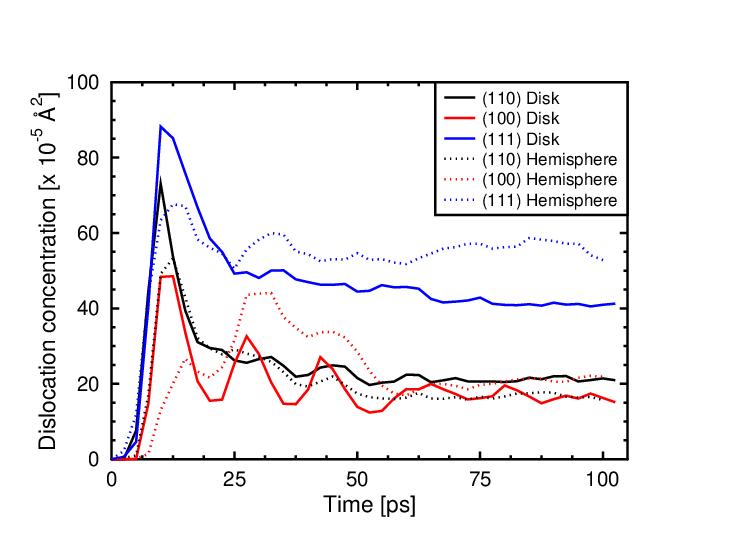}}
\end{center}
\caption{\label{fig:vold_dislocdensity} 
Evolution of the (a) change of volume of the void and (b) dislocation concentration in the differently oriented cells with a disk and hemisphere bubbles under \nhv=2.} 
\end{figure}

When comparing both bubble types containing the same pressure, we see observe several differences. In the volume change (Figure \ref{fig:vold_dislocdensity} (a)), we see that it is larger in the \hkl(110) case for both bubbles, however the hemisphere case has a slower increase leading to higher change eventually. The \hkl(100) cases show a considerable discrepancy between the two bubbles in their evolution. The disk shows a \textit{breathing} evolution of the volume and, on the other hand, the hemisphere shows a stable increase of the volume change. The \hkl(111)-oriented surfaces show less fluctuations but the highest discrepancy in the total change. We see that the dislocations grow on the border similarly in all the cases. However, the orientation of the surface plays a role absorbing them, being this effect greater in the \hkl(111) case, where some large stacking faults planes are in contact with the surface of the bubble and the surface of the cell, in the disk-bubble case.

In Figure \ref{fig:vold_dislocdensity} (b) we see the dislocation concentration in each case. We do not see clear differences between the two geometries. On the other hand, the \hkl(111) cases provide the highest concentration in both cases.

\section{Stress-strain Analysis}

We include information extracted from the stress and strain tensor.

\begin{table}[H]
\begin{tabular}{|l|l|l|l|l|l|}
\hline
 & Disk & Hemisphere &  & Disk & Hemisphere \\ \hline
e$^{\hkl(110)}_{shear}$ & $0.0610 \pm 2\times10^{-4}$ & $0.0879\pm 3\times10^{-4}$ & $\sigma_{shear}^{\hkl(110)}$ & $63.48 \pm 0.02$ & $64.85\pm 0.03$ \\ \hline
e$^{\hkl(100)}_{shear}$  & $0.0440 \pm 2\times10^{-4}$  & $0.0848 \pm 4\times10^{-4}$ & $\sigma_{shear}^{\hkl(100)}$ & $62.7 \pm 0.1$ & $65.27\pm 0.03$ \\ \hline
e$^{\hkl(111)}_{shear}$  & $0.0640 \pm 3\times10^{-4}$ & $0.0750 \pm 2\times10^{-4}$ & $\sigma_{shear}^{\hkl(111)}$ & $62.0 \pm 0.1$ & $65.19\pm 0.02$ \\ \hline
e$^{\hkl(110)}_{volumetric}$ & $0.0151\pm 6\times10^{-5}$ & $0.0112\pm 1\times10^{-4}$ & $\sigma_{hydrostatic}^{\hkl(110)}$ & $4.3 \pm 0.1$ & $-2.50\pm0.05$ \\ \hline
e$^{\hkl(100)}_{volumetric}$  & $0.0096\pm 6\times10^{-5}$ & $0.0070\pm 1\times10^{-4}$ & $\sigma_{hydrostatic}^{\hkl(100)}$ & $-0.035 \pm 0.2$ & $-4.1\pm0.1$ \\ \hline
e$^{\hkl(111)}_{volumetric}$  & $0.0067\pm 5\times10^{-5}$ & $0.0081\pm 5\times10^{-5}$ & $\sigma_{hydrostatic}^{\hkl(111)}$ & $-1.7 \pm 0.1$ & $-8.03\pm0.07$ \\ \hline

\end{tabular}
\caption{\label{tab:strain_stress_data} 
Average value (per Cu atom) of strain and stress (in kbar) components in the differently oriented cells with a disk and hemisphere bubbles under \nhv=2. OVITO was used to obtain shear and volumetric strain.} 
\end{table}

From the volumetric stress, we can see that the larger hydrostatic strain is observed in the \hkl(110) case, something that can be observed from the volume change in Figure \ref{fig:vold_dislocdensity} (a). Conversely, the change is smaller in the \hkl(100) and the smallest is for the \hkl(111). On the other hand, when we analyse the dislocation density, we observe that, regardless of the difference of values between the cases, all reach a maximum about 10 ps. From that moment the protrusion starts to be visible in all the cases. The reason of this is that the dislocations have reached the surface, shifting the atoms in the z-direction, however, differently depending on the surface orientation.

\subsection{Shear Strain analysis}

\begin{figure}[H]
\begin{center} 
\subfloat[\hkl(110), $x-y$]{\includegraphics[width=.33\linewidth]{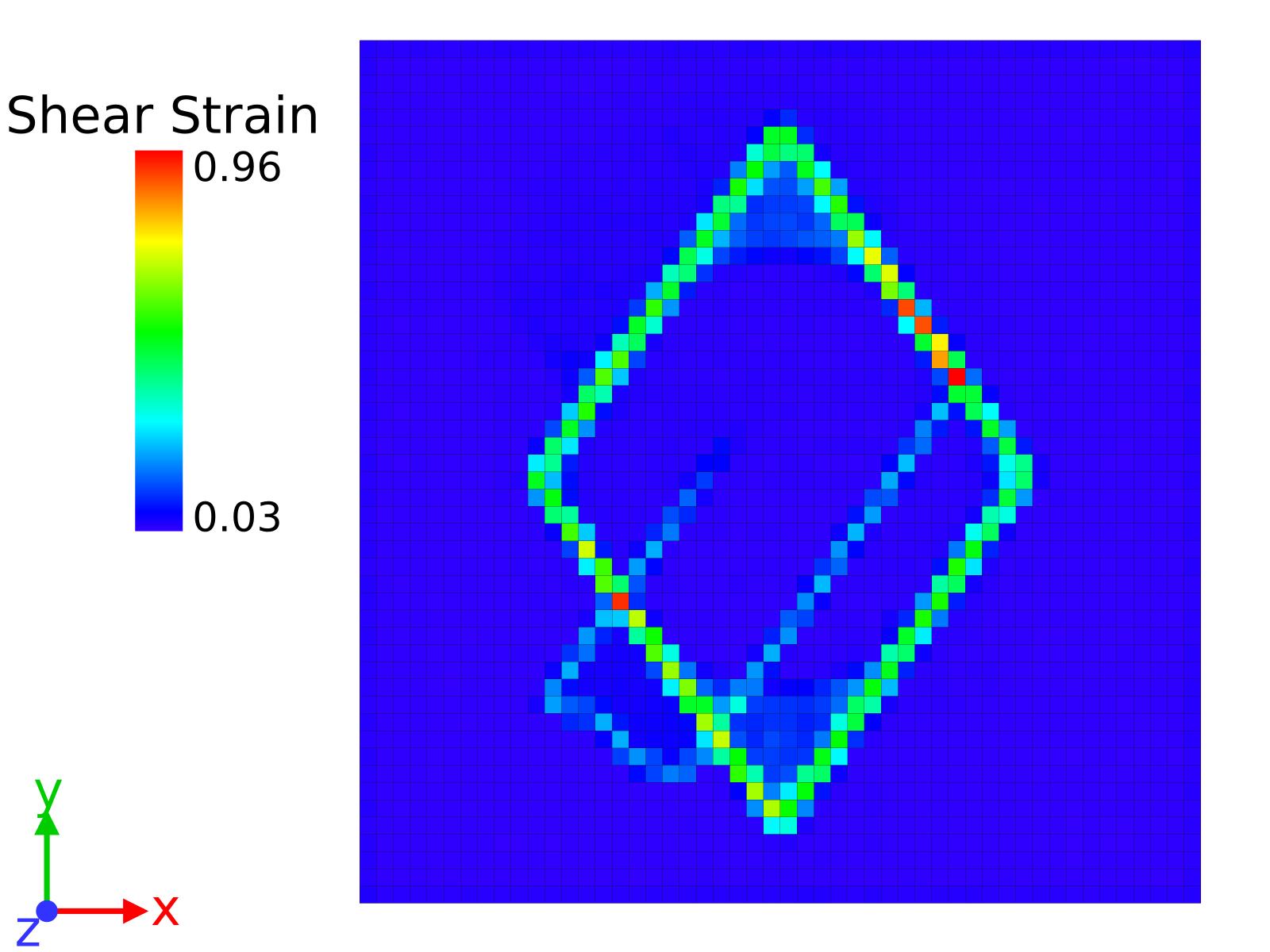}}
\subfloat[\hkl(110), $x-z$]{\includegraphics[width=.33\linewidth]{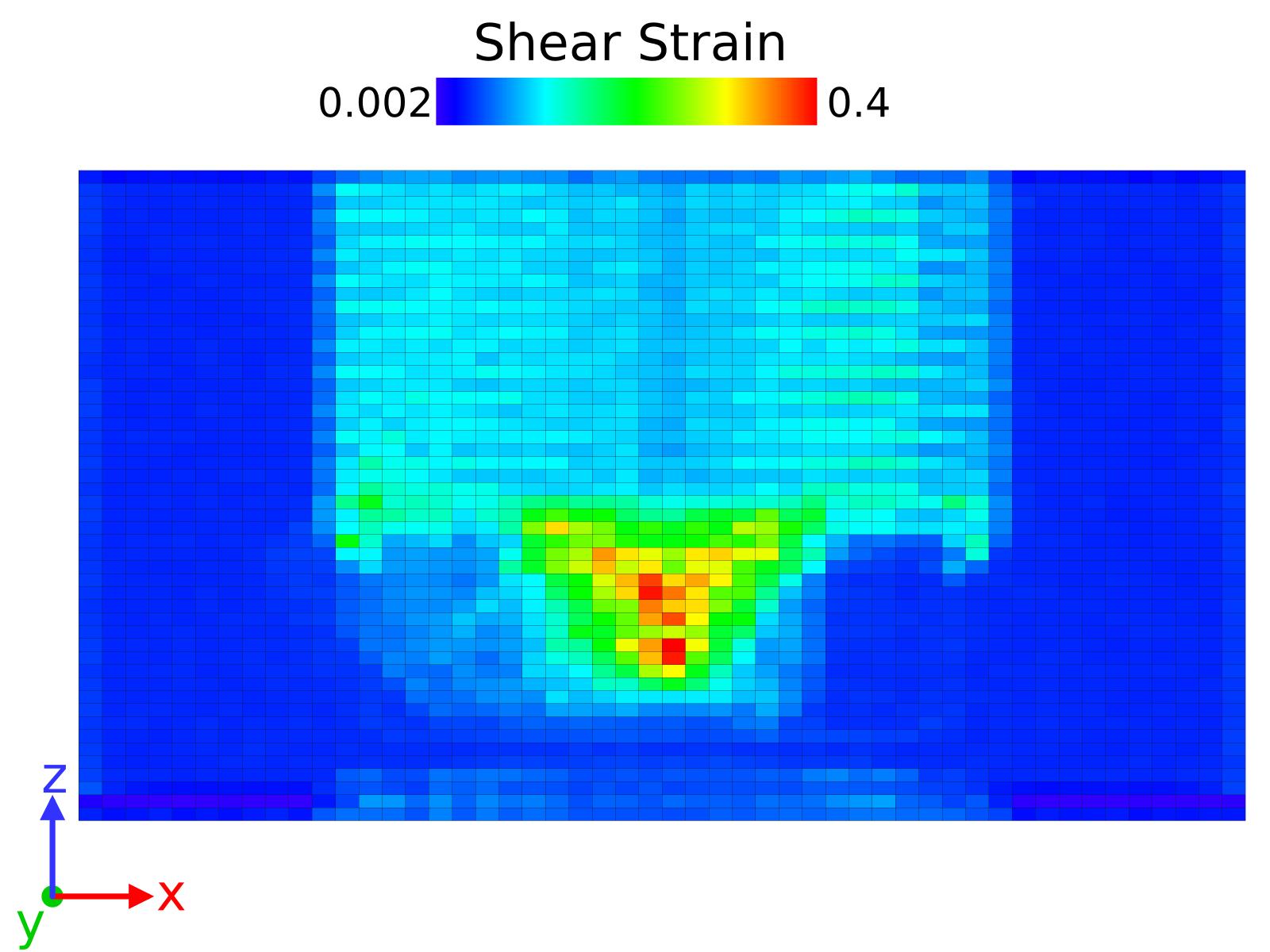}}
\subfloat[\hkl(110) $y-z$]{\includegraphics[width=.33\linewidth]{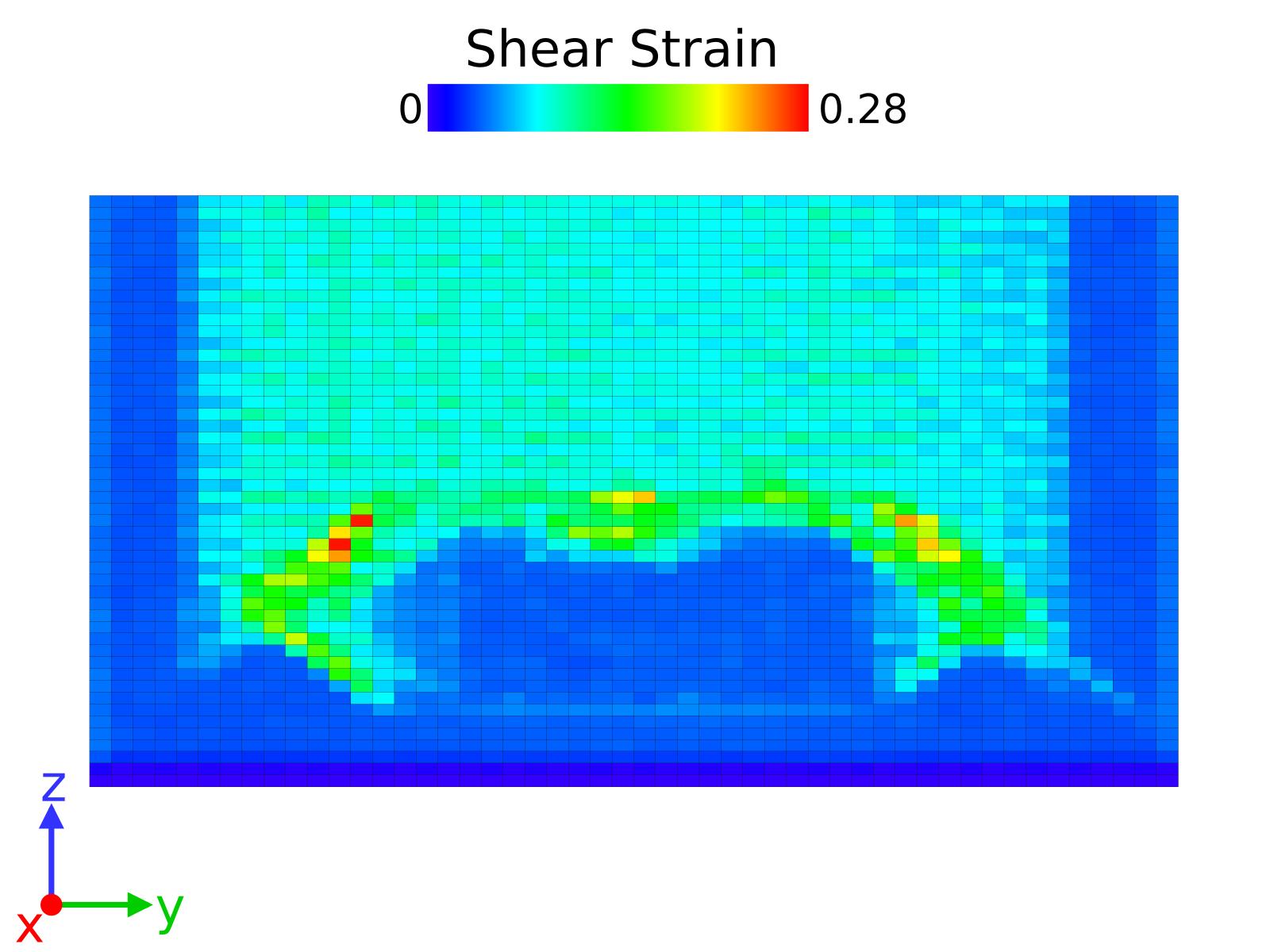}}\\
\subfloat[\hkl(100), $x-y$]{\includegraphics[width=.33\linewidth]{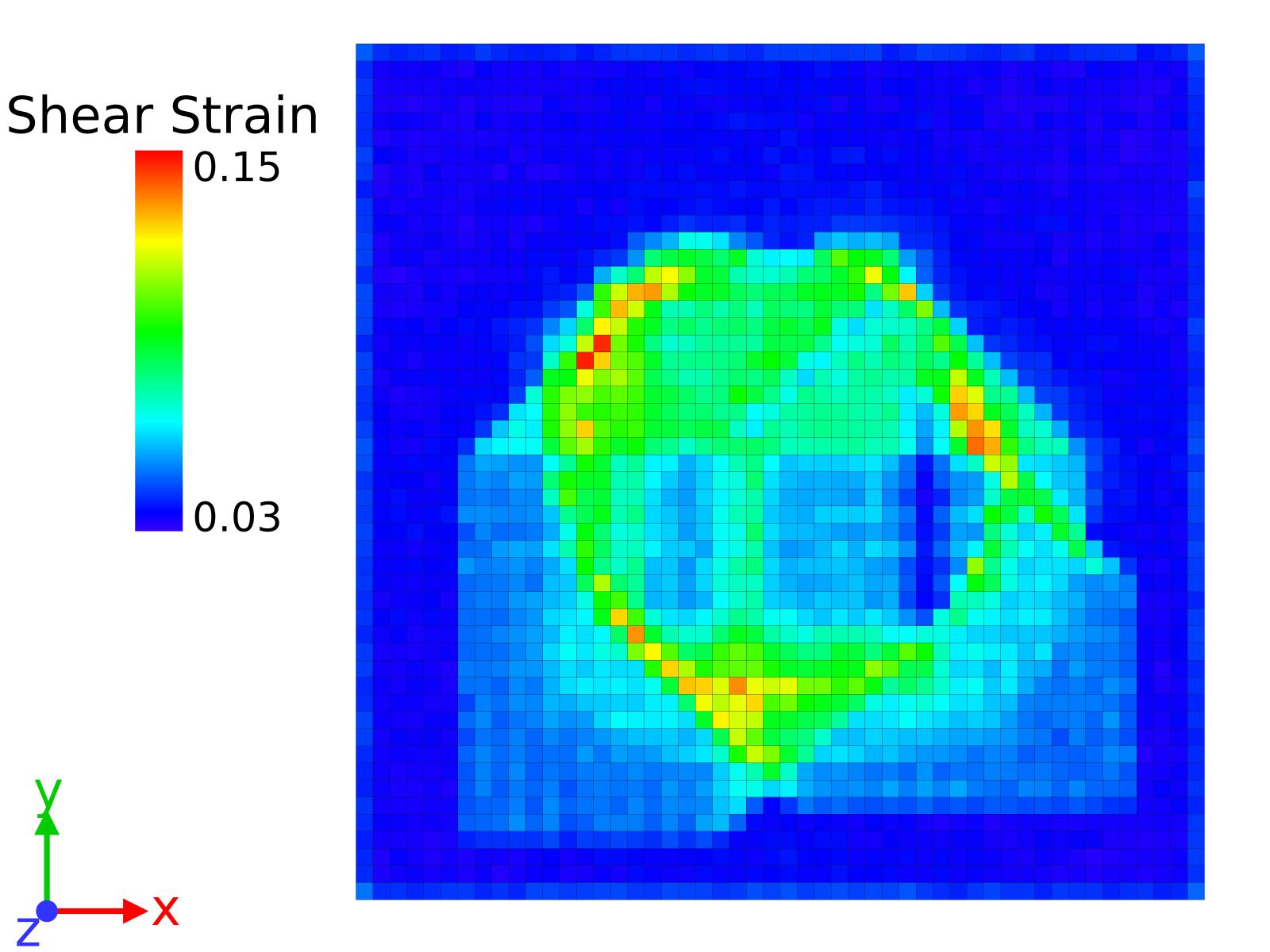}}
\subfloat[\hkl(100) $x-z$]{\includegraphics[width=.33\linewidth]{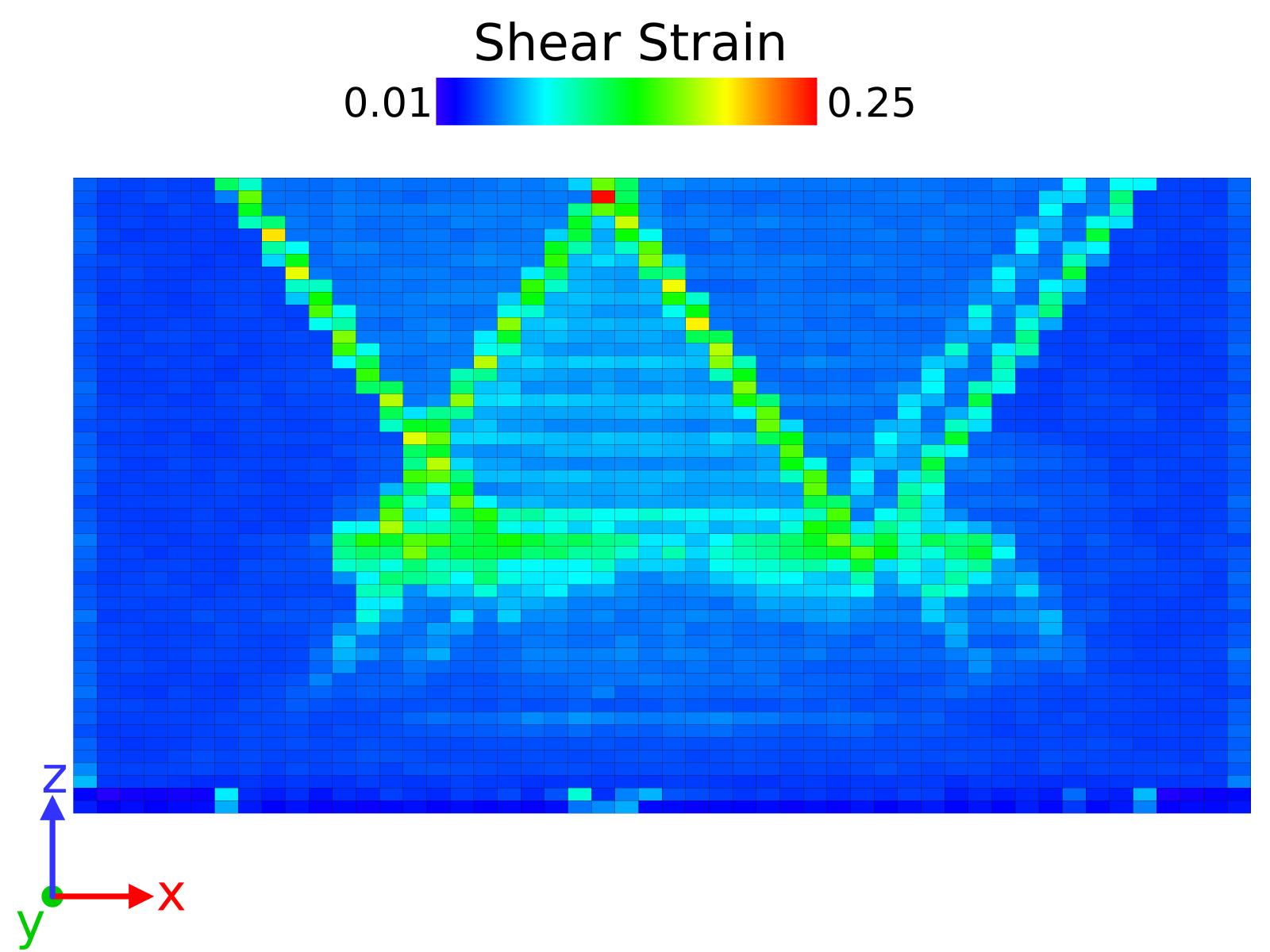}}
\subfloat[\hkl(100) $y-z$]{\includegraphics[width=.33\linewidth]{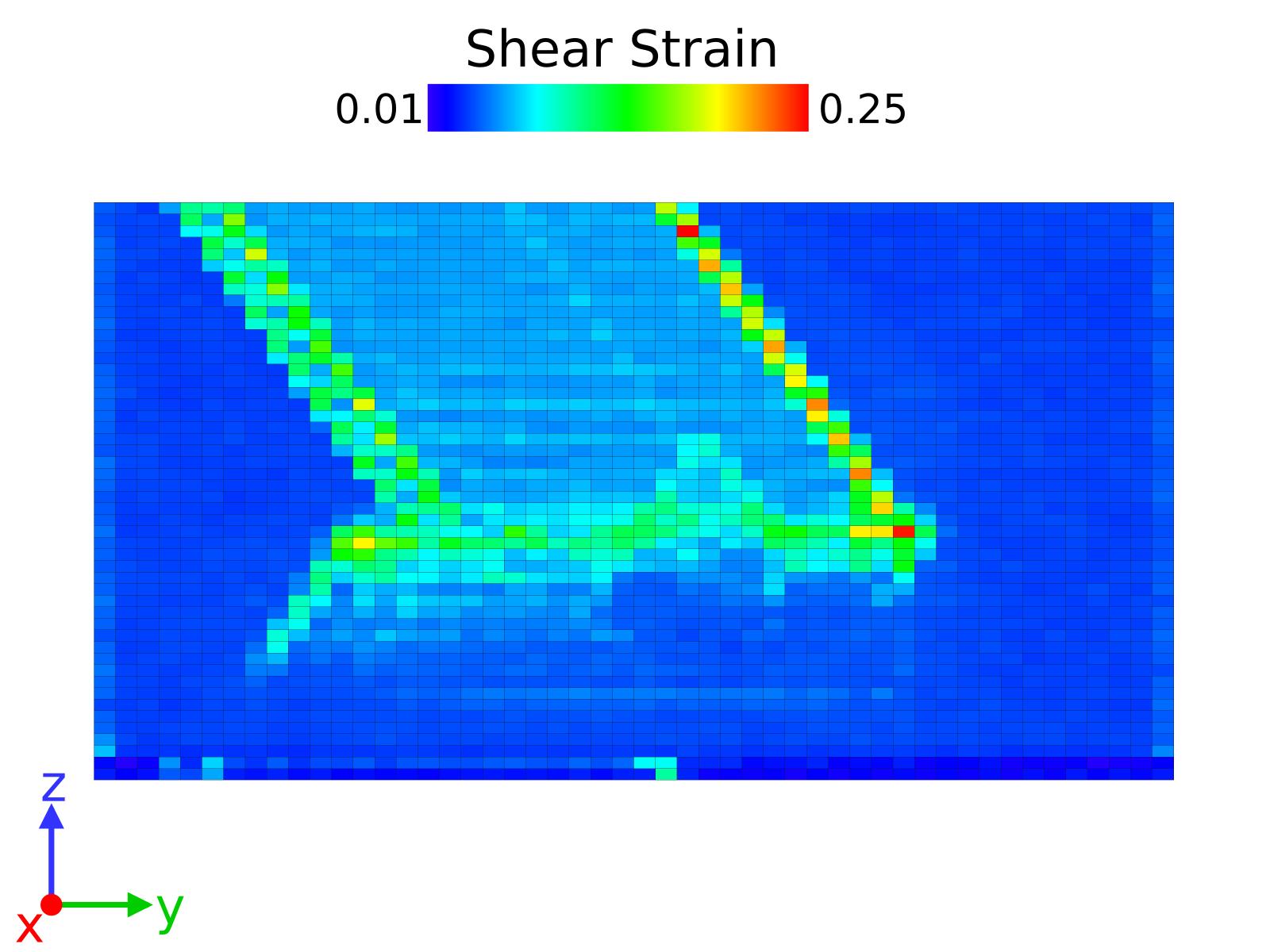}}\\
\subfloat[\hkl(111), $x-y$]{\includegraphics[width=.33\linewidth]{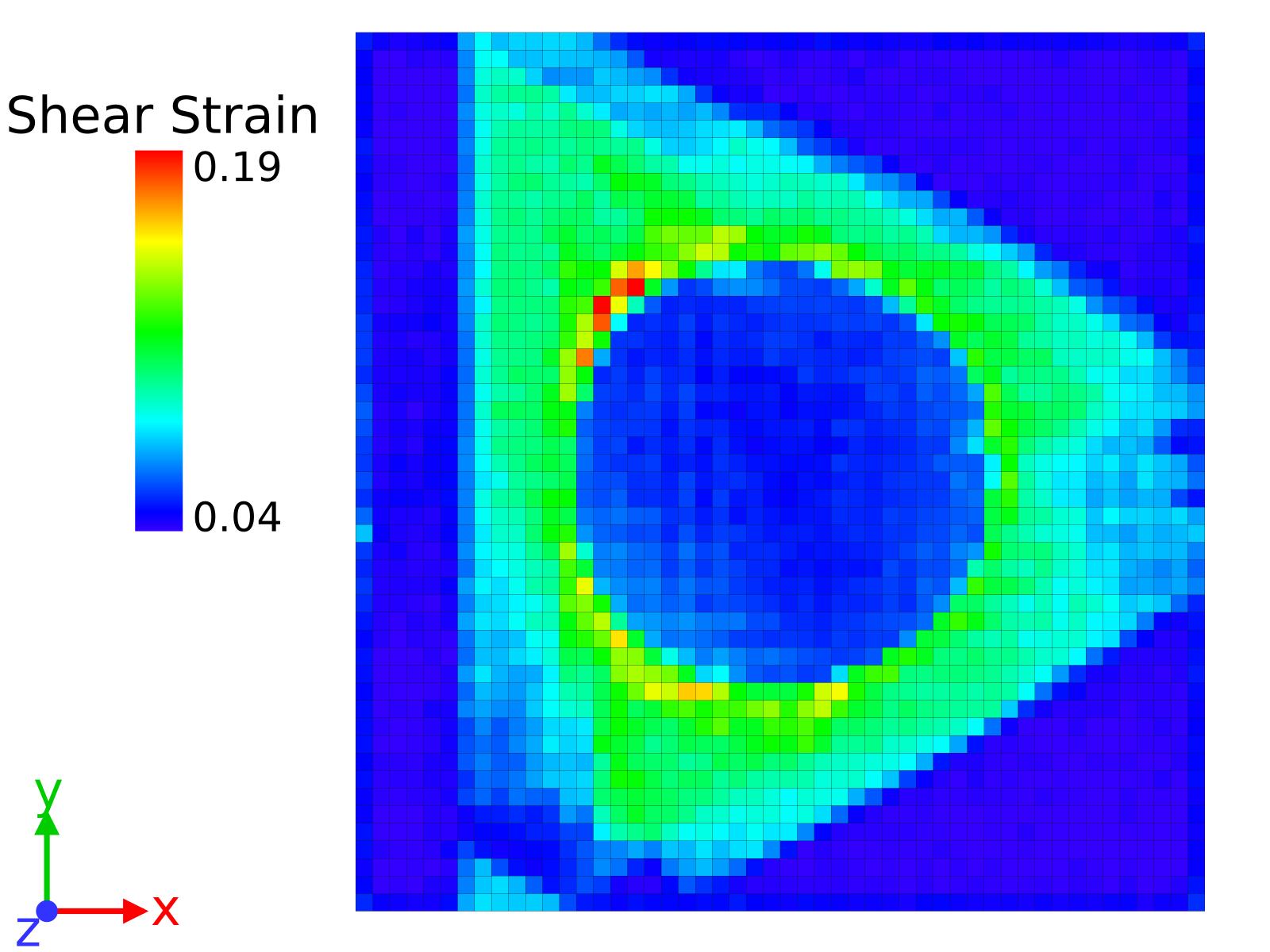}}
\subfloat[\hkl(111) $x-z$]{\includegraphics[width=.33\linewidth]{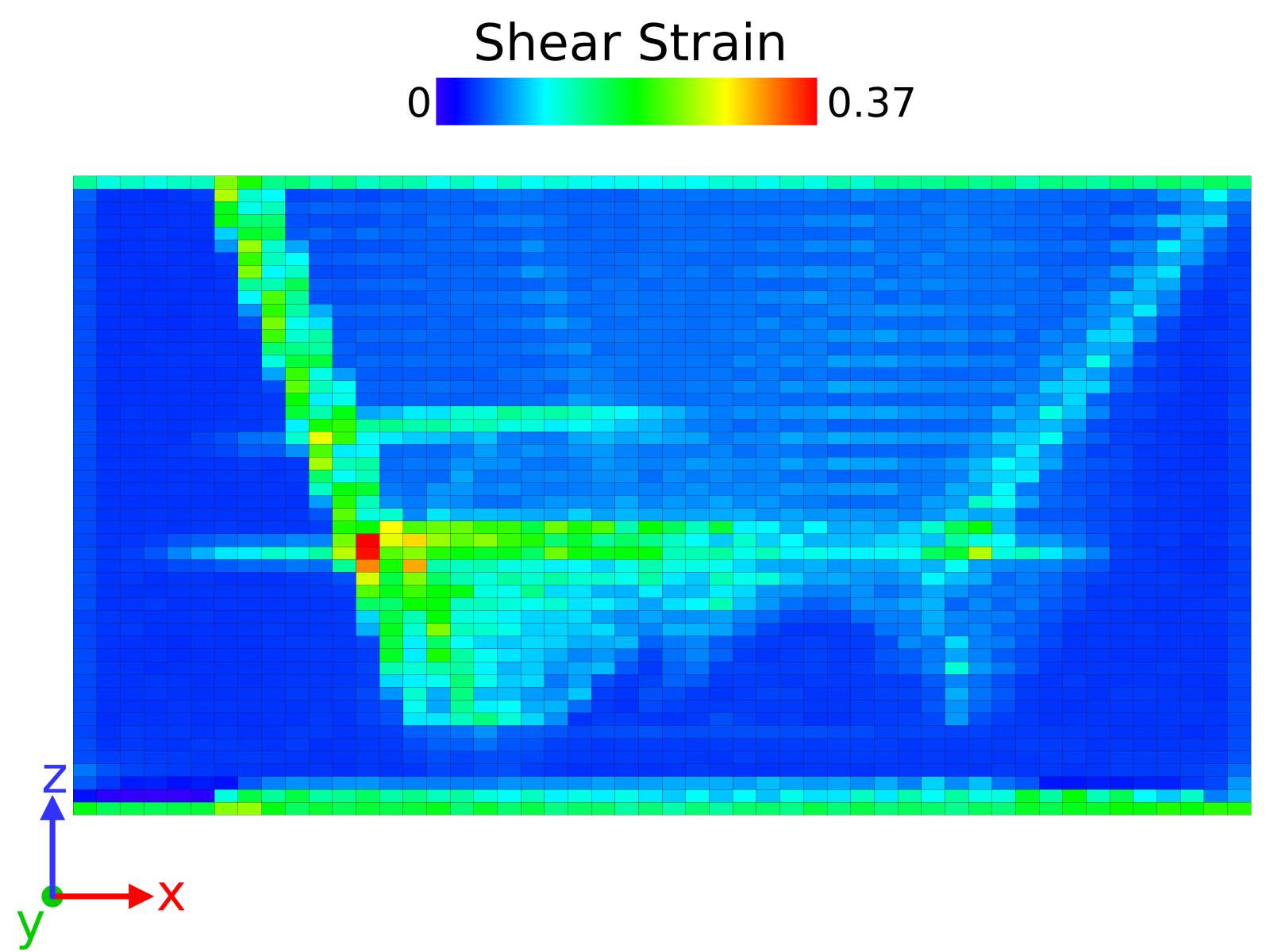}}
\subfloat[\hkl(111) $y-z$]{\includegraphics[width=.33\linewidth]{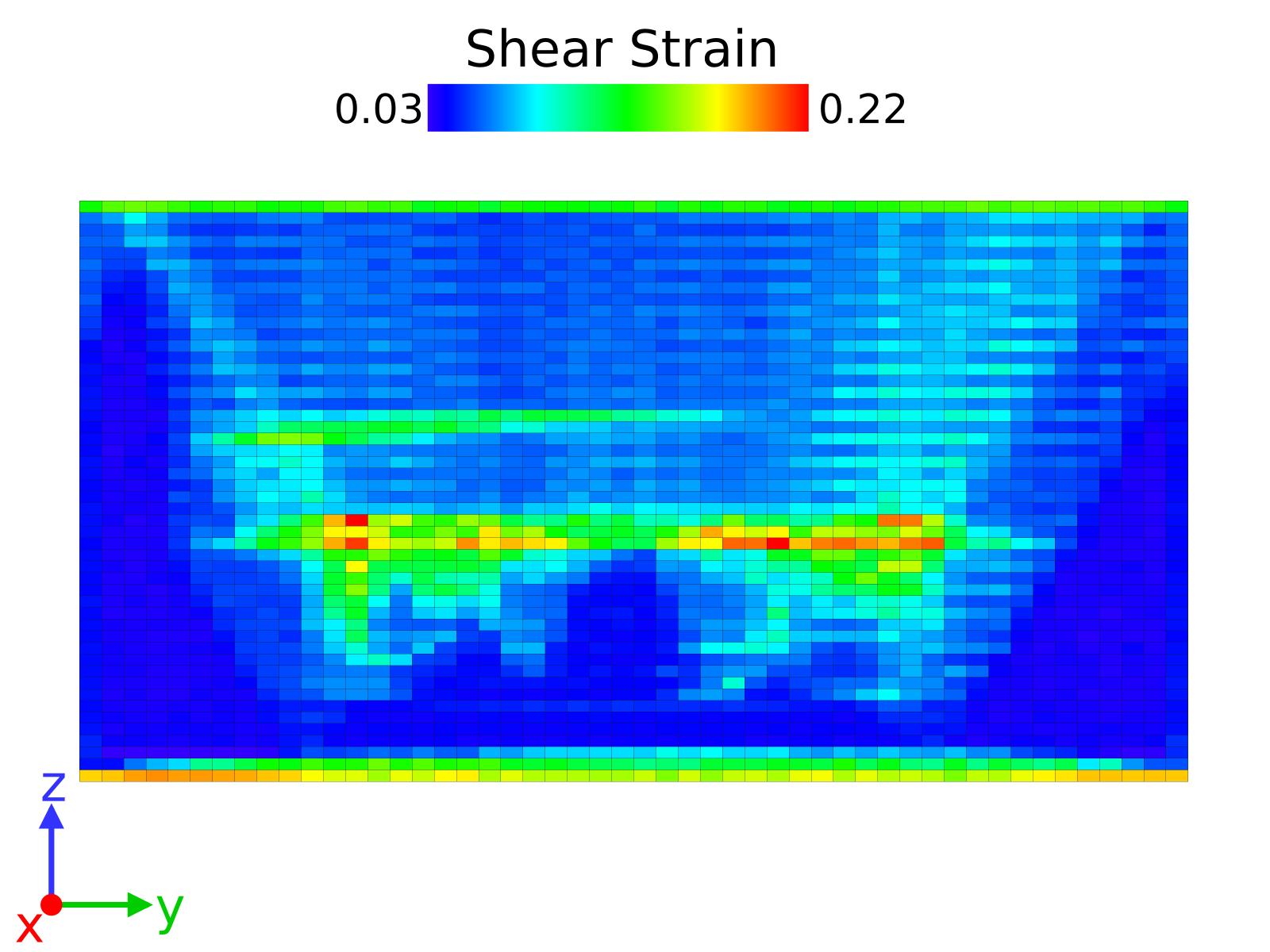}}
\end{center}
\caption{\label{fig:disk_shear_strain_all_histograms} 2-D histograms using 50$\times$50 bins in each direction representing the mean value of shear stress after 100 ps in every bin for \hkl(110) (a-c), \hkl(100) (d-f) and \hkl(111) (g-i) in the x-y, x-z and y-z planes respectively in each row. The cases for the disk bubble are presented.}
\end{figure}

\begin{figure}[H]
\begin{center} 
\subfloat[\hkl(110), $x-y$]{\includegraphics[width=.33\linewidth]{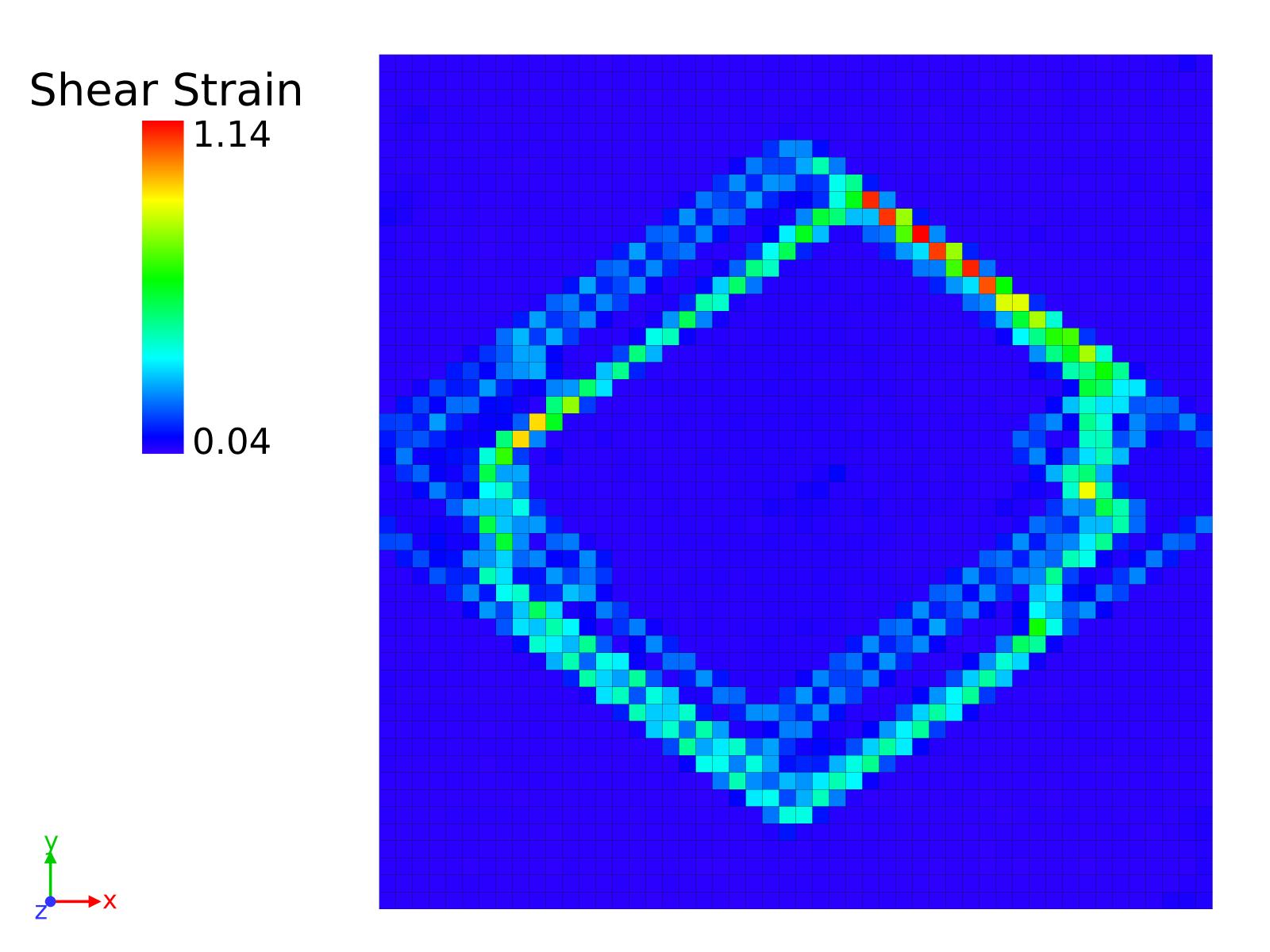}}
\subfloat[\hkl(110), $x-z$]{\includegraphics[width=.33\linewidth]{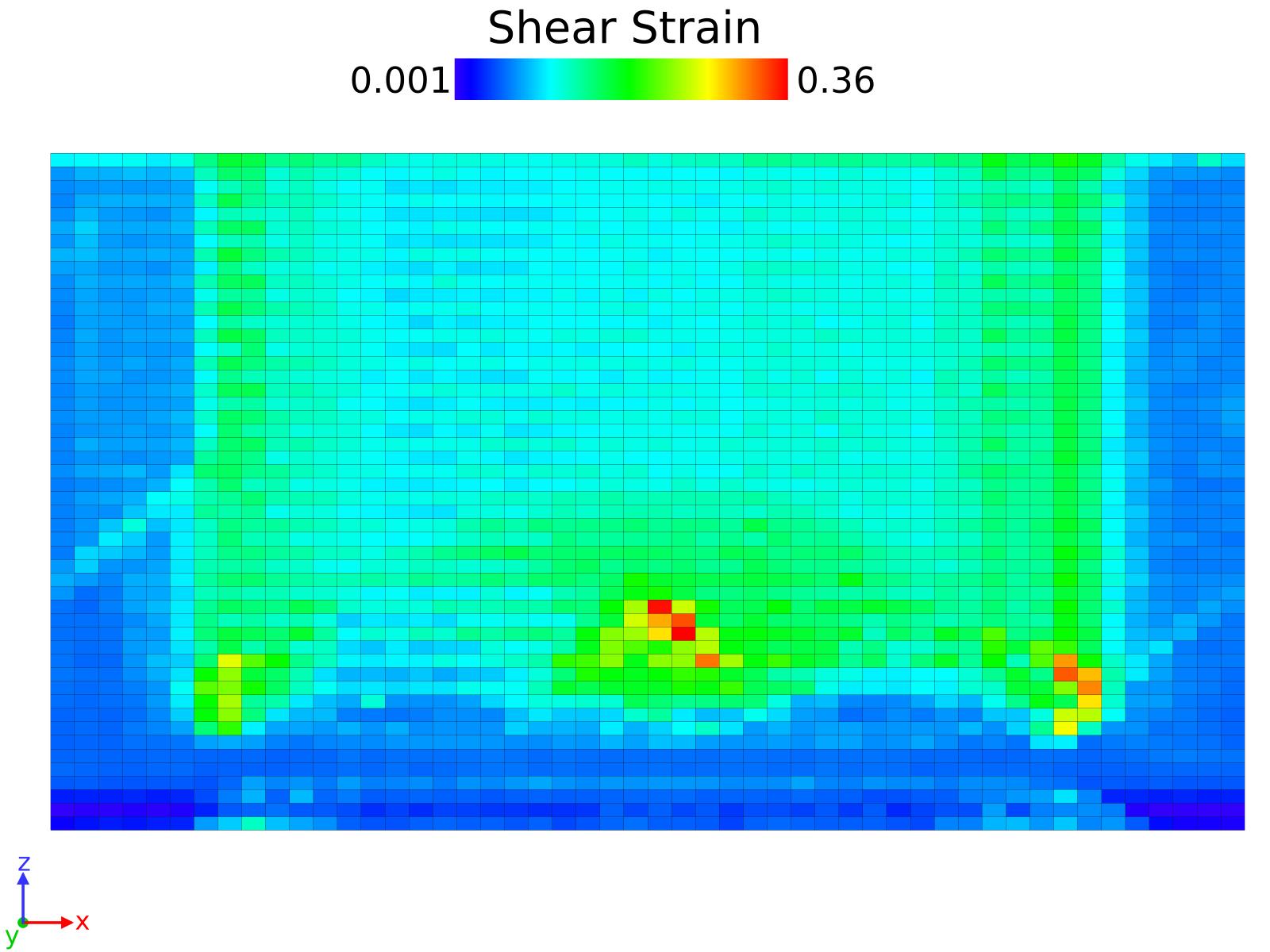}}
\subfloat[\hkl(110) $y-z$]{\includegraphics[width=.33\linewidth]{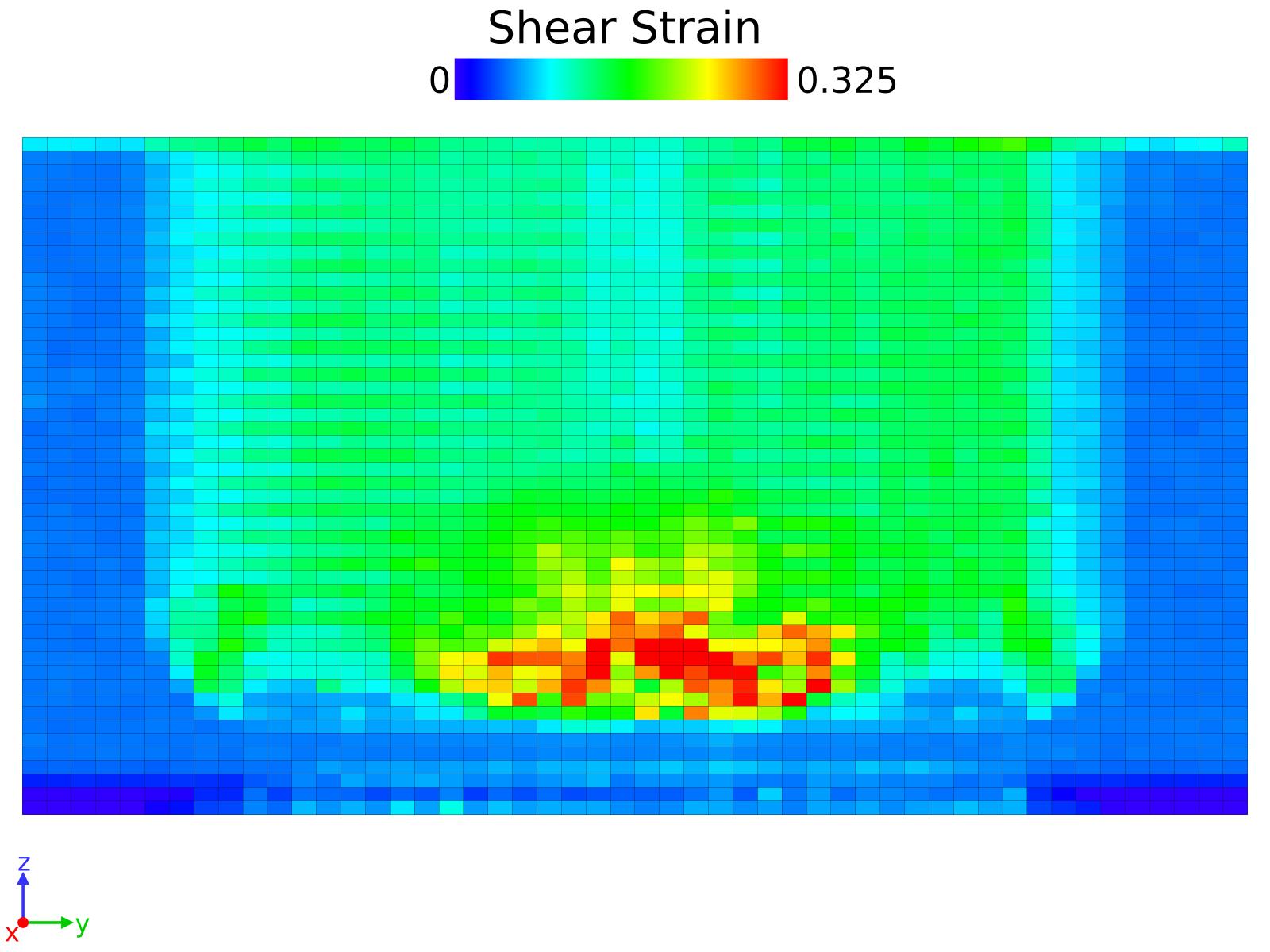}}\\
\subfloat[\hkl(100), $x-y$]{\includegraphics[width=.33\linewidth]{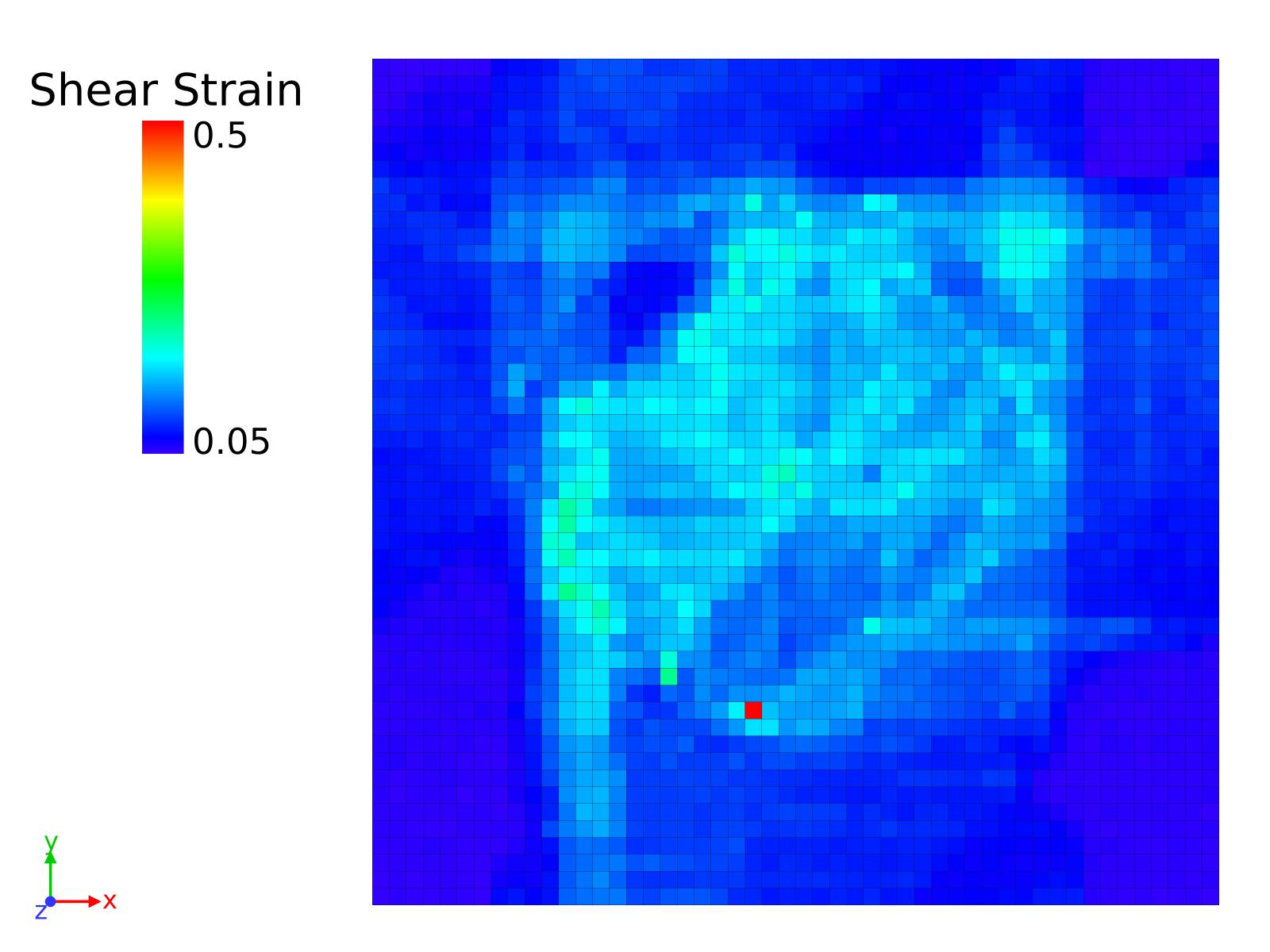}}
\subfloat[\hkl(100) $x-z$]{\includegraphics[width=.33\linewidth]{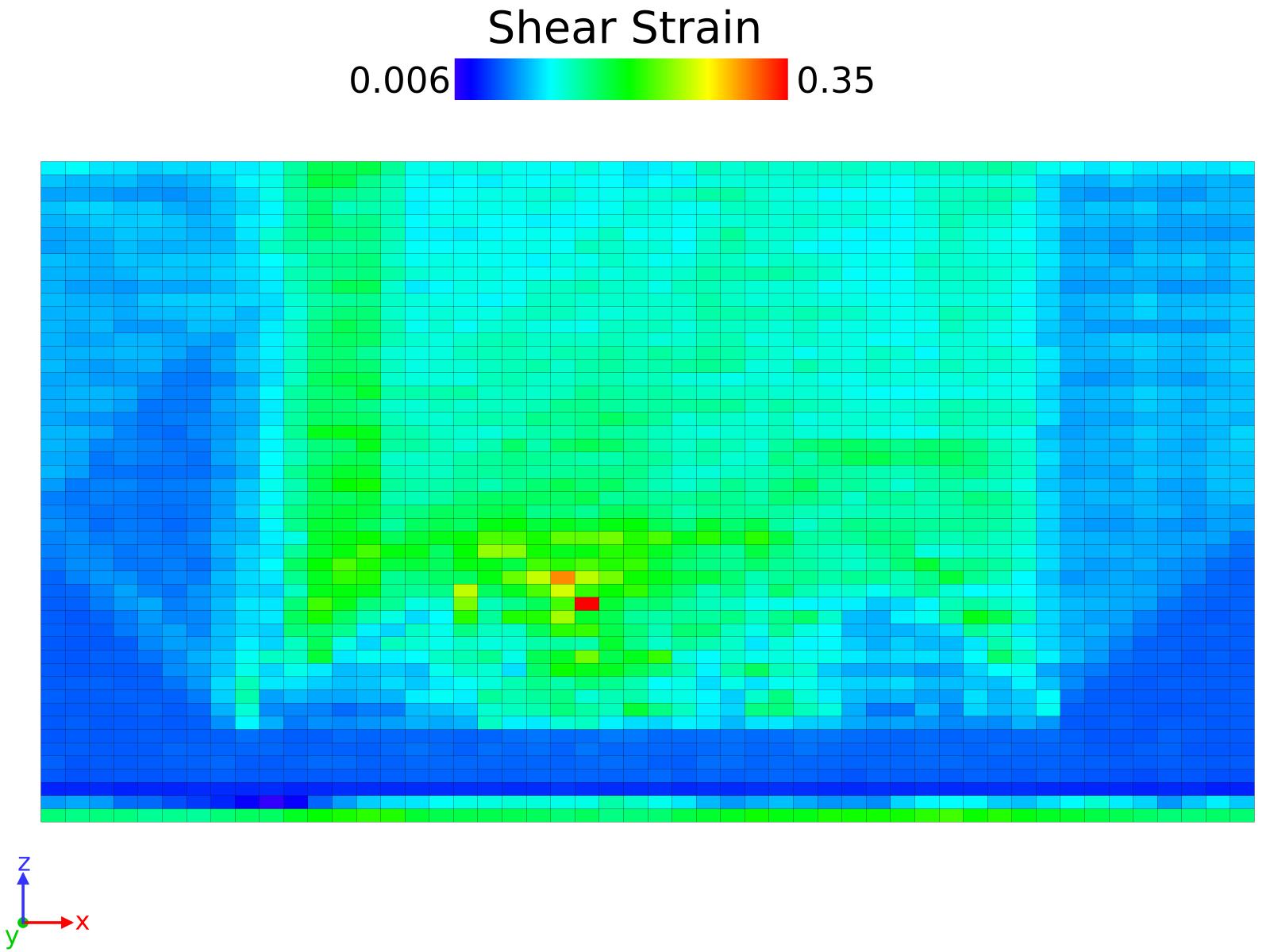}}
\subfloat[\hkl(100) $y-z$]{\includegraphics[width=.33\linewidth]{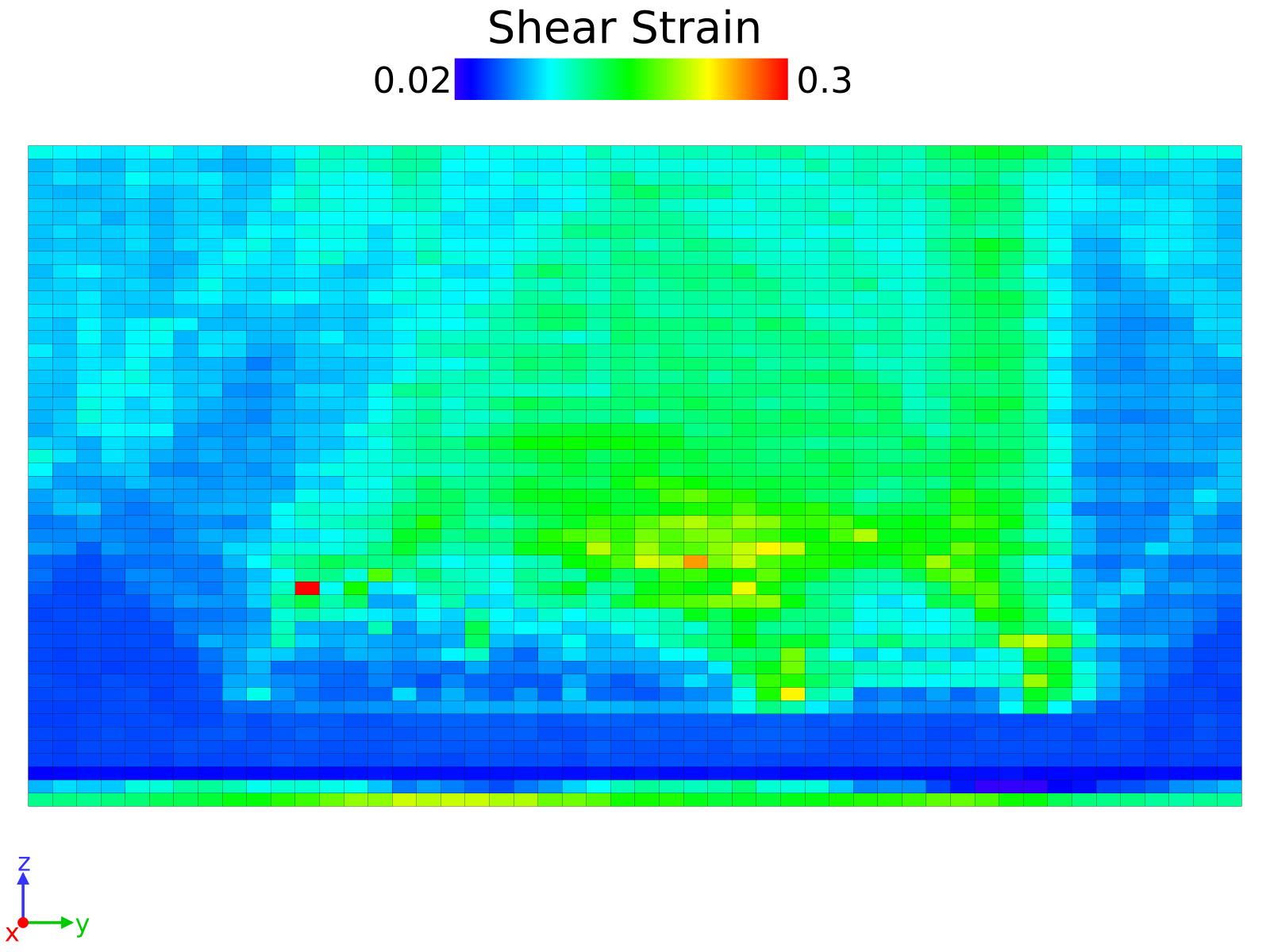}}\\
\subfloat[\hkl(111), $x-y$]{\includegraphics[width=.33\linewidth]{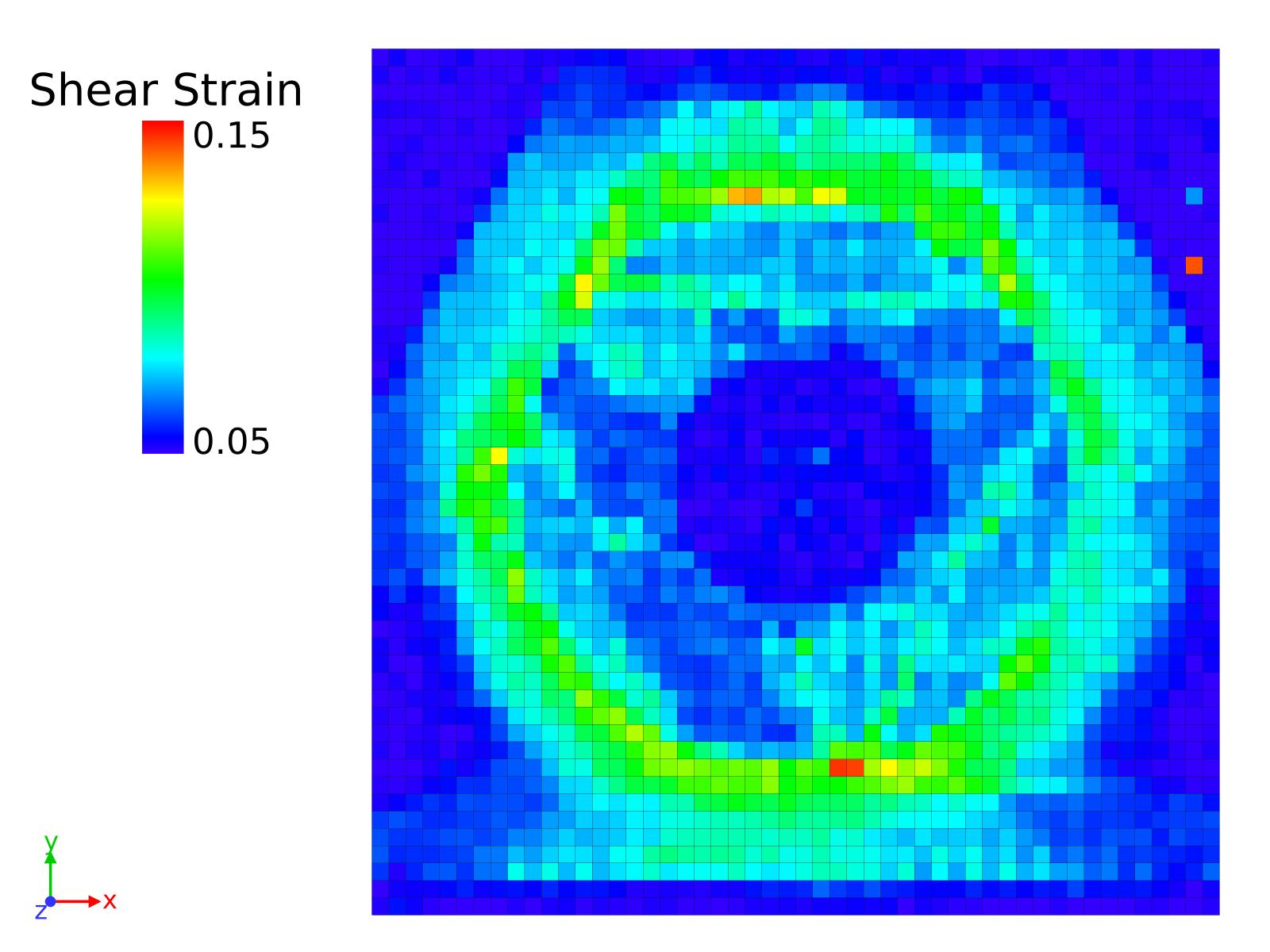}}
\subfloat[\hkl(111) $x-z$]{\includegraphics[width=.33\linewidth]{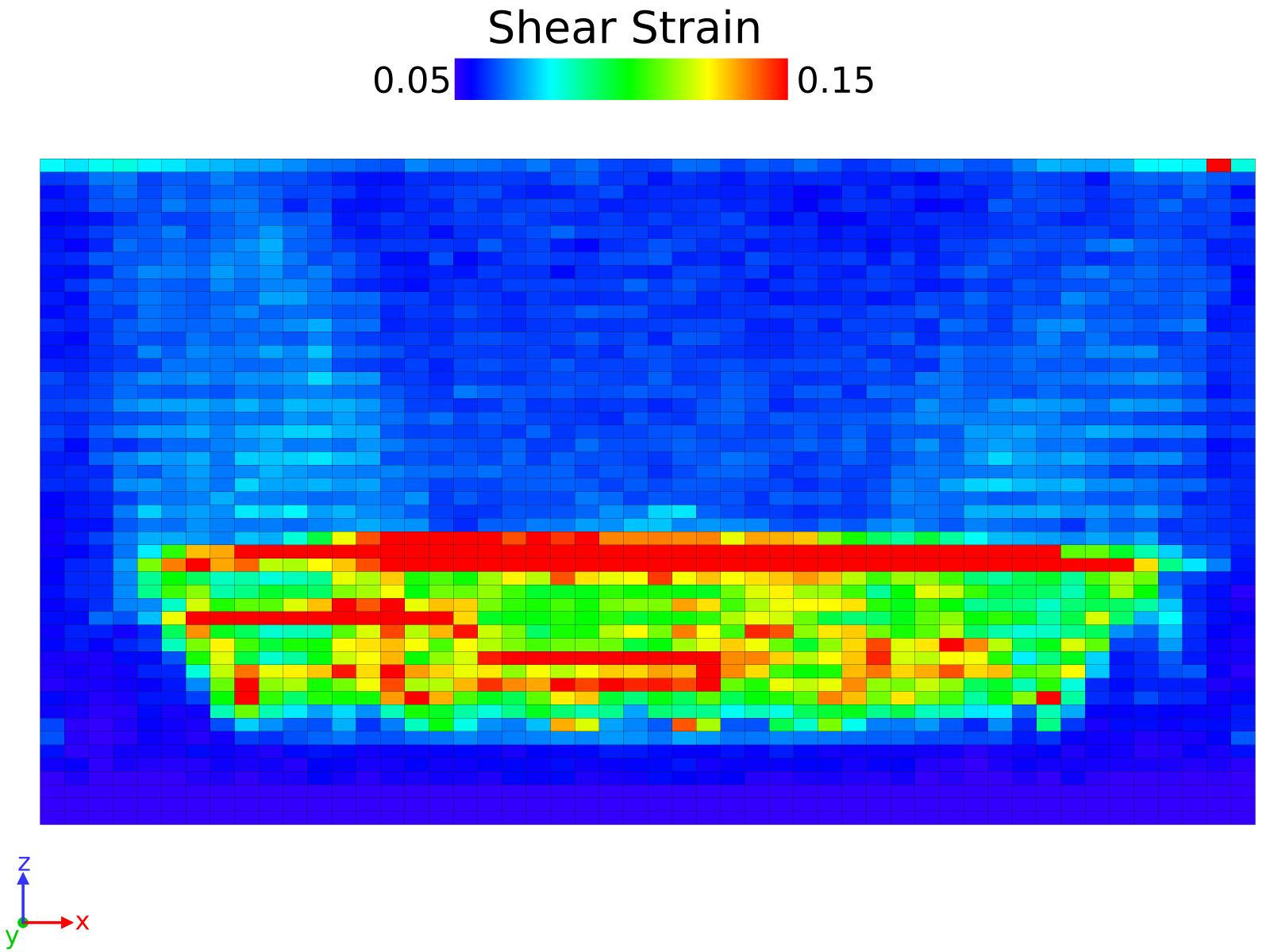}}
\subfloat[\hkl(111) $y-z$]{\includegraphics[width=.33\linewidth]{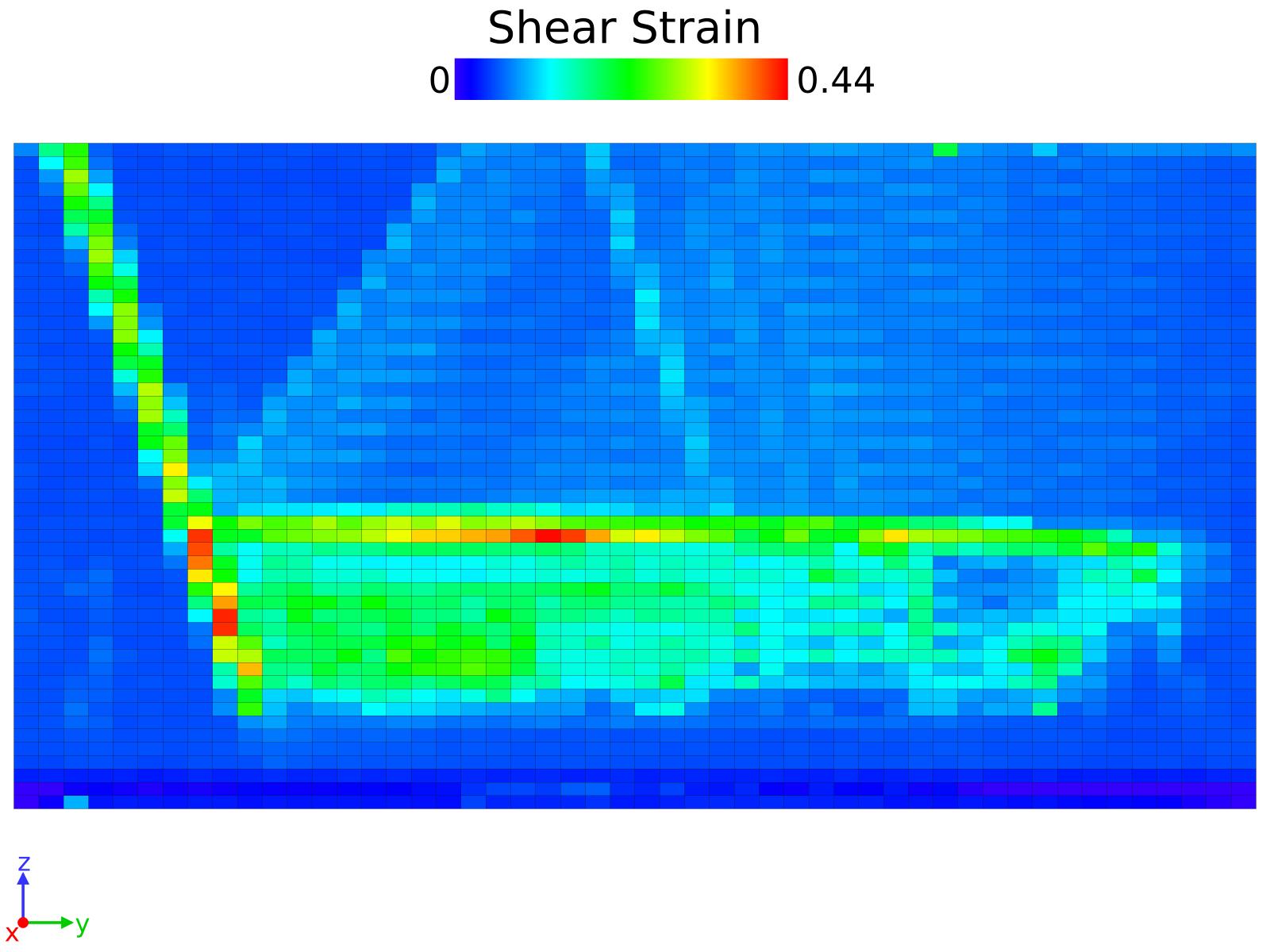}}
\end{center}
\caption{\label{fig:hemisphere_shear_strain_all_histograms} 2-D histograms using 50$\times$50 bins in each direction representing the mean value of shear stress after 100 ps in every bin for \hkl(110) (a-c), \hkl(100) (d-f) and \hkl(111) (g-i) in the x-y, x-z and y-z planes respectively in each row. The cases for the hemispheric bubble are presented.}
\end{figure}

\subsection{Volumetric Strain analysis}

\begin{figure}[H]
\begin{center} 
\subfloat[\hkl(110), $x-y$]{\includegraphics[width=.33\linewidth]{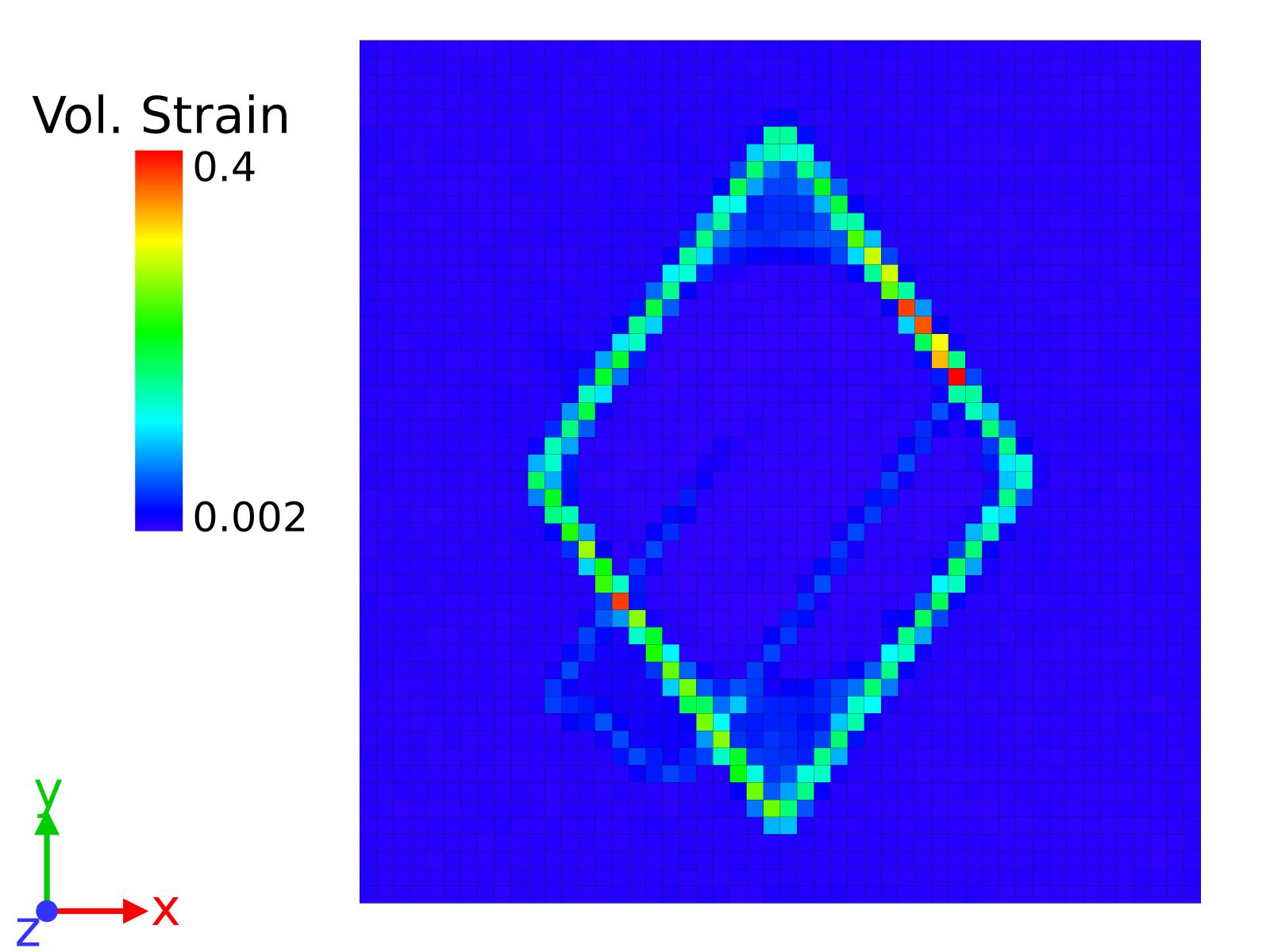}}
\subfloat[\hkl(110), $x-z$]{\includegraphics[width=.33\linewidth]{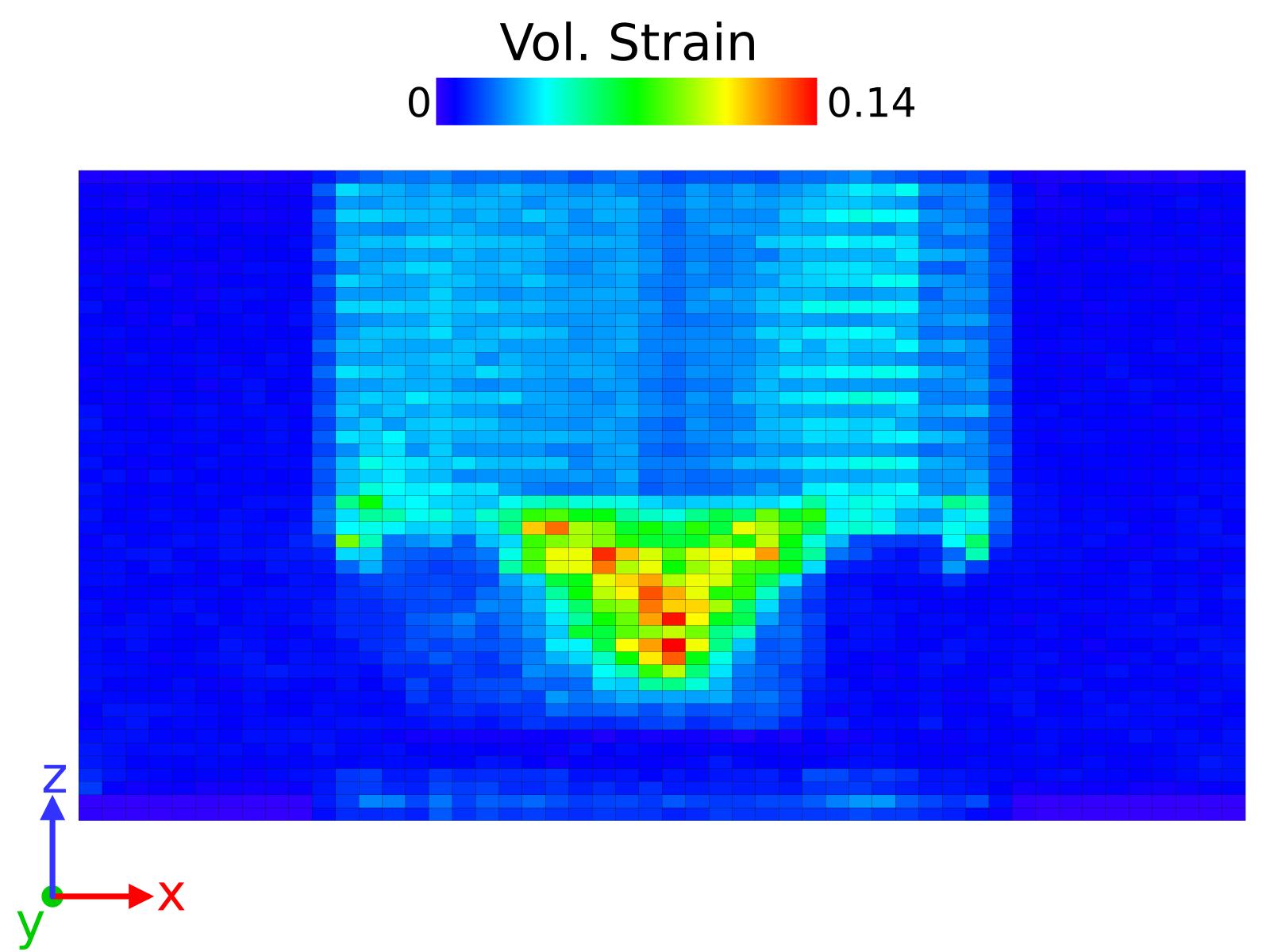}}
\subfloat[\hkl(110) $y-z$]{\includegraphics[width=.33\linewidth]{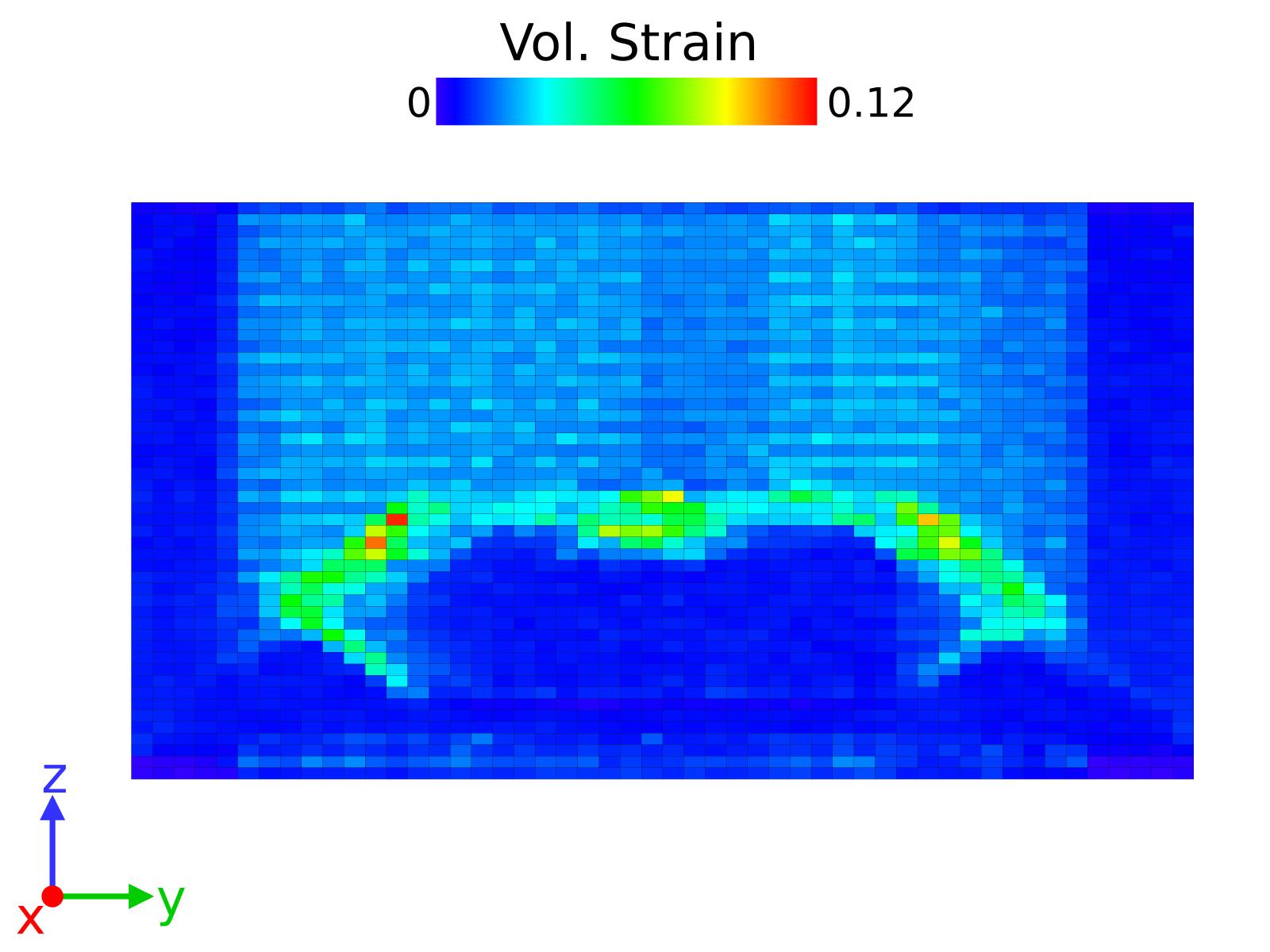}}\\
\subfloat[\hkl(100), $x-y$]{\includegraphics[width=.33\linewidth]{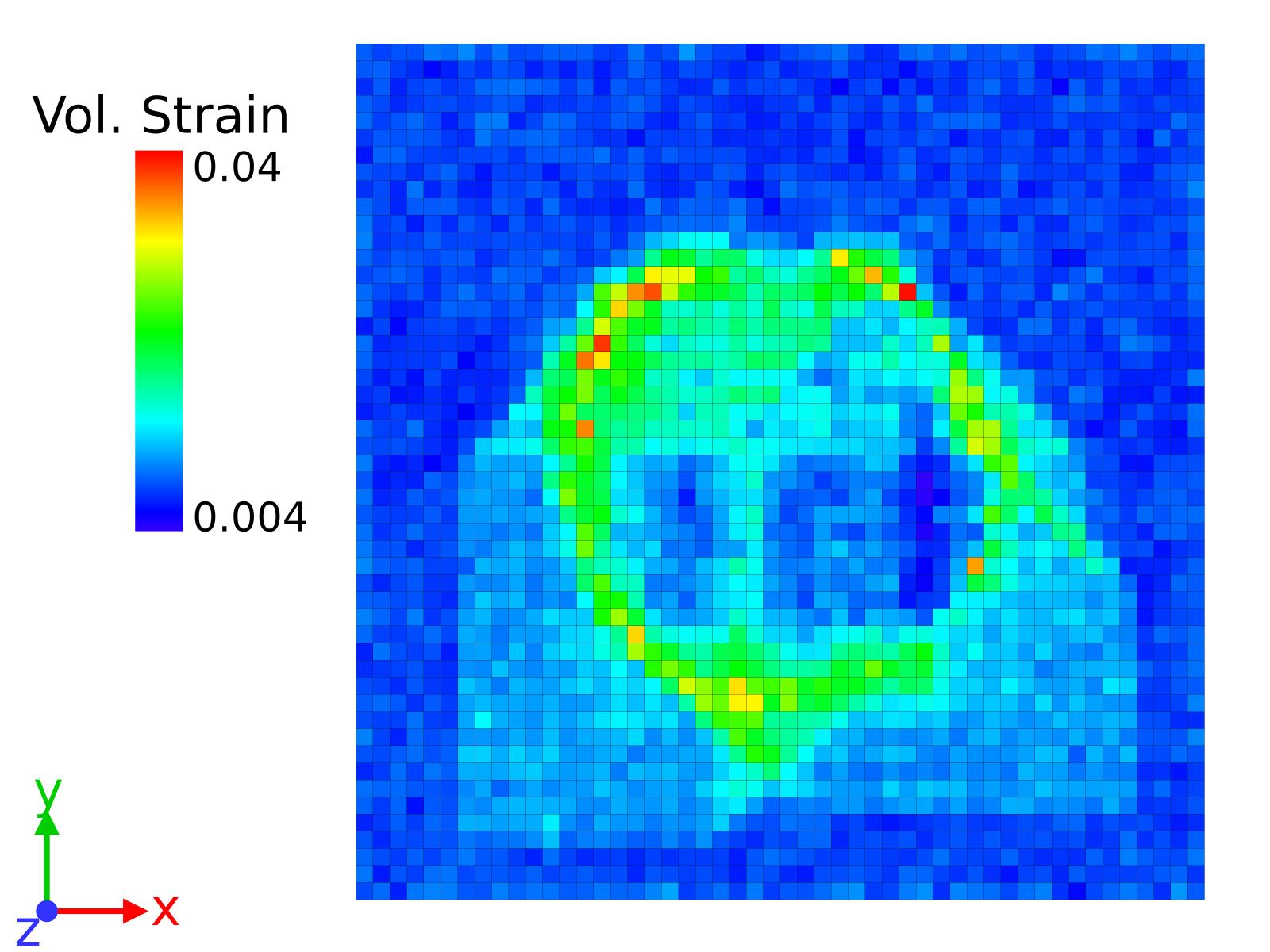}}
\subfloat[\hkl(100) $x-z$]{\includegraphics[width=.33\linewidth]{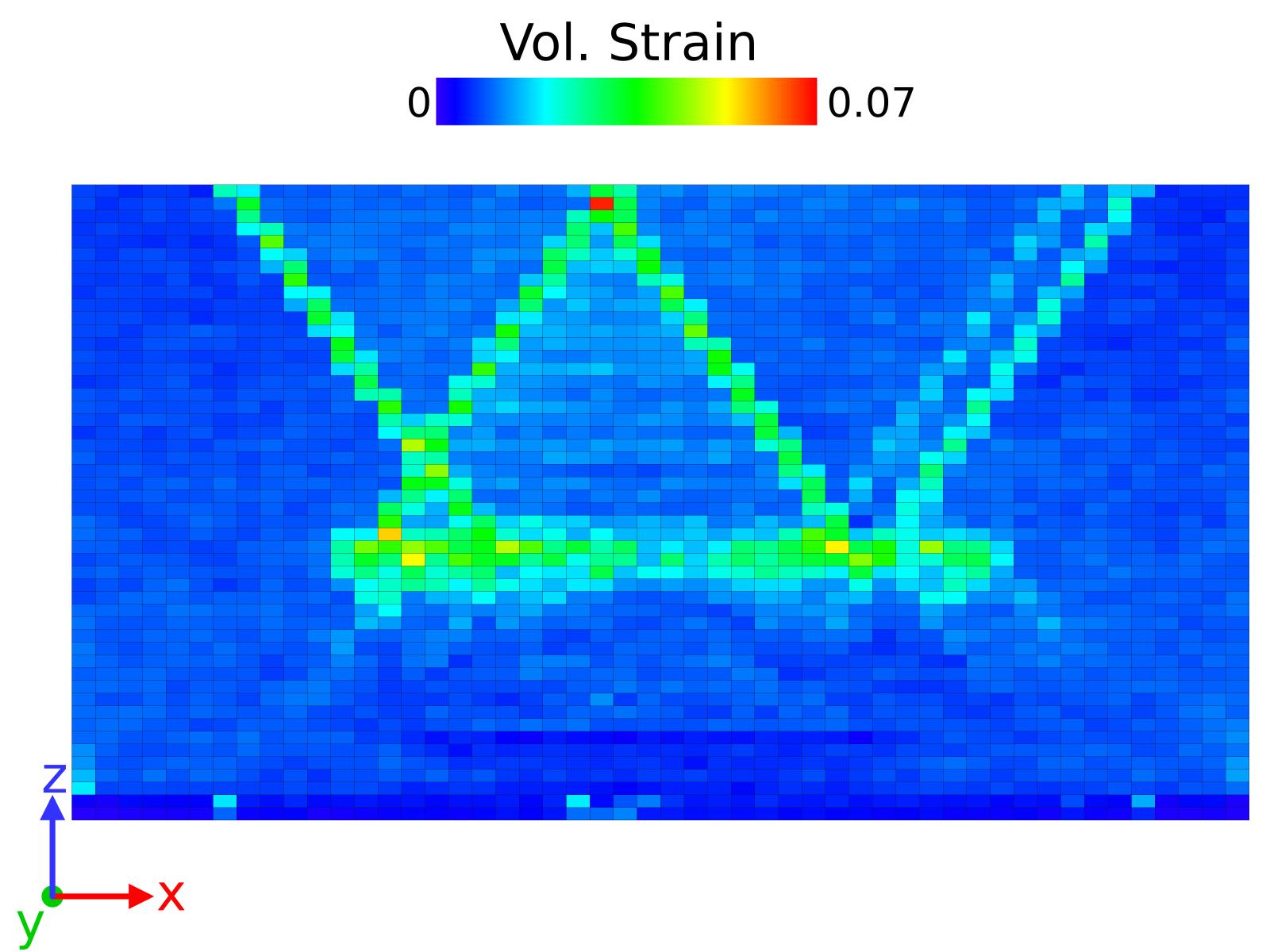}}
\subfloat[\hkl(100) $y-z$]{\includegraphics[width=.33\linewidth]{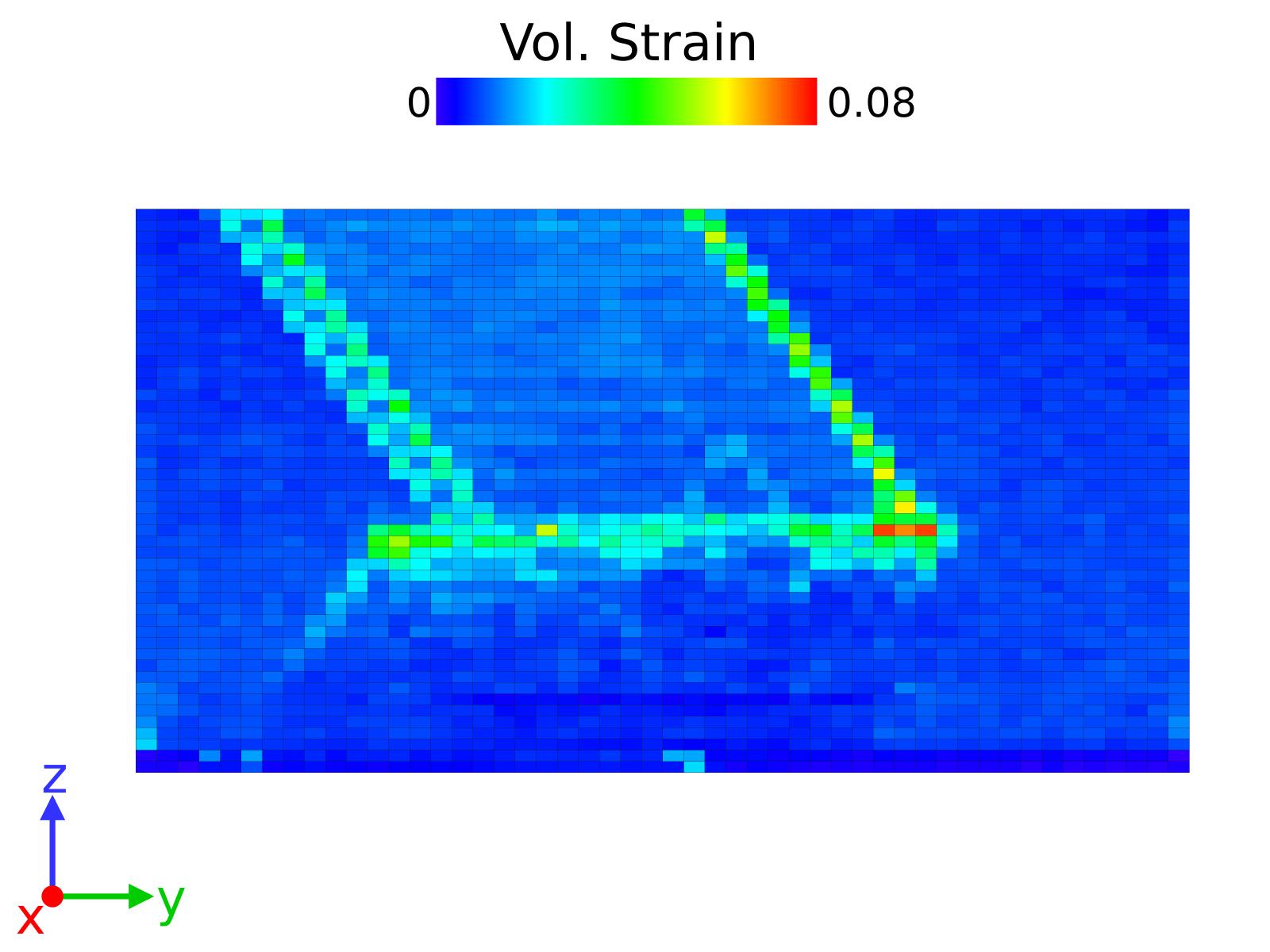}}\\
\subfloat[\hkl(111), $x-y$]{\includegraphics[width=.33\linewidth]{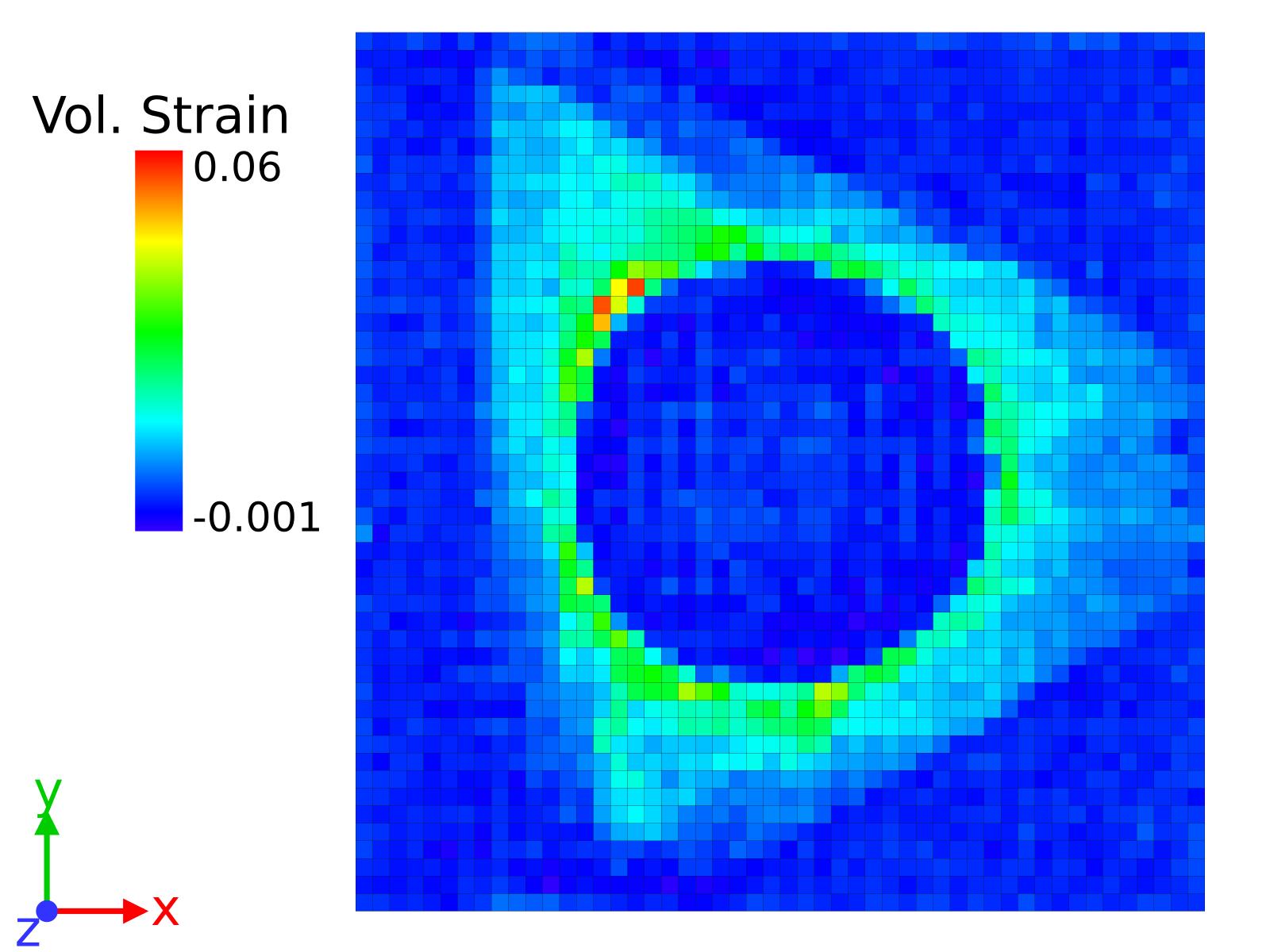}}
\subfloat[\hkl(111) $x-z$]{\includegraphics[width=.33\linewidth]{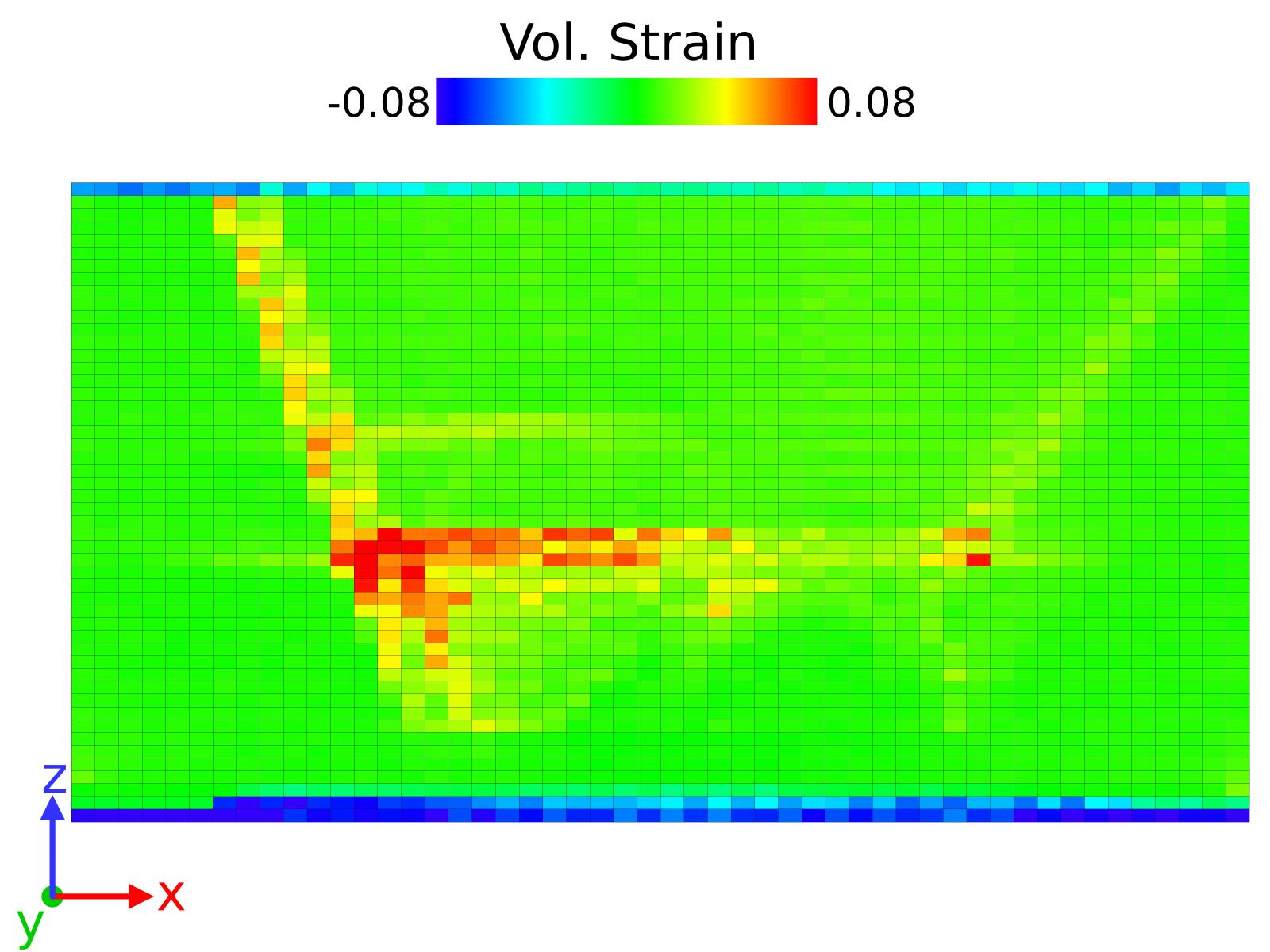}}
\subfloat[\hkl(111) $y-z$]{\includegraphics[width=.33\linewidth]{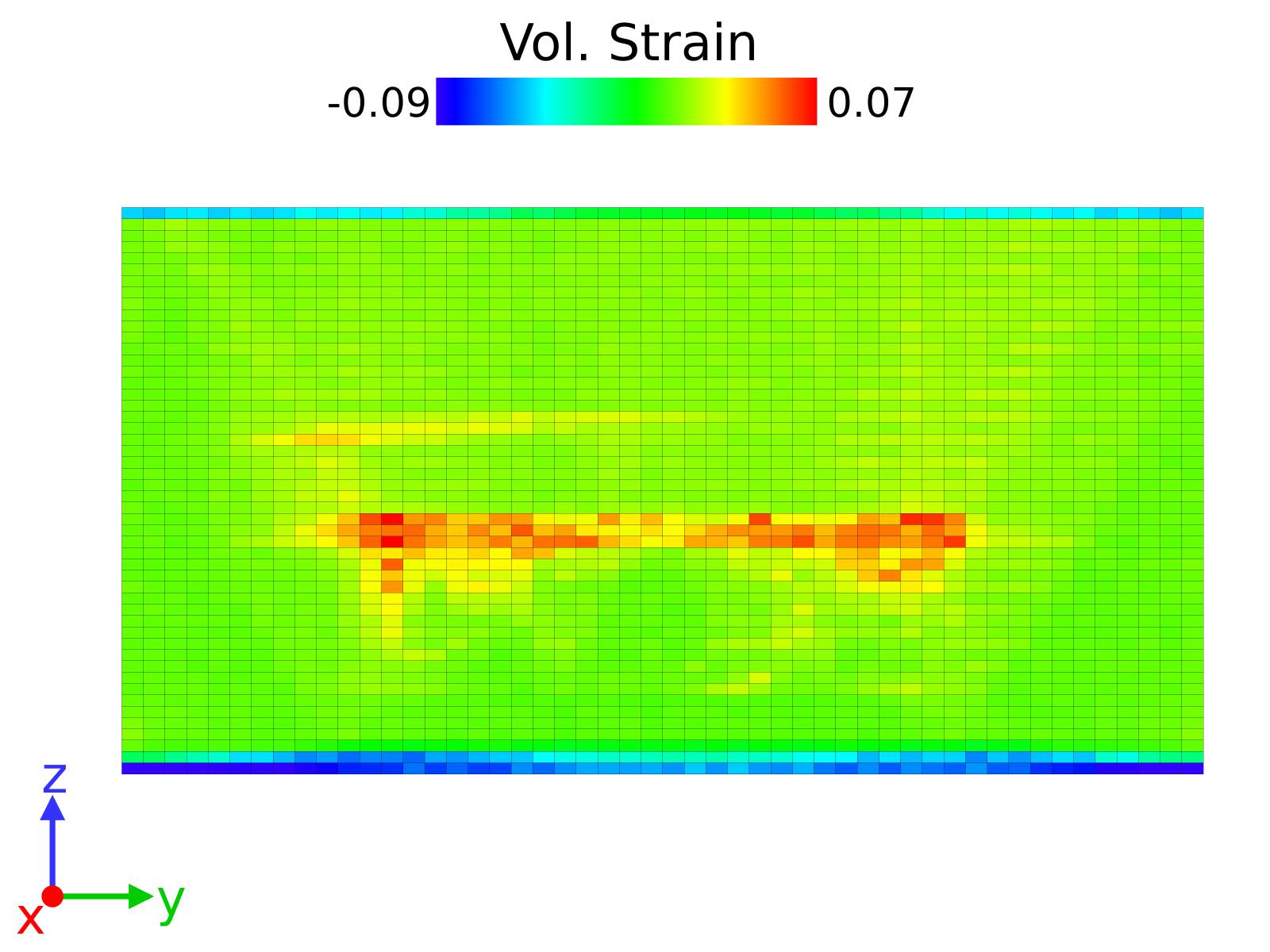}}
\end{center}
\caption{\label{fig:disk_vol_strain_all_histograms} 2-D histograms using 50$\times$50 bins in each direction representing the mean value of volumetric stress after 100 ps in every bin for \hkl(110) (a-c), \hkl(100) (d-f) and \hkl(111) (g-i) in the x-y, x-z and y-z planes respectively in each row. The cases for the disk bubble are presented.}
\end{figure}

\begin{figure}[H]
\begin{center} 
\subfloat[\hkl(110), $x-y$]{\includegraphics[width=.33\linewidth]{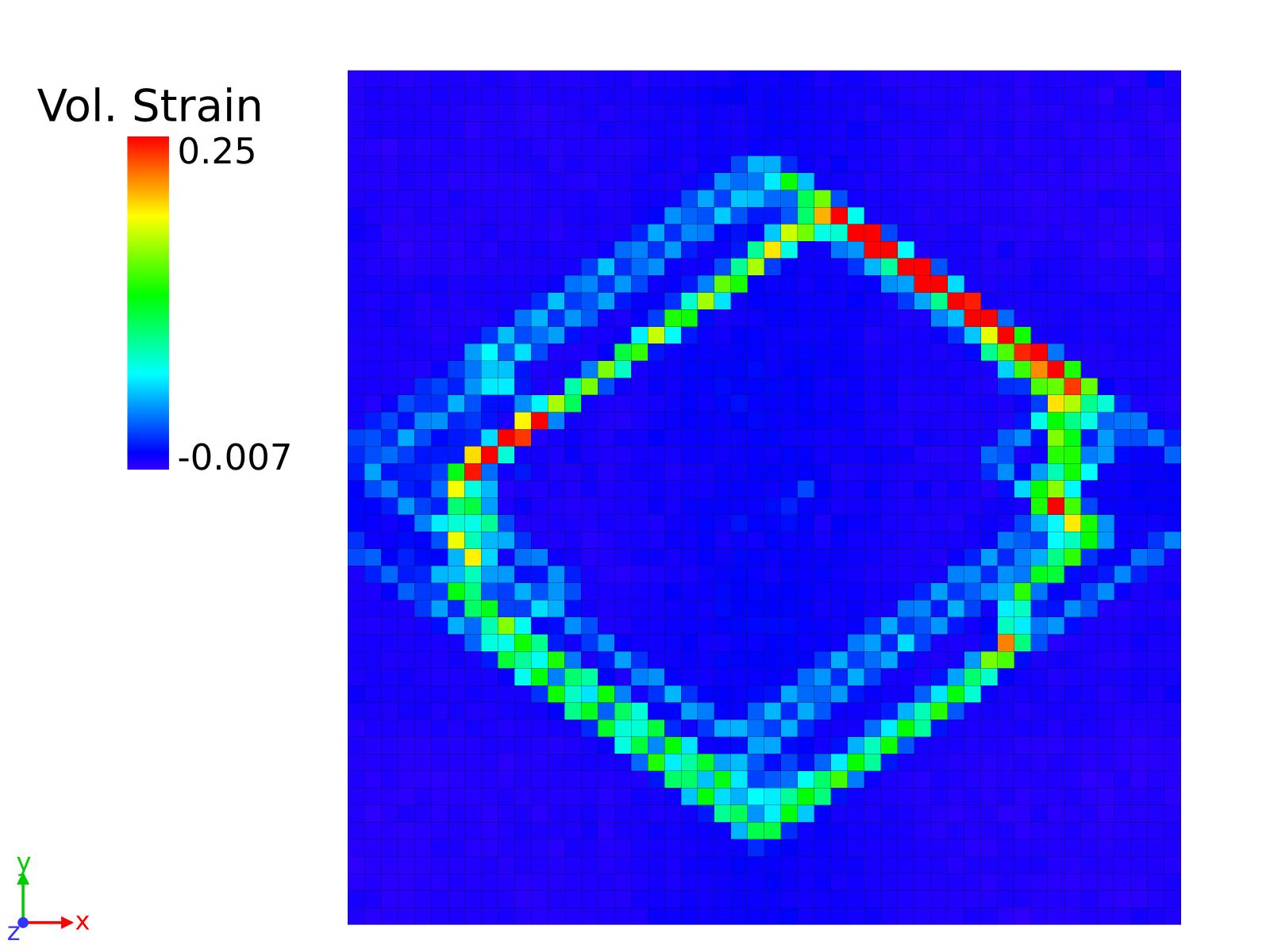}}
\subfloat[\hkl(110), $x-z$]{\includegraphics[width=.33\linewidth]{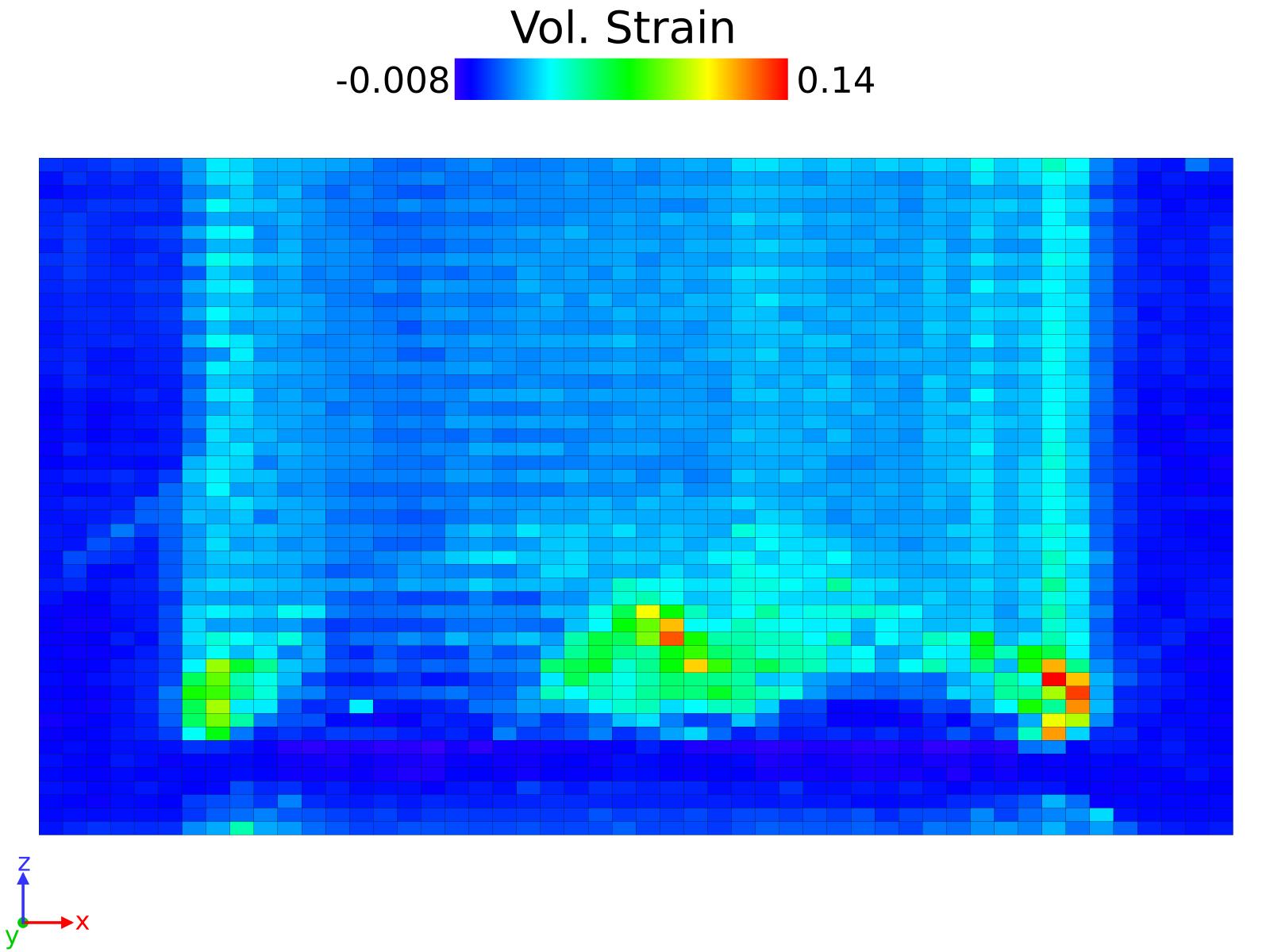}}
\subfloat[\hkl(110) $y-z$]{\includegraphics[width=.33\linewidth]{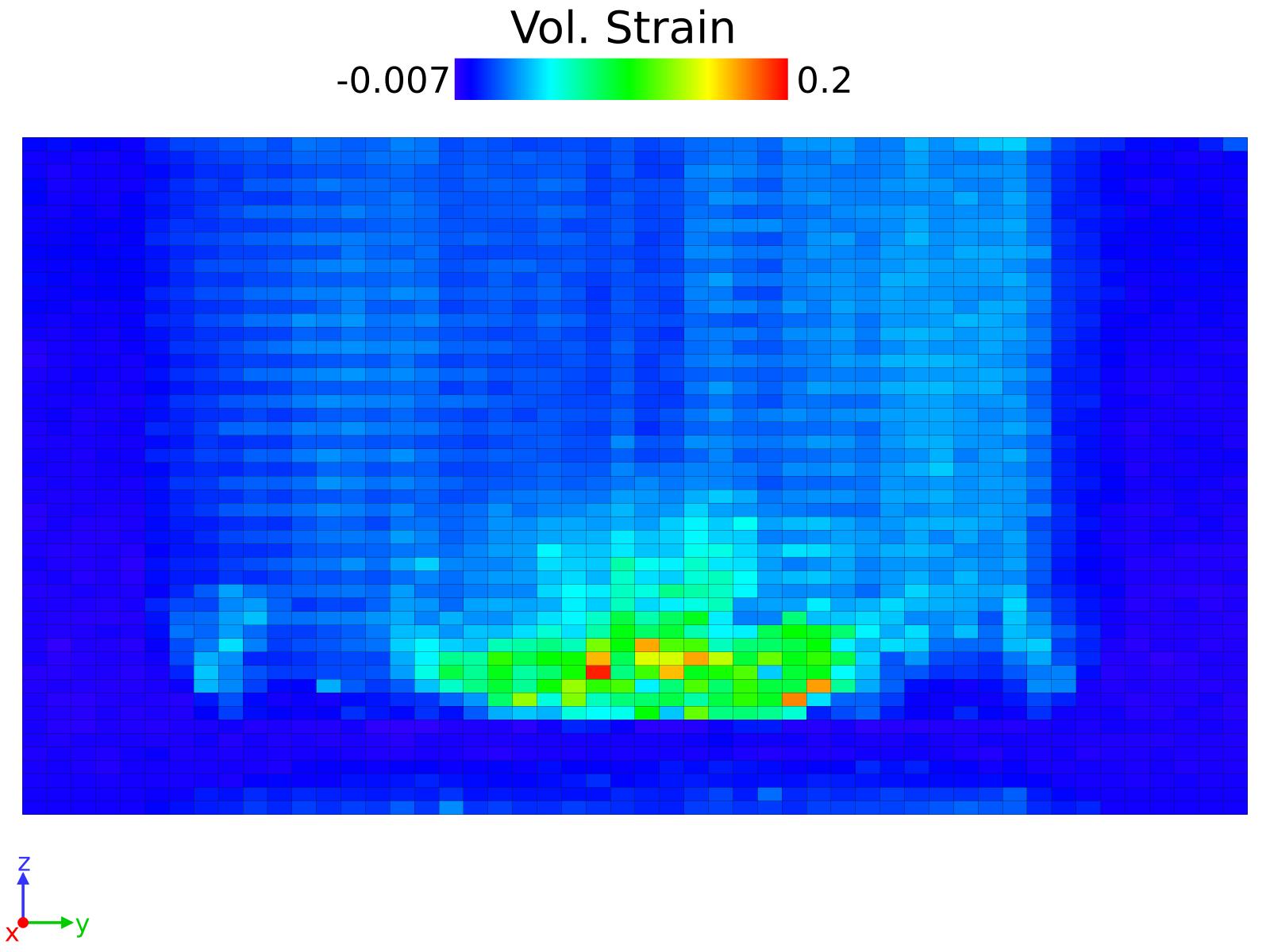}}\\
\subfloat[\hkl(100), $x-y$]{\includegraphics[width=.33\linewidth]{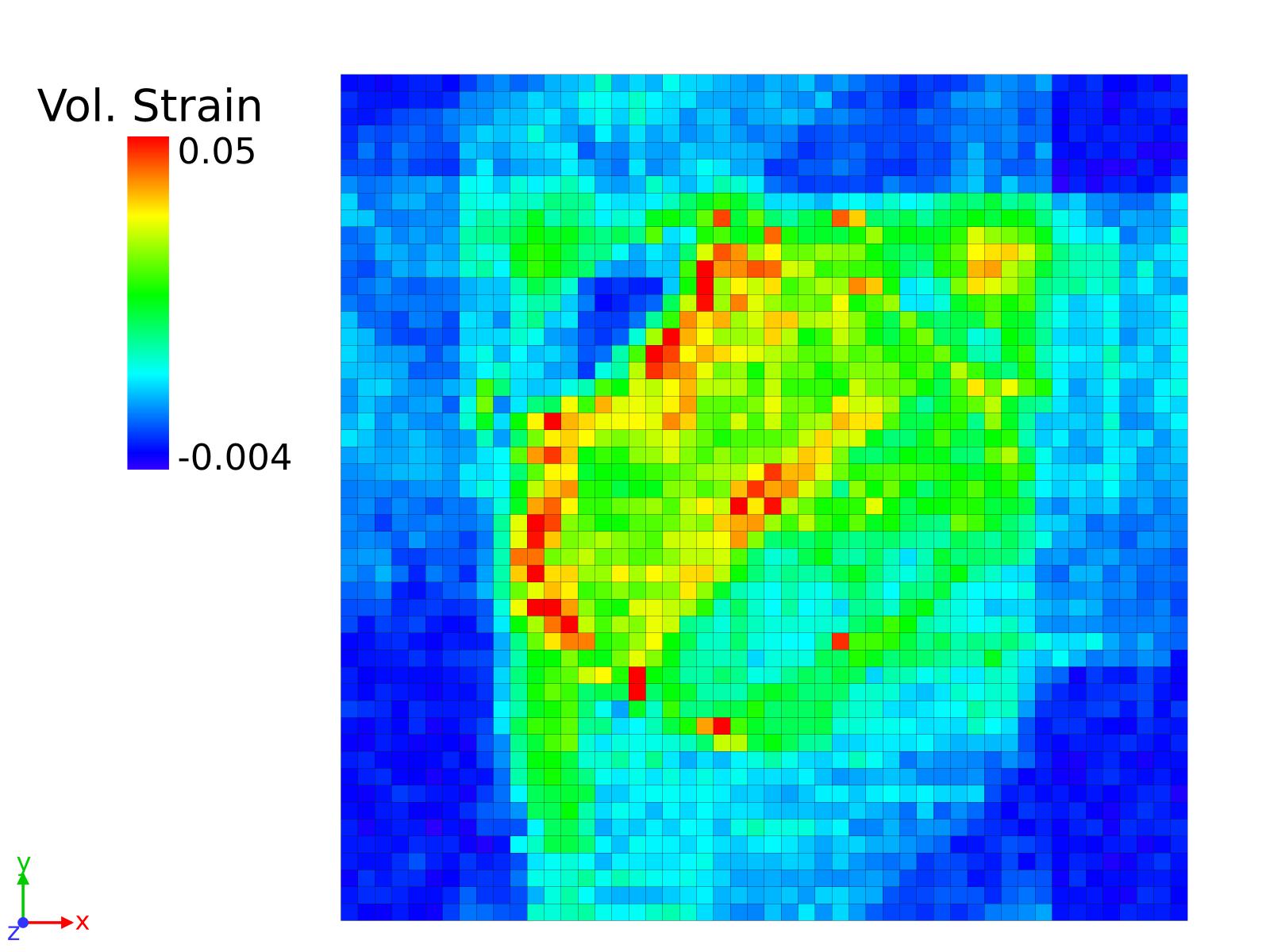}}
\subfloat[\hkl(100) $x-z$]{\includegraphics[width=.33\linewidth]{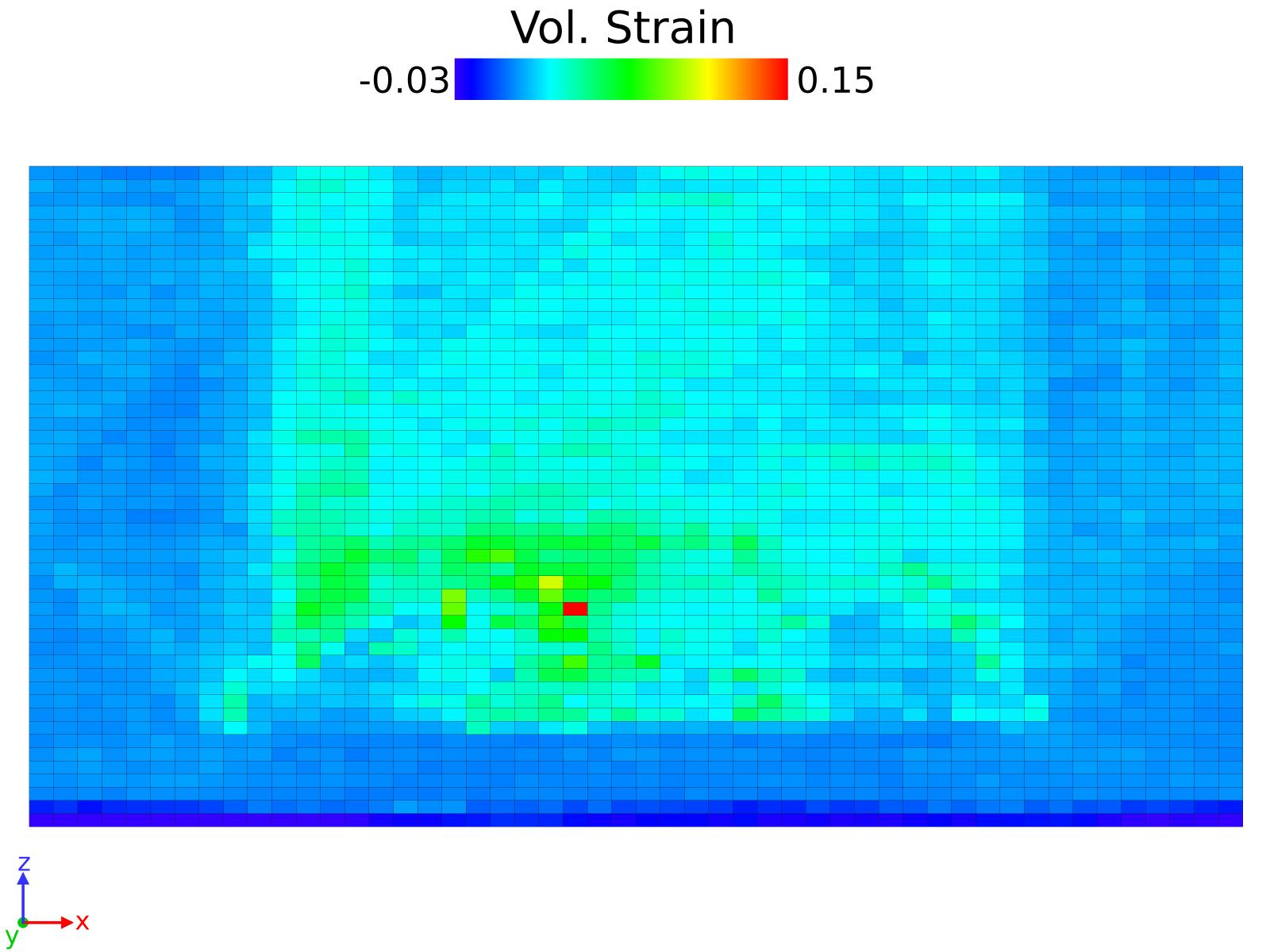}}
\subfloat[\hkl(100) $y-z$]{\includegraphics[width=.33\linewidth]{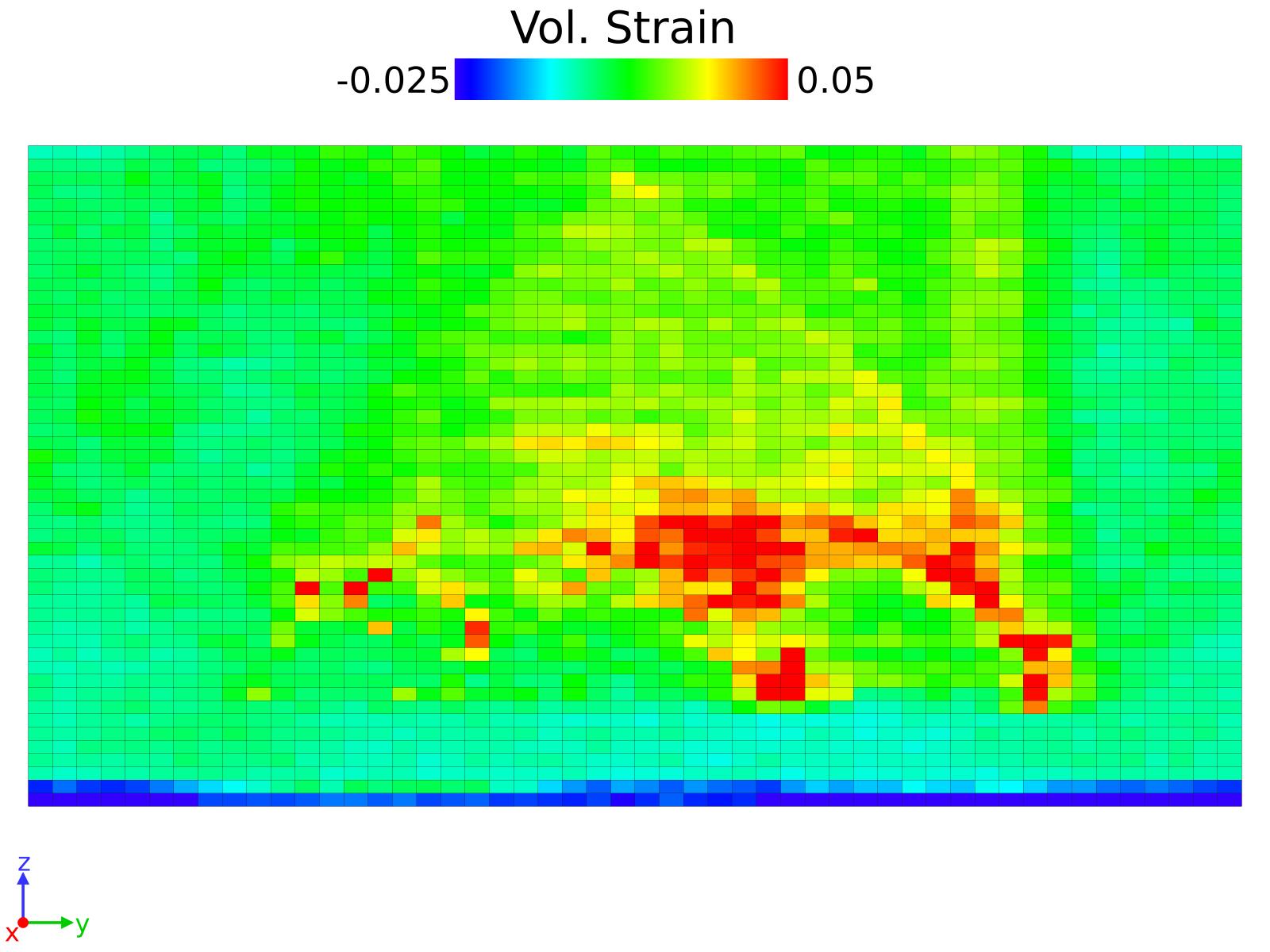}}\\
\subfloat[\hkl(111), $x-y$]{\includegraphics[width=.33\linewidth]{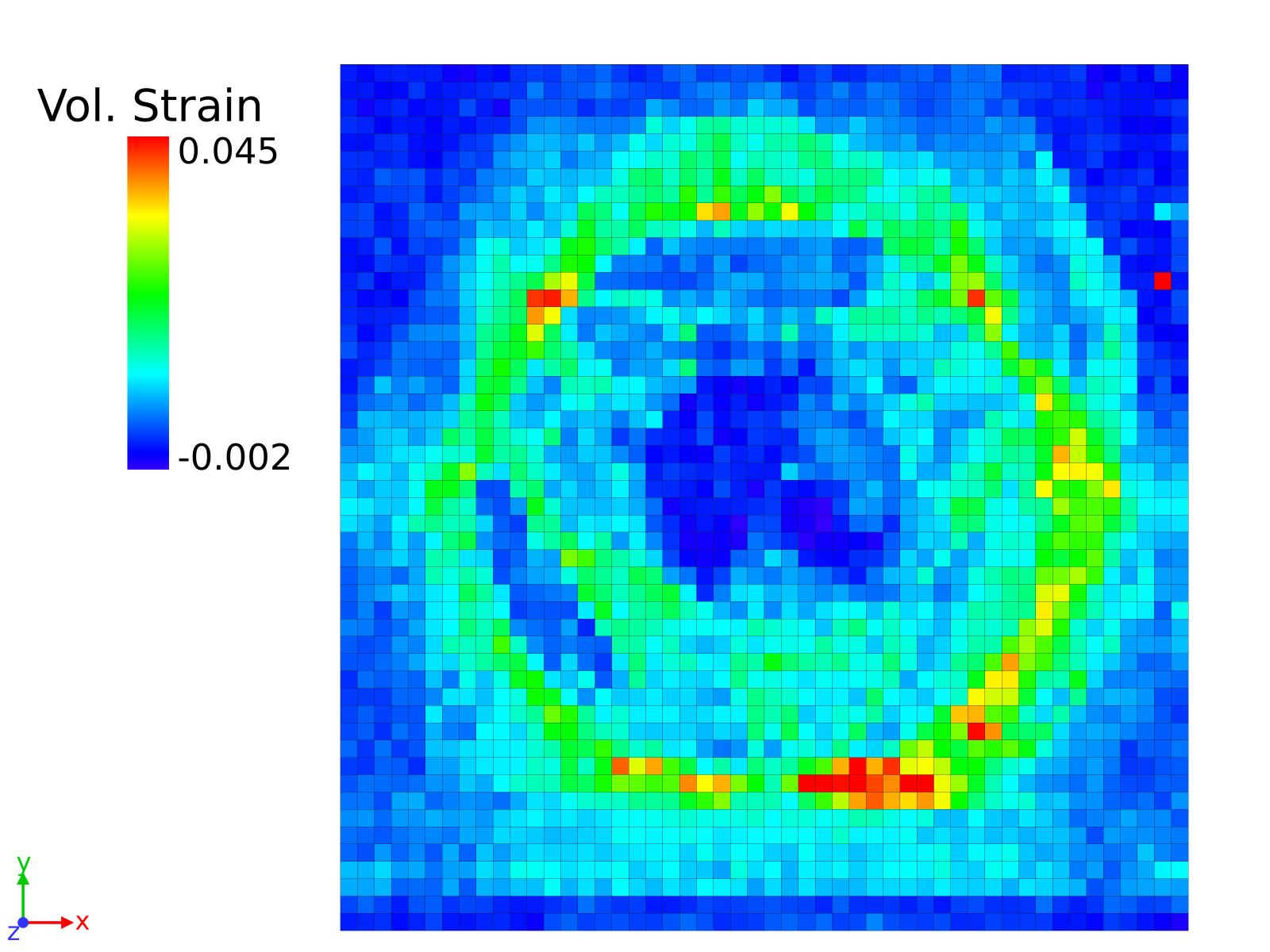}}
\subfloat[\hkl(111) $x-z$]{\includegraphics[width=.33\linewidth]{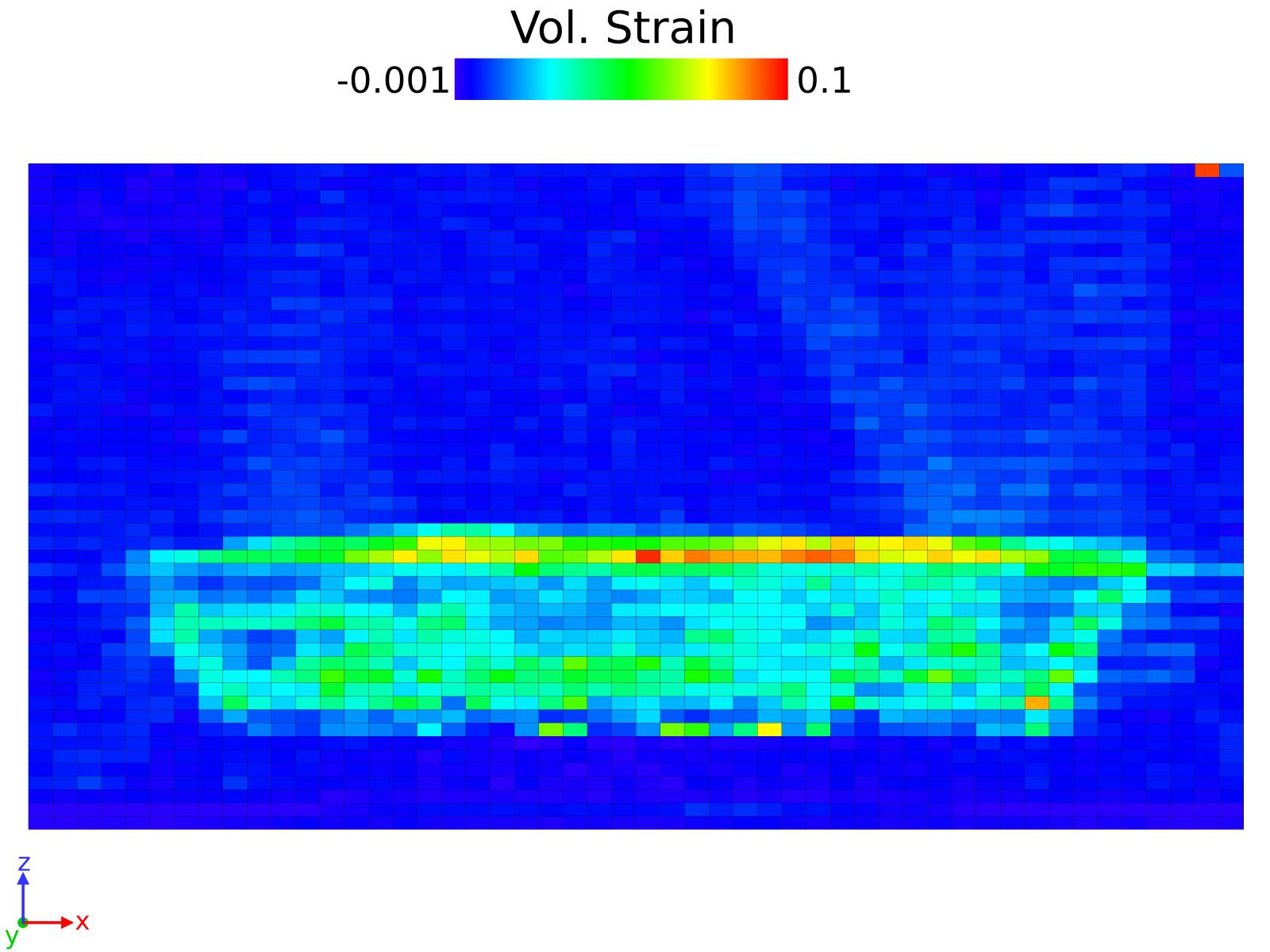}}
\subfloat[\hkl(111) $y-z$]{\includegraphics[width=.33\linewidth]{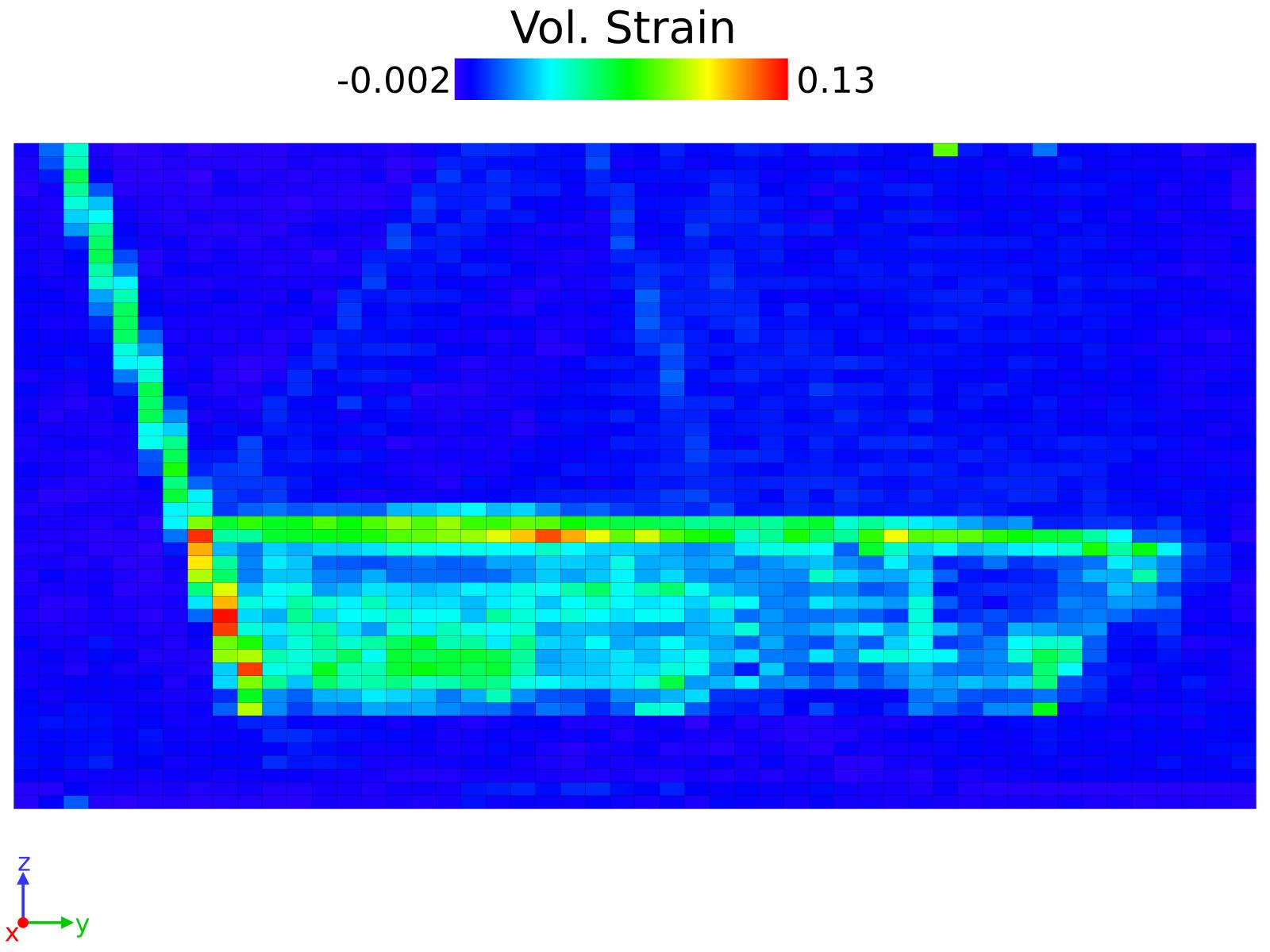}}
\end{center}
\caption{\label{fig:hemisphere_vol_strain_all_histograms} 2-D histograms using 50$\times$50 bins in each direction representing the mean value of volumetric stress after 100 ps in every bin for \hkl(110) (a-c), \hkl(100) (d-f) and \hkl(111) (g-i) in the x-y, x-z and y-z planes respectively in each row. The cases for the hemispheric bubble are presented.}
\end{figure}

\end{document}